\documentclass[10pt]{book}
\usepackage{pgstylestandalone}

\title{Political Geometry}
\author{  }
\date{ }

\begin{document}
\Urlmuskip=0mu plus 1mu\relax 


\chapter[Redistricting Algorithms -- Becker and Solomon]{Redistricting Algorithms}\label{BeckerSolomon}

\chapterauthor{AMARIAH BECKER \\ JUSTIN SOLOMON}
\authorheader{BECKER \& SOLOMON}
\titleheader{Redistricting Algorithms}

\chaptersummary{Why not have a computer just draw a map? This is something you hear a lot when people talk about gerrymandering, and it’s easy to think at first that this could solve redistricting altogether. But there are more than a couple problems with this idea. In this chapter, two computer scientists survey what's been done in algorithmic redistricting, discuss what doesn't work and highlight approaches that show promise. This preprint was prepared as a chapter in the forthcoming edited volume Political Geometry, an interdisciplinary collection of essays on redistricting.  (\url{mggg.org/gerrybook})}

\section{Introduction}\label{sec:intro}

Given frustrations with human behavior when drawing and assessing district boundaries, technologically-minded citizens often call for \emph{algorithms} to serve as unbiased arbiters in the redistricting process.  Projecting this optimism about the objectivity of computers, popular science articles regularly trumpet a programmer who has ``solved gerrymandering in his spare time''~\cite{ingraham2014this} or claim that a ``tech revolution... could fix America's broken voting districts''~\cite{sankin2016the}; one blogger even opined that Google could ``quickly create a neutral, non-gerrymandered election map... in a few weeks''~\cite{grace2017how}.

Enthusiasm for algorithmic redistricting dates back at least to the 1960s.  Even in the early days of computer technology, multiple authors recognized its potential value for redistricting.  In 1961, Vickrey wrote in favor of what he called ``procedural fairness''~\cite{vickrey1961prevention}:
\begin{quote}
	If there is thus no available criterion of substantive fairness, it is necessary, if there is to be any attempt at all to purify the electoral machinery in this respect, to resort to some kind of procedural fairness. This means, in view of the subtle possibilities for favoritism, that the human element must be removed as completely as possible from the redistricting process. In part, this means that the process should be completely mechanical, so that once set up, there is no room at all for human choice. More than this, it means that the selection of the process, which must itself be at least initiated by human action, should be as far removed from the results of the process as possible, in the sense that it should not be possible to predict in any detail the outcome of the process.
\end{quote}

Writing  at nearly the same time, Edward Forrest~\cite{forrest1964apportionment} advocated for using computers for unbiased redistricting in a behavioral science journal in 1964: 
\begin{quote}
    Since the computer doesn't know how to gerrymander---because two plus two always equals four---the electronically generated map can't be anything but unbiased. Its validity is immediately acceptable to responsible political leaders and the courts.
\end{quote}  

Since that time, many algorithms have been designed to support redistricting.  Software has become a ubiquitous partner in the design and analysis of districting plans, including sophisticated tools that leverage census data, electoral returns, and highly detailed maps.

The introduction of computer technology brings a new set of ethical and philosophical---as well as mathematical---challenges to redistricting.  Modern algorithms make it possible to engineer districts with remarkable precision and control, providing opportunities to gerrymander in subtle ways.  Questions of allowable data and procedures are complicating long-standing conversations about ``traditional districting criteria."  On the technical side, fundamental limits involving the computational complexity of certain redistricting problems reveal that Vickrey's and Forrest's dream of perfect redistricting through algorithms may be practically unrealizable.

On the other hand, recent progress has made algorithms into very promising partners in redistricting reform.  
This chapter explores the ways in which computing has been used in redistricting and presents a survey of redistricting algorithms.  We categorize algorithms into those that are used to \emph{generate} new plans and those that are used to \emph{assess} proposed plans.  
We conclude with general discussion about best practices for algorithmic tools in the redistricting process.

\subsection{What is an Algorithm?}\label{sec:prelims}

An \emph{algorithm} is a procedure or set of instructions that a computer uses to solve a problem.  Generally, algorithms take \emph{input} data and produce an \emph{output} solution.  For example, an algorithm that generates a districting plan might take as input the populations and geographies of census units (e.g., precincts, census blocks, census block groups) as well as the desired number of districts and produce as output a plan listing which census units are assigned to each district.  

While a computer may ultimately be carrying out an algorithm, \emph{humans} write the instructions and make the algorithmic design decisions.  For example, for computers to help identify \emph{good} districting plans, a human must first define what it means for one plan to be better than another.  A computer has no built-in method for assessing plans or anything else; it simply follows the user's instructions.  In this sense, an algorithm or piece of software easily can inherit the biases and assumptions of its human designer.

Our chapter focuses on techniques that generate district boundaries, and we leave the discussion of how to use those for other parts of the book \cite{book}. 
\emph{Enumeration} algorithms list every possible way to district a given piece of geography.  These algorithms have the advantage that no stone is left unturned, but even small municipalities can have unfathomably huge numbers of possible plans, putting this hope out of practical reach.  So this section will be brief.
We divide the subsequent discussion into two variants of this problem:
\begin{itemize}
    \item \emph{Sampling} algorithms also generate collections of districting plans, but the intention is not to be exhaustive.   When carried out well, these methods can provide an overview of the properties of possible plans.
    \item \emph{Optimization} algorithms attempt to identify a single plan that extremizes some quality score.  These automated redistricting tools are effective in some scenarios but require everyone to agree on a single scoring function---a difficult task since so many metrics are used to evaluate proposed districting plans. 
\end{itemize}
We will examine various algorithms aimed at sampling or optimization, including scenarios where they are likely to perform well or poorly.

When it comes to quality, some of the algorithms described in this chapter 
are specifically built around particular measures (e.g., Plan $A$ is better than Plan $B$ if and only if it has a smaller population deviation), while others allow the user to specify a score function (e.g., $\alpha\cdot \text{county splits} + \beta\cdot\text{population deviation}$ where $\alpha$ and $\beta$ are left as user parameters).  The decision of how to weight various measures when comparing plans is expressly human, and the consequences of adjusting the weighting even a little can be drastic. For the rest of this chapter, unless otherwise noted, comparative terms like \emph{better plan} and \emph{best plan} are assumed to be with respect to whichever plan score is being used, and should not be interpreted as 
endorsements of fairness.

\subsection{Comparing Algorithms}

Algorithms are primarily analyzed by two considerations: the \emph{quality} of the solutions they identify and their \emph{efficiency}.  The former addresses how good or usable an output is, and the latter addresses how long it takes to generate output.  Usually there is a trade-off between quality and runtime: it takes more time to find better solutions.  
Typically, however, one algorithm will outperform another for some problem instances but not for others.  Moreover, many of the algorithms we present are designed to accomplish slightly different objectives from each other and may not be suitable for direct comparison.  Ultimately, which algorithm is \emph{right} or \emph{best} depends on the priorities of the user. 

In redistricting, major properties of interest for sampling algorithms include whether the algorithm is efficient enough to be practical on (large) real-world instances, whether it is actually used as a sampler in practice, whether it generates or can generate every valid plan, whether it tends to generate more compact plans (with nicely shaped districts), and whether it targets a known probability distribution (i.e., if it can be tuned to weight a certain plan more than than another by a factor that we control).  Table~\ref{table:sampling_summary} summarizes several sampling algorithms presented in this chapter along these axes. For some algorithms the given property is true with a caveat (indicated with yellow squares), explained in the caption.

\begin{table}[ht]
\centering
 \begin{tabular}{r|c|c|c|c|c|c|}
 \multicolumn{1}{c}{}& 
 \multicolumn{1}{c}{\rlap{\rotatebox{45}{Generates \emph{every} valid plan}}}&
 \multicolumn{1}{c}{\rlap{\rotatebox{45}{\emph{Can} generate any valid plan}}}&
 \multicolumn{1}{c}{\rlap{\rotatebox{45}{Can sample real-world instances}}}&
 \multicolumn{1}{c}{\rlap{\rotatebox{45}{Used as sampler in practice}}}&
 \multicolumn{1}{c}{\rlap{\rotatebox{45}{Promotes compact plans}}}&
 \multicolumn{1}{c}{\rlap{\rotatebox{45}{Targets a known distribution}}}\\
 \cline{2-7}
Enumeration  &\cellcolor{dollarbill}\checkmark&\cellcolor{dollarbill}\checkmark&\cellcolor{tablered}--&\cellcolor{tablered}--&\cellcolor{tablered}--&\cellcolor{dollarbill}\checkmark\\
\cline{2-7}
Random-Unit Assignment&\cellcolor{tablered}--&\cellcolor{dollarbill}\checkmark&\cellcolor{tablered}--&\cellcolor{tablered}--&\cellcolor{tablered}--&\cellcolor{dollarbill}\checkmark\\
\cline{2-7}
Flood Fill &\cellcolor{tablered}--&\cellcolor{dollarbill}\checkmark&\cellcolor{tablered}--&\cellcolor{tablered}--&\cellcolor{bananayellow}\checkmark&\cellcolor{tablered}--\\\cline{2-7}
Iterative Merging &\cellcolor{tablered}--&\cellcolor{tablered}--&\cellcolor{dollarbill}\checkmark&\cellcolor{dollarbill}\checkmark&\cellcolor{dollarbill}\checkmark&\cellcolor{tablered}--\\\cline{2-7}
Flip Step Walk &\cellcolor{tablered}--&\cellcolor{bananayellow}\checkmark&\cellcolor{dollarbill}\checkmark&\cellcolor{dollarbill}\checkmark&\cellcolor{bananayellow}\checkmark&\cellcolor{bananayellow}\checkmark\\\cline{2-7}
Recombination Walk &\cellcolor{tablered}--&\cellcolor{bananayellow}\checkmark&\cellcolor{dollarbill}\checkmark&\cellcolor{dollarbill}\checkmark{}&\cellcolor{dollarbill}\checkmark{}&\cellcolor{dollarbill}\checkmark\\\cline{2-7}
Power Diagrams &\cellcolor{tablered}--&\cellcolor{tablered}--&\cellcolor{bananayellow}\checkmark&\cellcolor{tablered}--&\cellcolor{dollarbill}\checkmark&\cellcolor{tablered}--\\\cline{2-7}
 \end{tabular}
 \caption{
 General properties of redistricting sampling algorithms; note each of these methods admits many variations that may disagree with this table.
Caveats are indicated in yellow: power diagrams can handle large instances but generate geometric partitions rather than plans built from census units; the ability of flip step walks and recombination walks to generate \emph{any} valid plan depends on the particular redistricting constraints and underlying geography; some flood fill variants promote compactness; and flip step walks can target particular distributions (including ones that favor compactness) but often lack evidence of convergence. 
 \label{table:sampling_summary}}
 \end{table}
\begin{table}[t]
\centering
 \begin{tabular}{r|c|c|c|c|}
 \multicolumn{1}{c}{}& 
 \multicolumn{1}{c}{\rlap{\rotatebox{45}{Practical at scale}}}&
 \multicolumn{1}{c}{\rlap{\rotatebox{45}{Finds\ global\ optima}}}&
 \multicolumn{1}{c}{\rlap{\rotatebox{45}{Finds\ local\ optima}}}&
 \multicolumn{1}{c}{\rlap{\rotatebox{45}{Customizable}}} \\
 \cline{2-5}
Enumeration  &\cellcolor{tablered}--&\cellcolor{dollarbill}\checkmark&\cellcolor{dollarbill}\checkmark&\cellcolor{dollarbill}\checkmark\\\cline{2-5}
Power Diagrams &\cellcolor{bananayellow}\checkmark&\cellcolor{tablered}--&\cellcolor{dollarbill}\checkmark&\cellcolor{tablered}--\\\cline{2-5}
Metaheuristics/Random Walks &\cellcolor{bananayellow}\checkmark&\cellcolor{tablered}--&\cellcolor{dollarbill}\checkmark&\cellcolor{dollarbill}\checkmark\\\cline{2-5}
Integer Programming &\cellcolor{tablered}--&\cellcolor{dollarbill}\checkmark&\cellcolor{dollarbill}\checkmark&\cellcolor{dollarbill}\checkmark\\
\cline{2-5}
 \end{tabular}
 \caption{ General properties of redistricting optimization algorithms; note each of these methods admits many variations that may disagree with this table.
 Caveats: power diagrams are geometric and do not directly generate plans built from census units; and metaheuristics are not guaranteed to be practical, but are often efficient in practice.
 \label{table:opt_summary}}
 \end{table}

For optimization algorithms, in addition to whether it is efficient enough to be practical for real-world instances, we are further interested in whether the algorithm identifies global optima or only local optima and whether the algorithm can handle customized scoring (objective) functions. Table~\ref{table:opt_summary} summarizes these considerations for several optimization algorithms presented in this chapter. A key takeaway highlighted in this table is that there are no algorithms that are efficient enough to use in practice that can identify \emph{best possible} plans (global optima) for non-trivial scoring functions.  This makes it difficult to assess solution quality: If we do not know the best possible score, we have no yardstick with which to measure other solutions.

\subsubsection{Benchmarks}

\begin{figure} 
\centering
    \begin{subfigure}[h]{0.23\textwidth}
      \includegraphics[width=\textwidth,trim={2cm 1.5cm 2cm 1.5cm},clip]{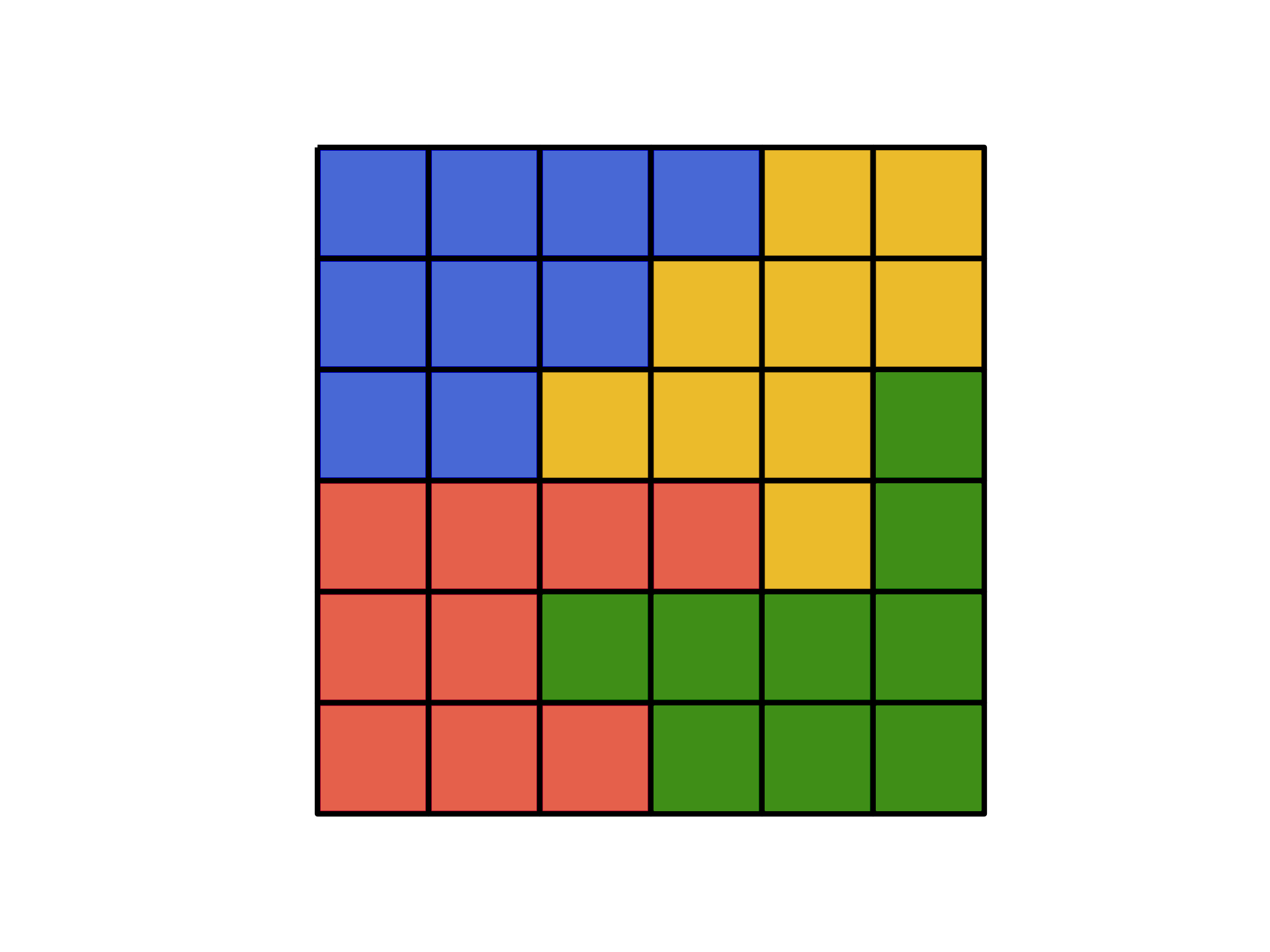}
    \end{subfigure}
    \begin{subfigure}[h]{0.23\textwidth}
      \includegraphics[width=\textwidth,trim={2cm 1.5cm 2cm 1.5cm},clip]{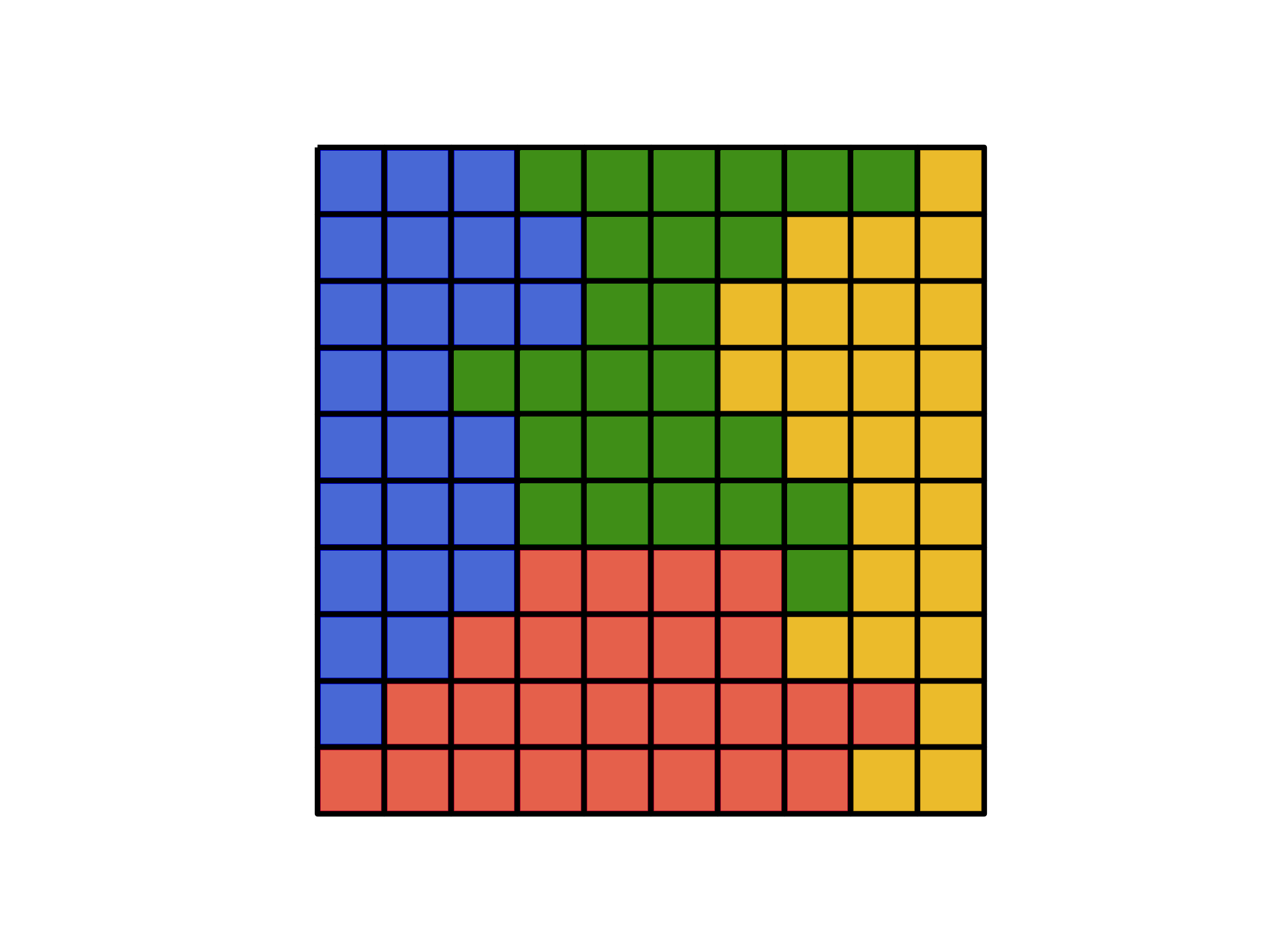}
    \end{subfigure}
    \begin{subfigure}[h]{0.23\textwidth}
      \includegraphics[width=\textwidth,trim={2cm 2cm 2cm 2cm},clip]{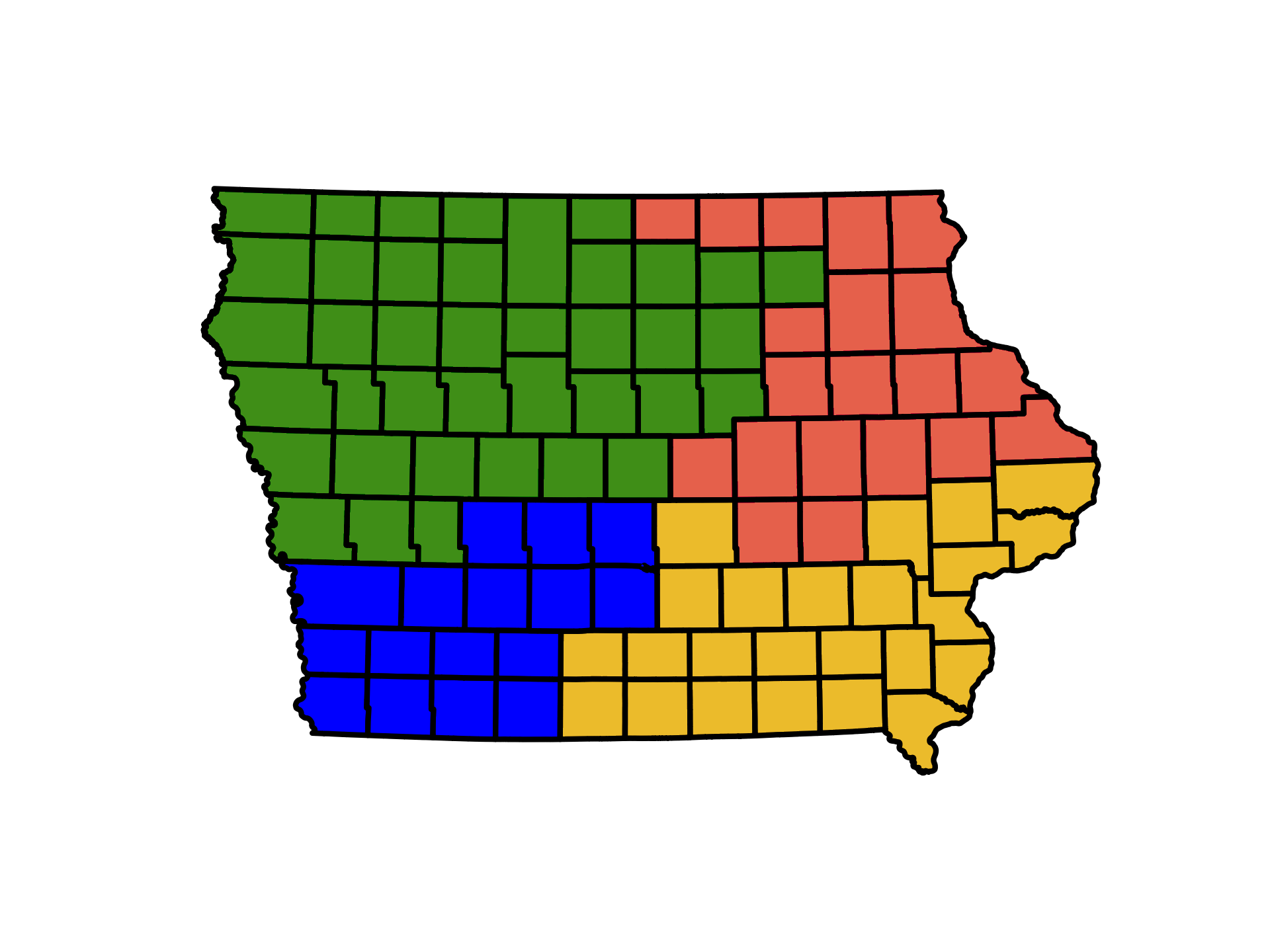}
    \end{subfigure}
    \begin{subfigure}[h]{0.23\textwidth}
      \includegraphics[width=\textwidth,trim={2cm 1.5cm 2cm 1.5cm},clip]{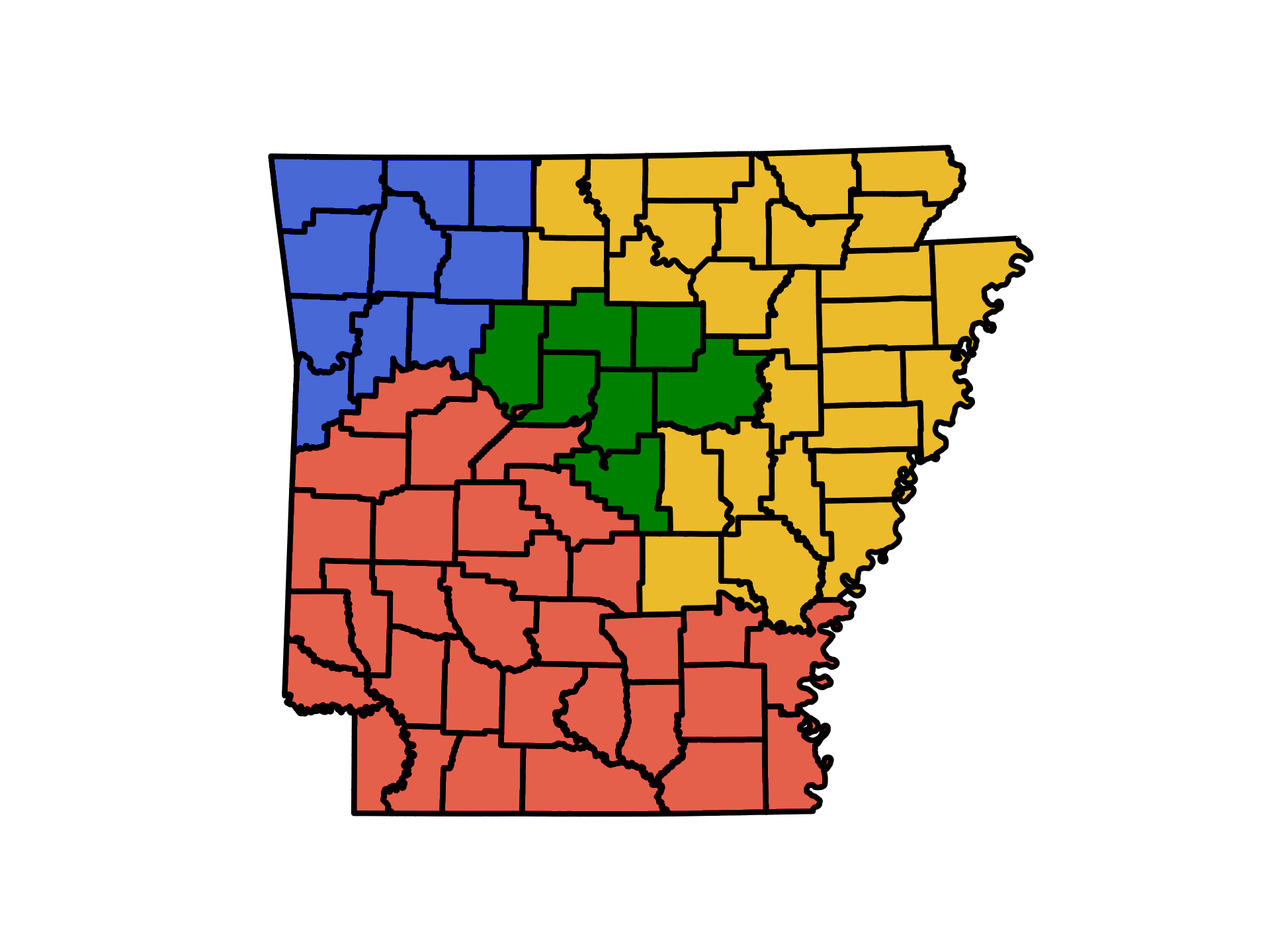}
    \end{subfigure}
\caption{\label{fig:benchmarks} Example districting plans for our four test cases.}  
\end{figure}

The remainder of this chapter explores these algorithmic properties in detail.  Throughout the chapter, we illustrate algorithms using figures and experiments.  In some cases, we rely on simplified (``toy'') examples, like partitioning the cells of a small grid into contiguous pieces.  Our four most frequent test cases will be 
the $6\times 6$ grid, the $10\times 10$ grid, the 99 counties of Iowa,
and the 75 counties of Arkansas (see Figure~\ref{fig:benchmarks}).

Iowa is a particularly useful choice because it is grid-like but has variable population in the units and a manageable number of pieces.\footnote{Unlike most states, which build plans out of much smaller census units (like census blocks), Iowa, by law, builds congressional districts out of whole counties.}  

As an objective function, we often use a compactness metric called \emph{cut edges} to compare algorithmic techniques.  By definition, the cut edge score counts how many pairs of neighboring units are assigned to different districts in a given plan.  

\begin{figure}[ht]
\centering
\includegraphics[width=3in,trim={1cm 3cm 1cm 3cm},clip]{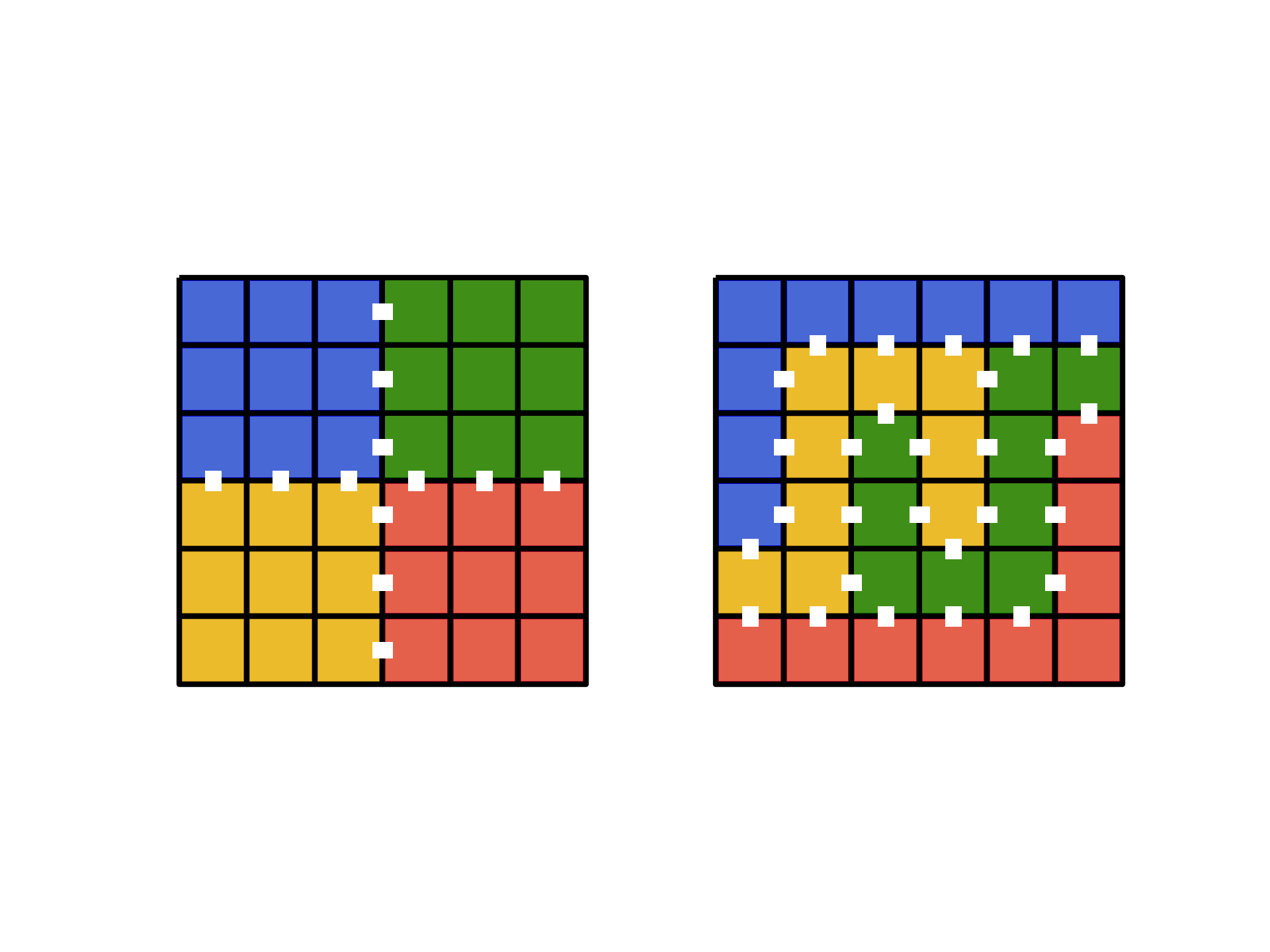}
\caption{A $6\times 6$ grid has a total of 60 pairs of neighboring units, and 
we could imagine drawing an edge between neighbors.
A plan dividing the grid into quadrants would cut just 12 of these edges.
On the other hand, a plan with  winding borders could cut up to 
28 edges out of 60. The white markers indicate the cut edges for the two plans shown.}
\end{figure}

\clearpage


\begin{figure} 
\centering
     \includegraphics[width=0.5\textwidth]{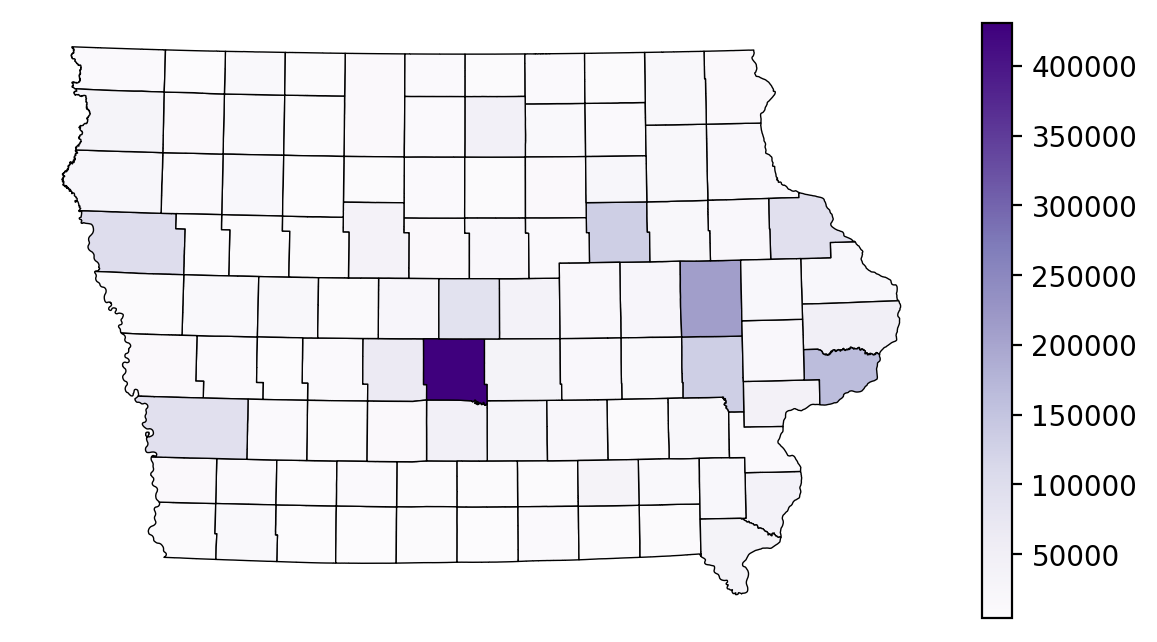}
    \caption{\label{fig:iowa_population_map} Iowa population map.}    
\end{figure}

\begin{figure} 
  \centering
    \includegraphics[width=0.8\textwidth]{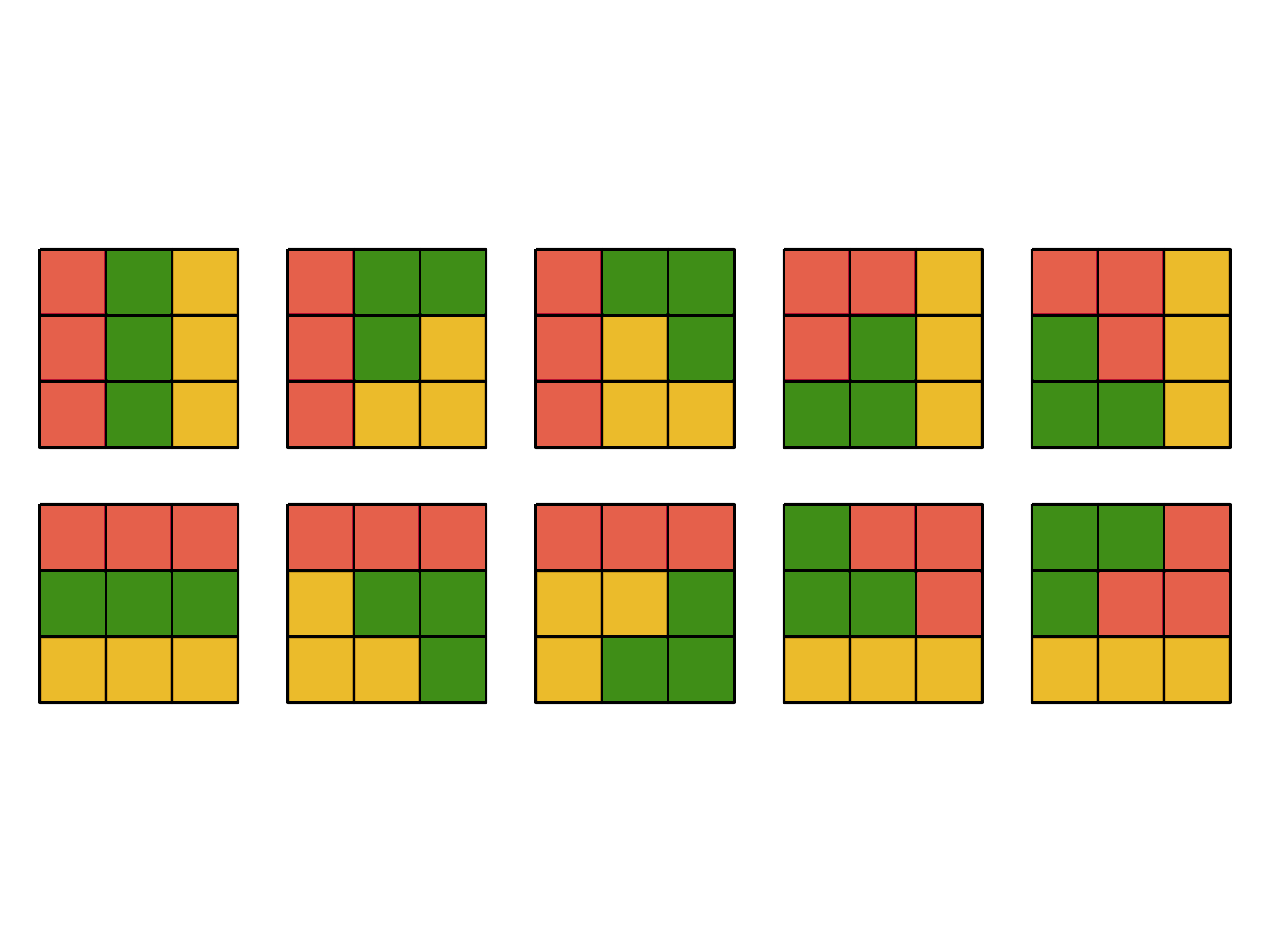}
    \caption{Complete enumeration of ways to divide a $3 \times 3$ grid into three equal-sized, rook-contiguous districts. Note each of the ten plans has the same number of cut edges (6 out of the 12 neighboring units are cut ).}\label{fig:complete-enum}
\end{figure}

\begin{figure} 
  \centering
    \includegraphics[width=0.5\textwidth, trim={2cm 2cm 2cm 2cm}, clip]{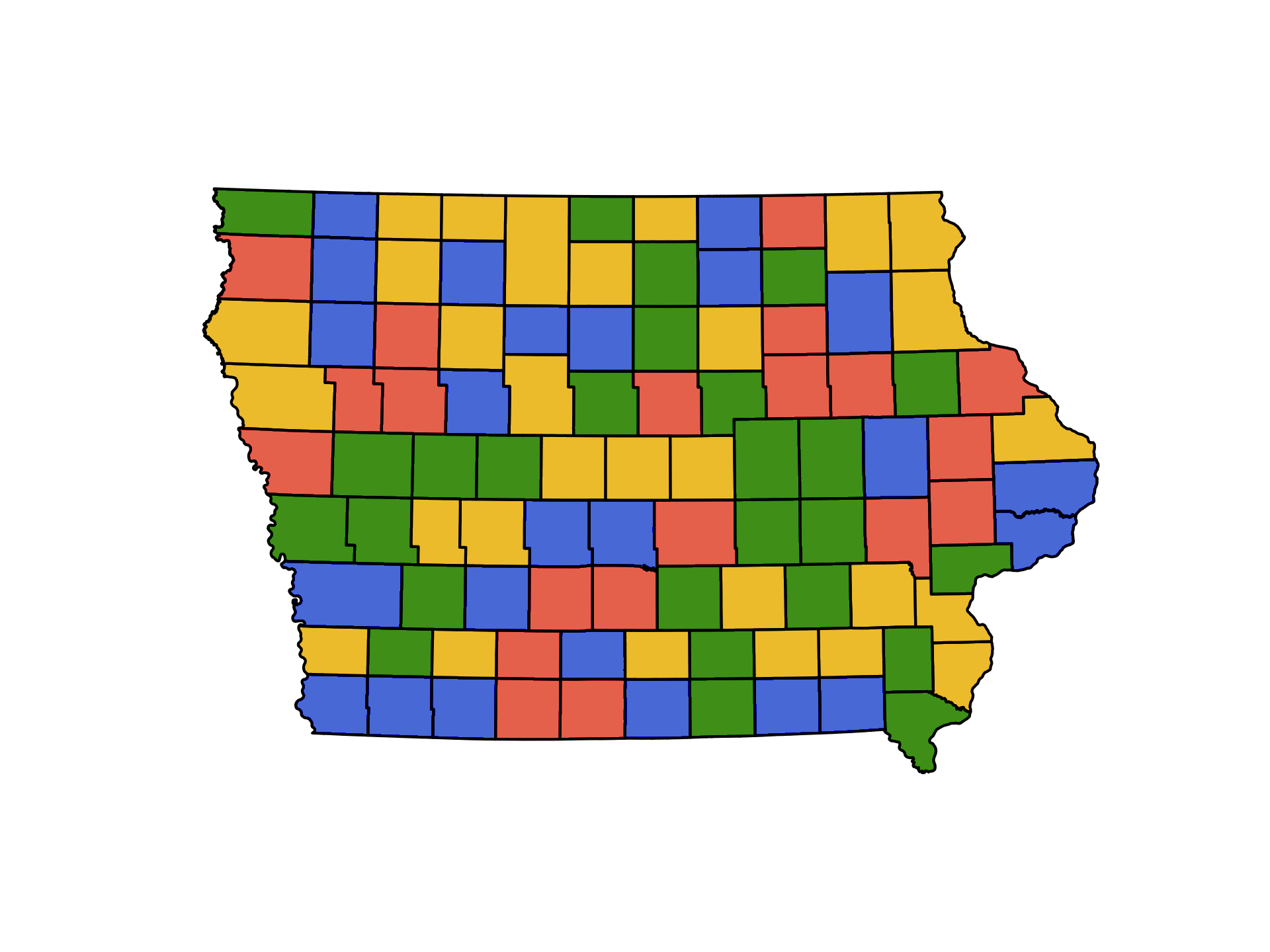}
    \caption{Random assignment of Iowa counties; each color represents a different district.  Unsurprisingly, the resulting plan has disconnected districts and balances neither the number of counties nor the district populations.}\label{fig:random}
\end{figure}

\begin{figure} 
\centering
    \begin{subfigure}[h]{0.3\textwidth}
      \includegraphics[width=\textwidth]{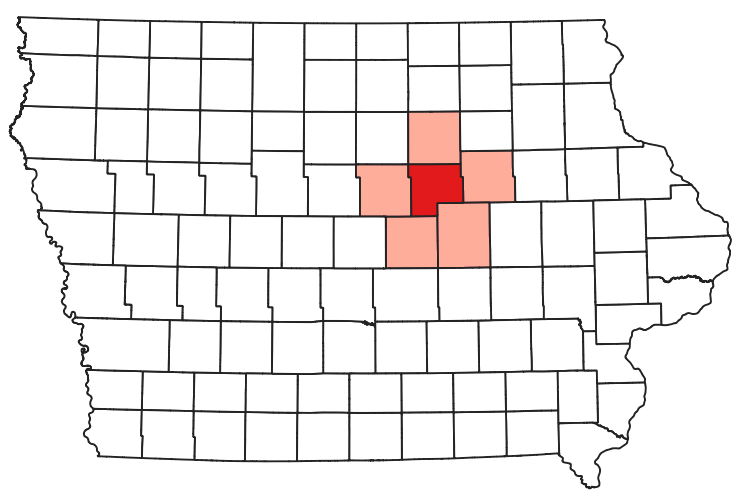}
    \end{subfigure}\ \ \ 
    \begin{subfigure}[h]{0.3\textwidth}
      \includegraphics[width=\textwidth]{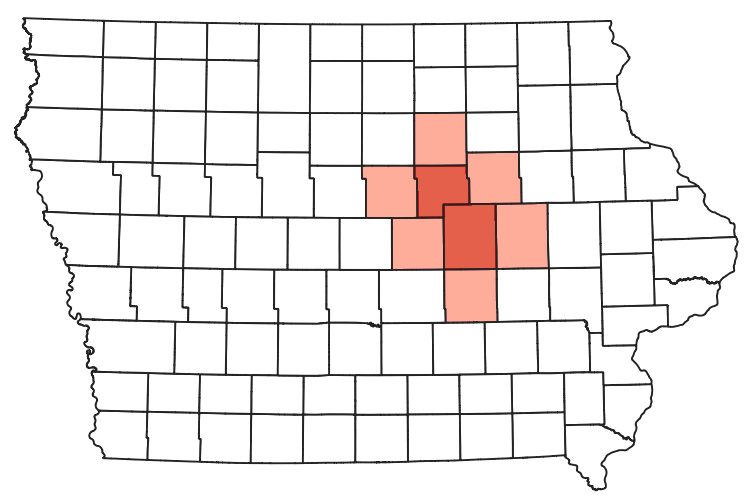}
    \end{subfigure}\ \ \ 
    \begin{subfigure}[h]{0.3\textwidth}
      \includegraphics[width=\textwidth]{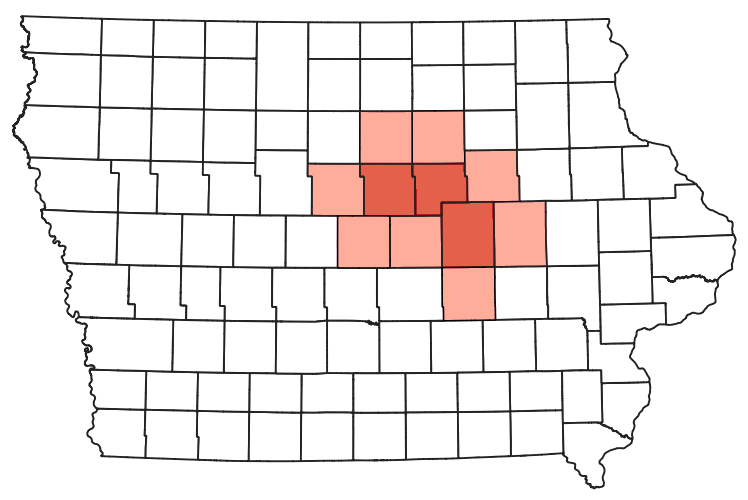}
    \end{subfigure}
\caption{\label{fig:flood_fill}An example of flood fill on Iowa counties.  Counties are colored red as they are added to the growing district.  The pink counties indicate candidate neighboring counties to annex at each step.}    
\end{figure}

\begin{figure} 
\centering
    \begin{subfigure}[h]{0.23\textwidth}
      \includegraphics[width=\textwidth,trim={1cm 1cm 1cm 1cm},clip]{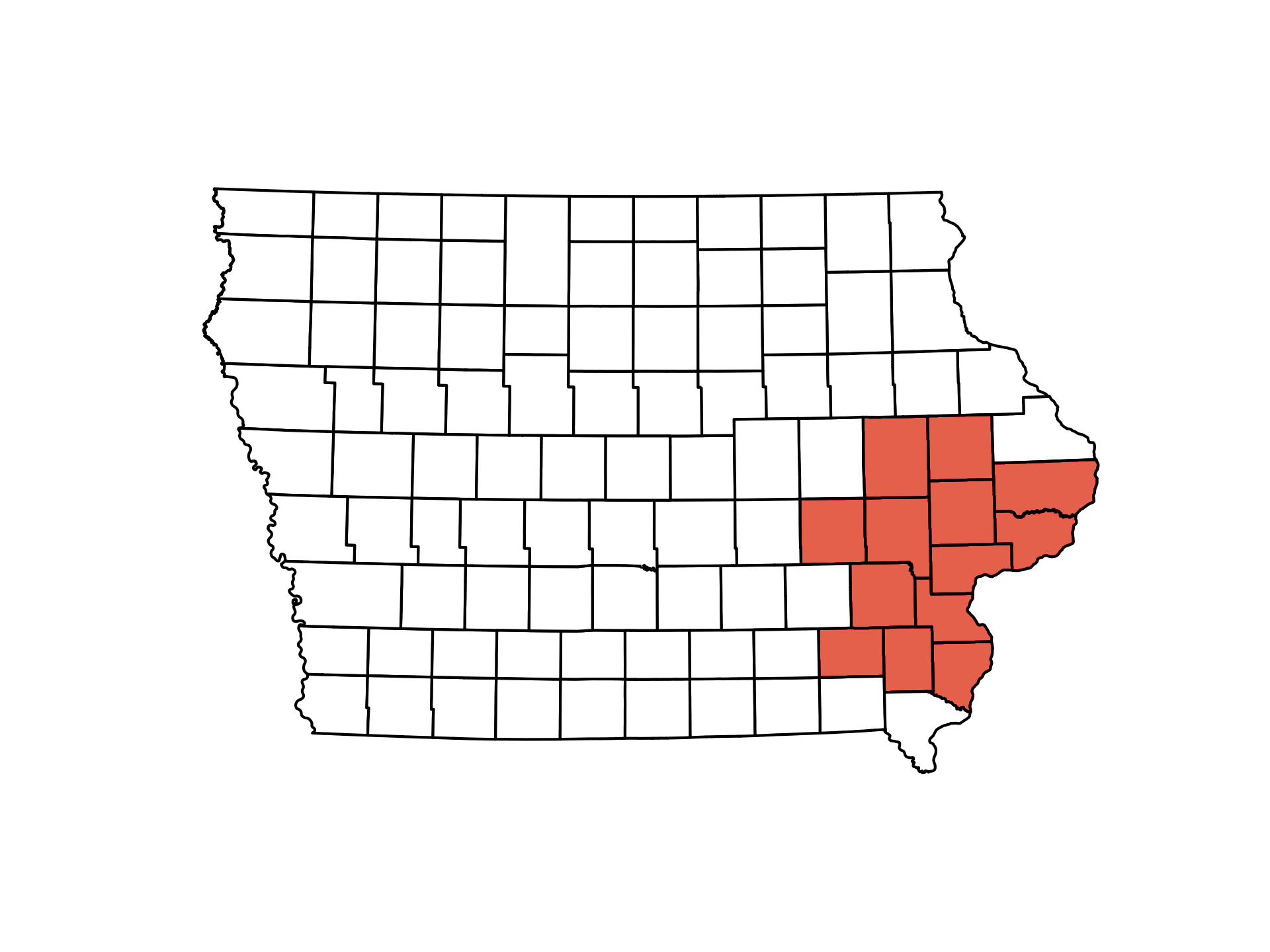}
    \end{subfigure}
    \begin{subfigure}[h]{0.23\textwidth}
      \includegraphics[width=\textwidth,trim={1cm 1cm 1cm 1cm},clip]{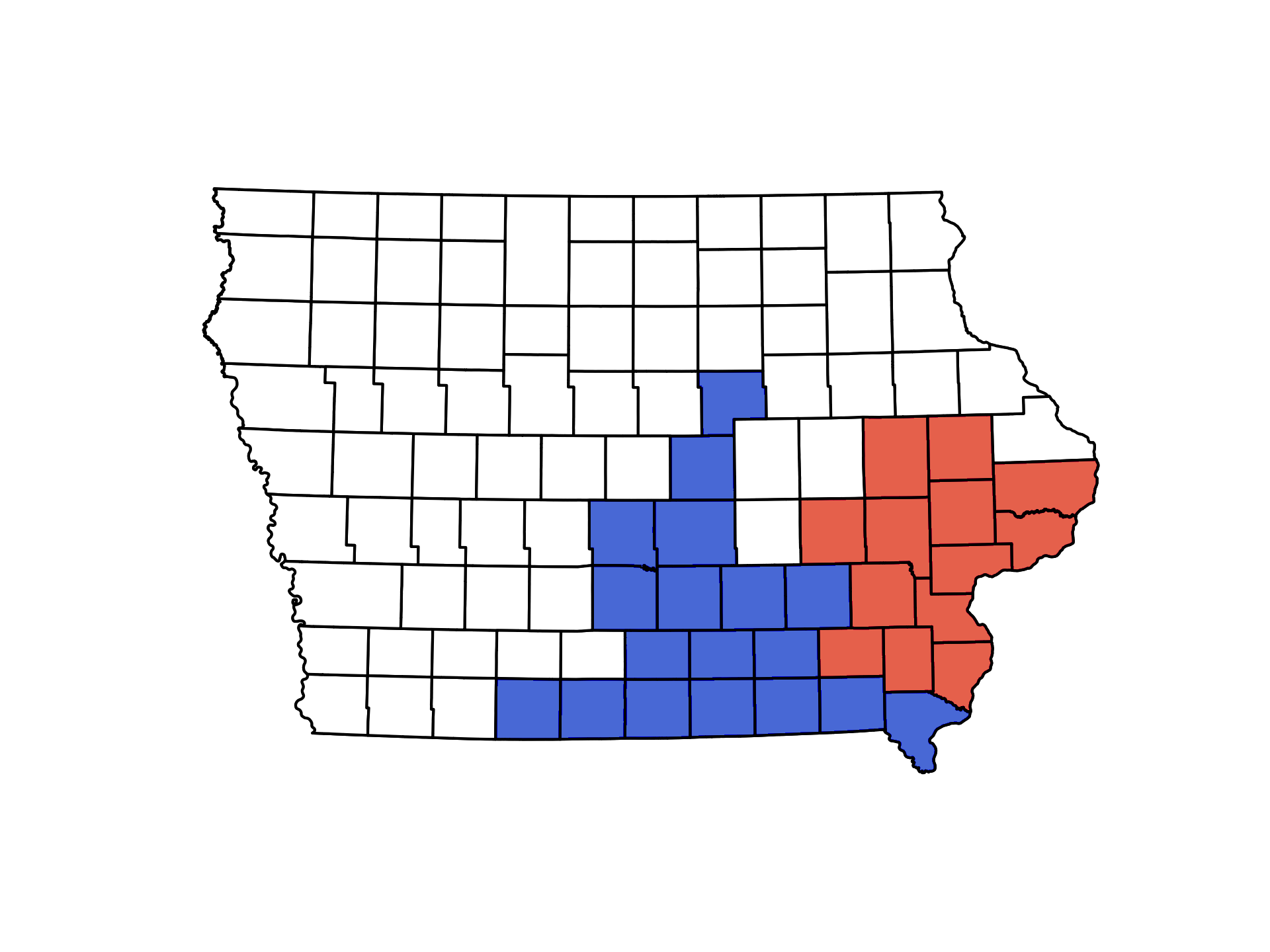}
    \end{subfigure}
    \begin{subfigure}[h]{0.23\textwidth}
      \includegraphics[width=\textwidth,trim={1cm 1cm 1cm 1cm},clip]{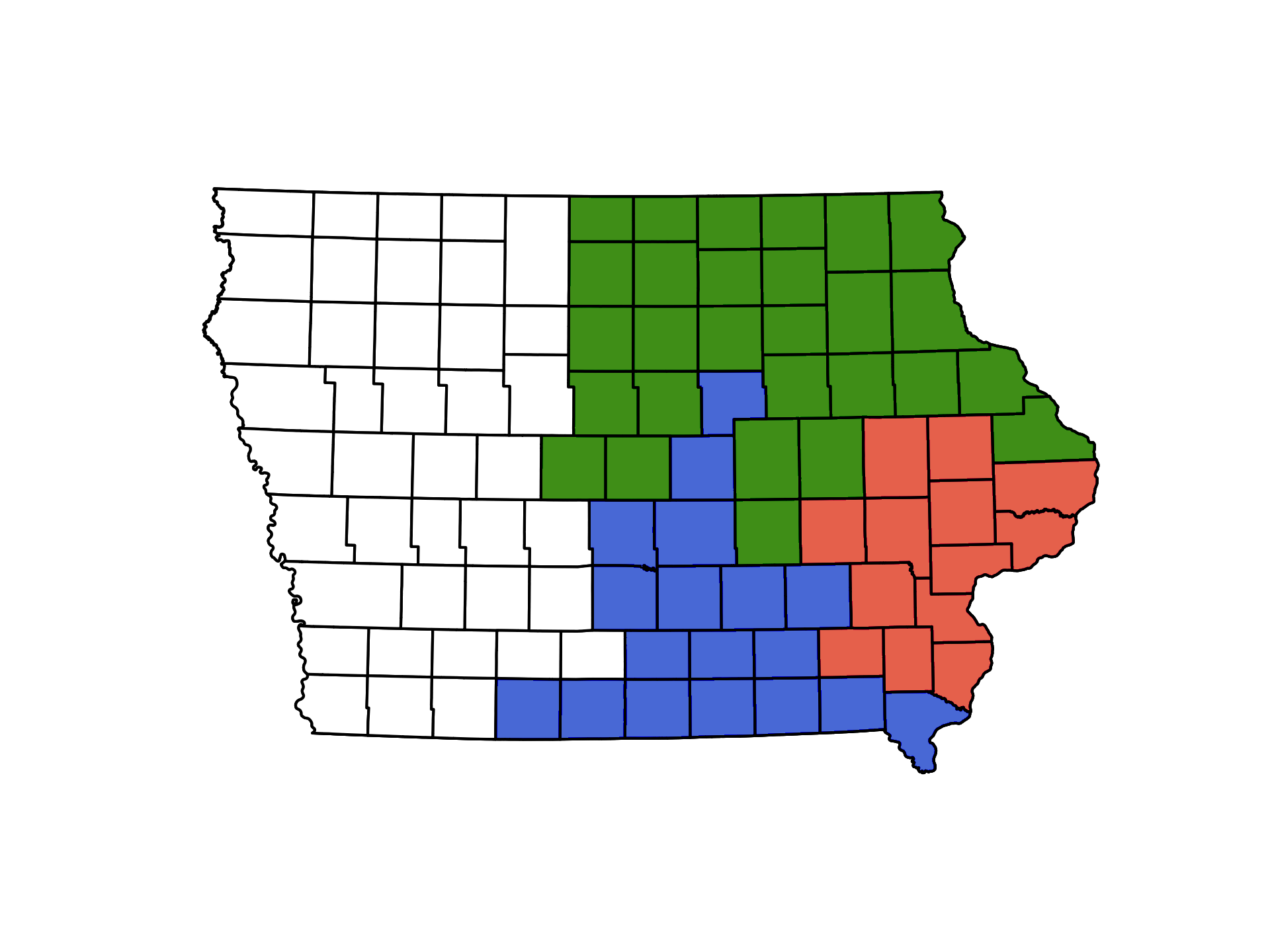}
    \end{subfigure}
    \begin{subfigure}[h]{0.23\textwidth}
      \includegraphics[width=\textwidth,trim={1cm 1cm 1cm 1cm},clip]{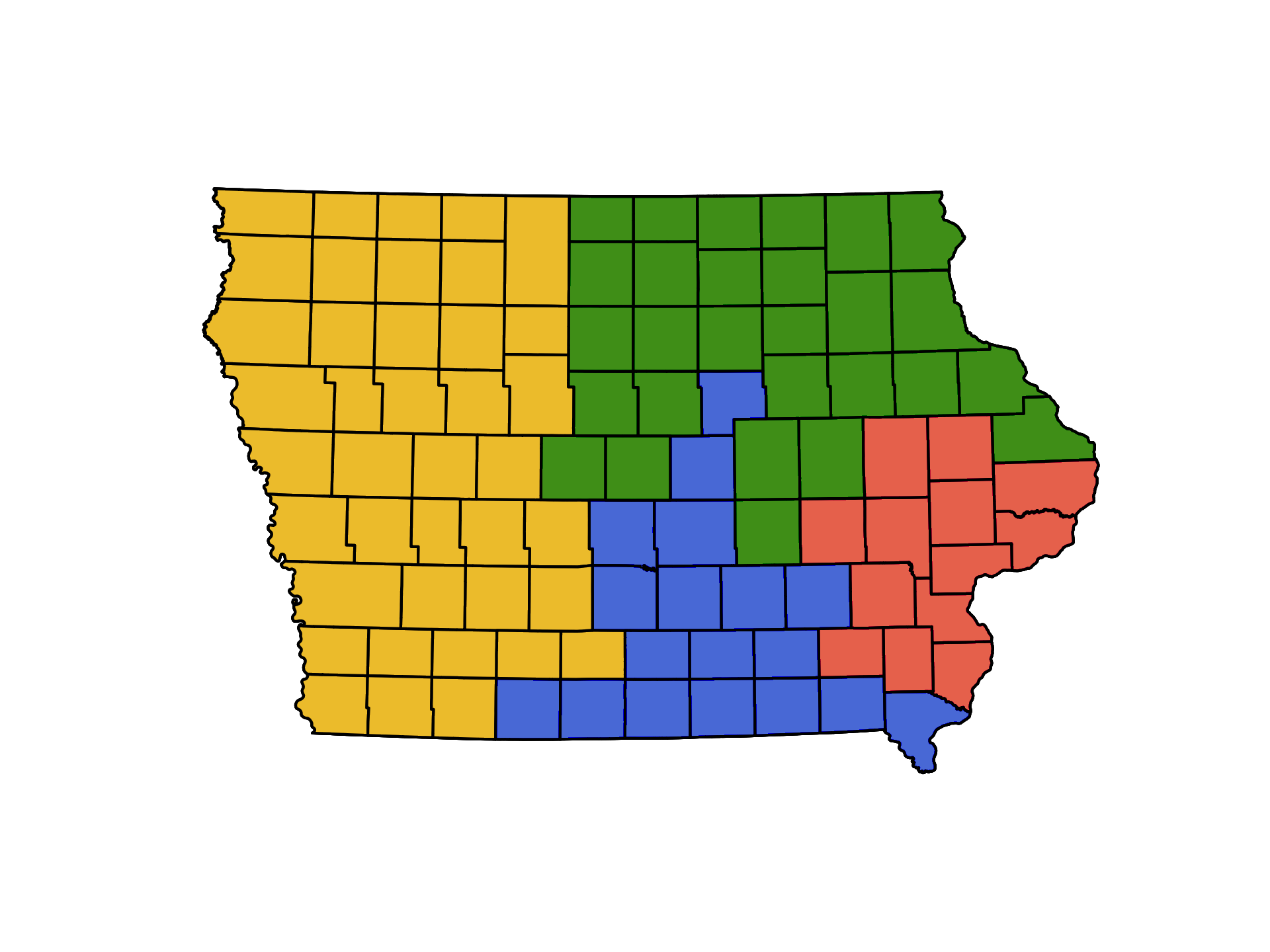}
    \end{subfigure}
    
 \rule{\textwidth}{.02cm}
        \begin{subfigure}[b]{0.23\textwidth}
      \includegraphics[width=\textwidth,trim={1cm 1cm 1cm 1cm},clip]{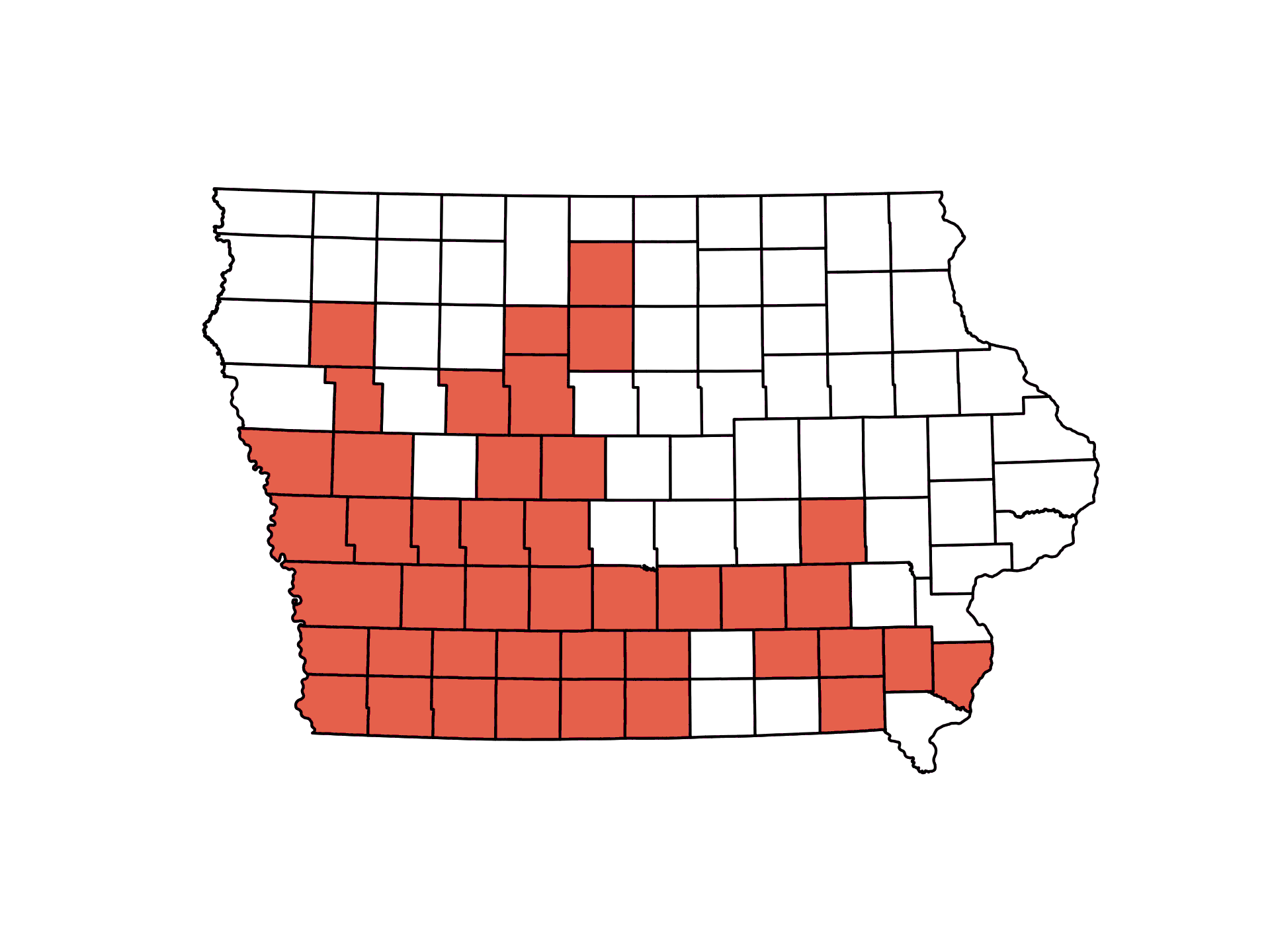}
    \end{subfigure}
    \begin{subfigure}[b]{0.23\textwidth}
      \includegraphics[width=\textwidth,trim={1cm 1cm 1cm 1cm},clip]{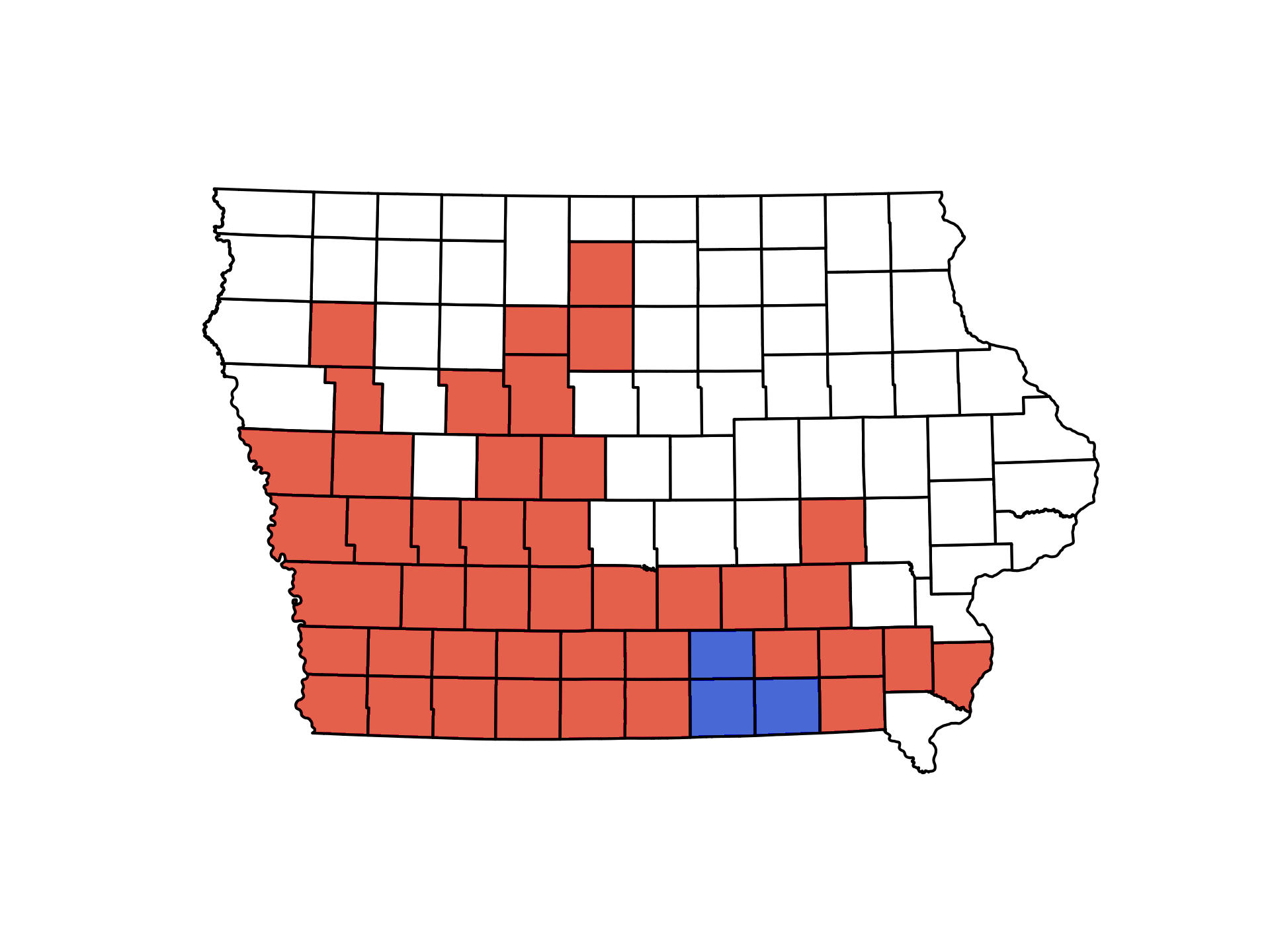}
    \end{subfigure}
    \vline
    \begin{subfigure}[b]{0.23\textwidth}
      \includegraphics[width=\textwidth,trim={1cm 1cm 1cm 1cm},clip]{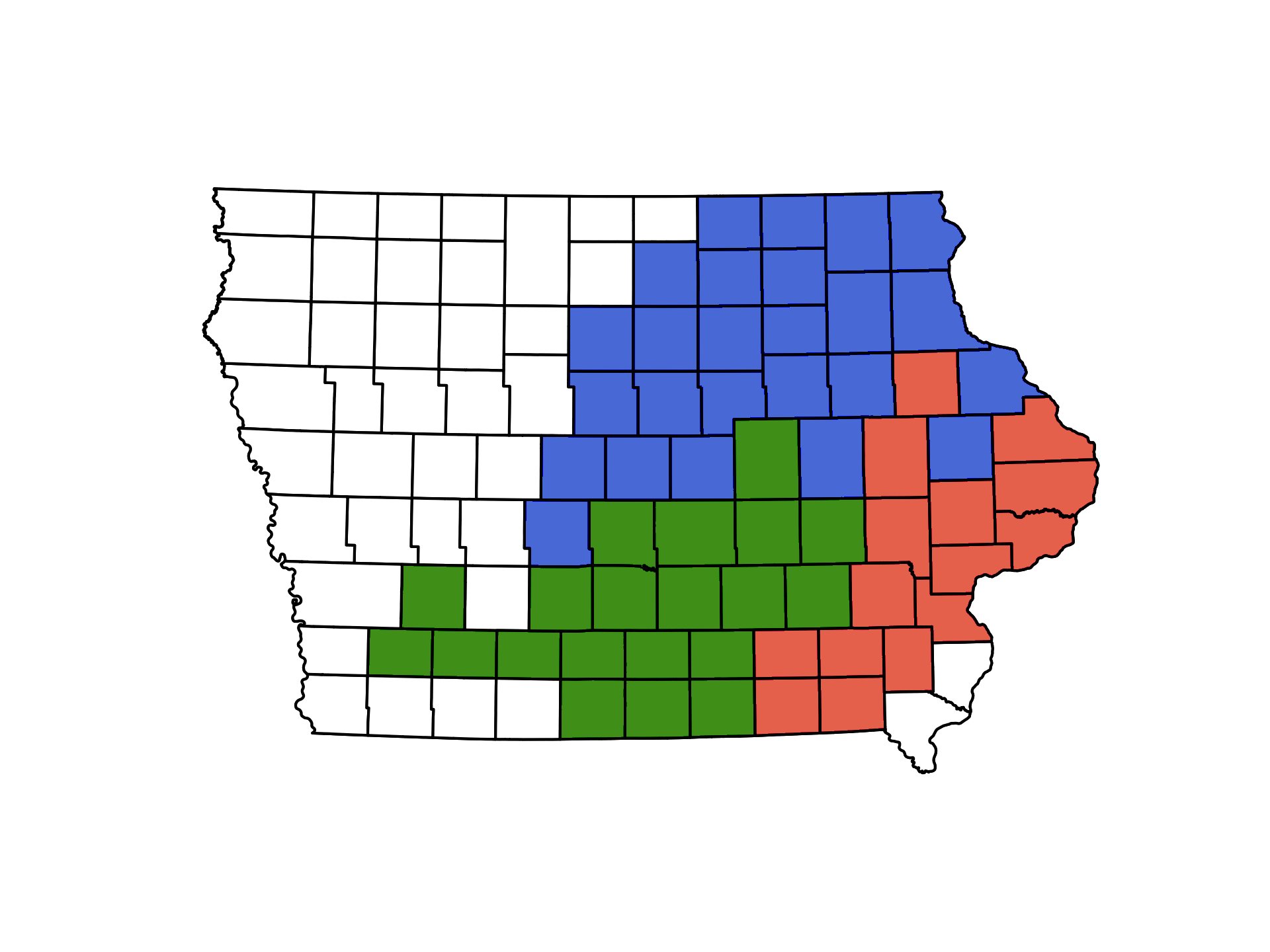}
    \end{subfigure}
    \begin{subfigure}[b]{0.23\textwidth}
      \includegraphics[width=\textwidth,trim={1cm 1cm 1cm 1cm},clip]{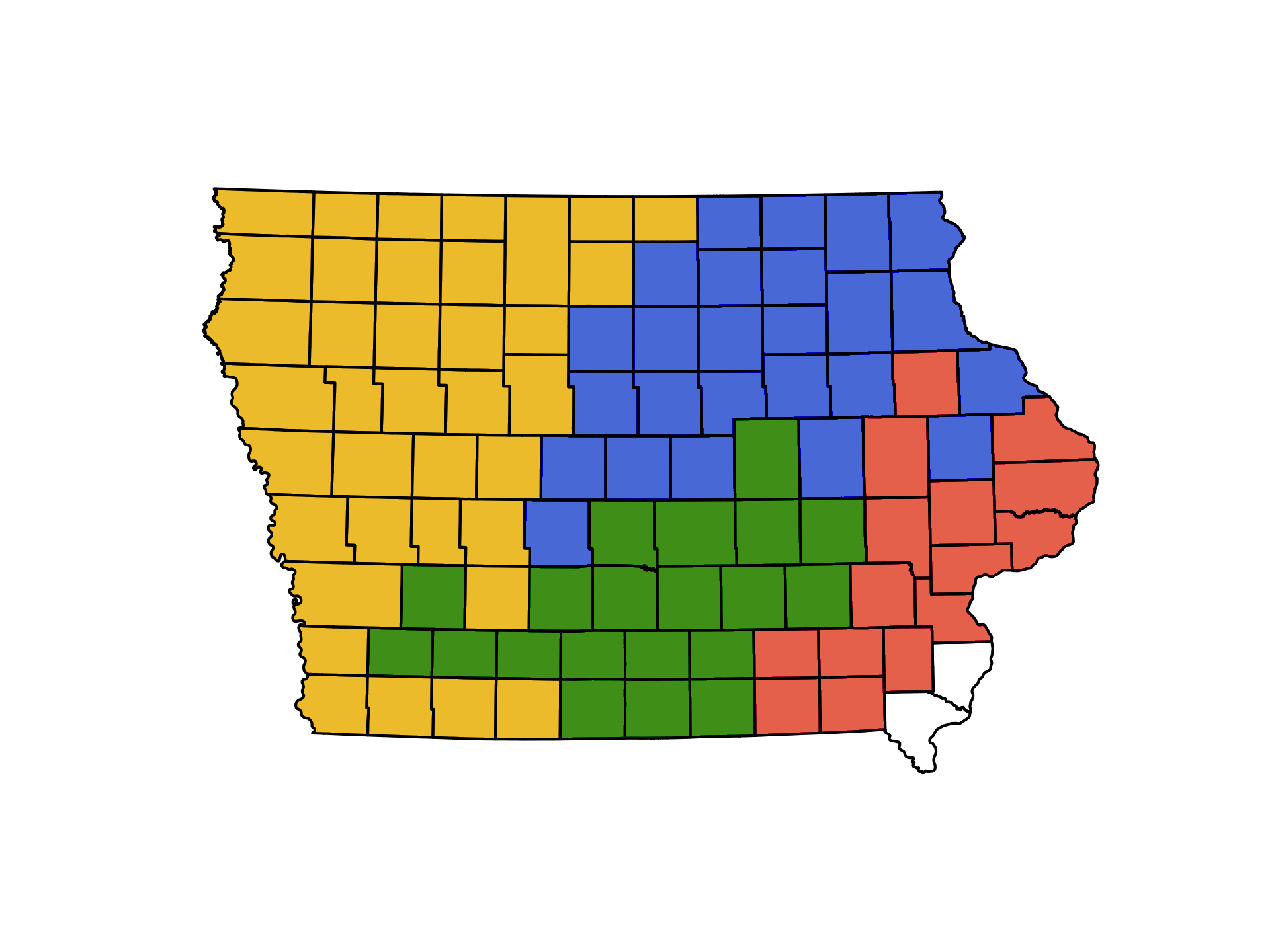}
    \end{subfigure}
\caption{\label{fig:flood_fill_success} \label{fig:flood_fill_fails}
The flood fill algorithm grows the districts one at a time. The top example arrives at a complete plan.  The bottom two examples lead to rejection because there is no way to complete the plan with contiguous districts and population balance.}    \end{figure}

\begin{figure} 
\centering
    \begin{subfigure}[h]{0.4\textwidth}
      \includegraphics[width=\textwidth]{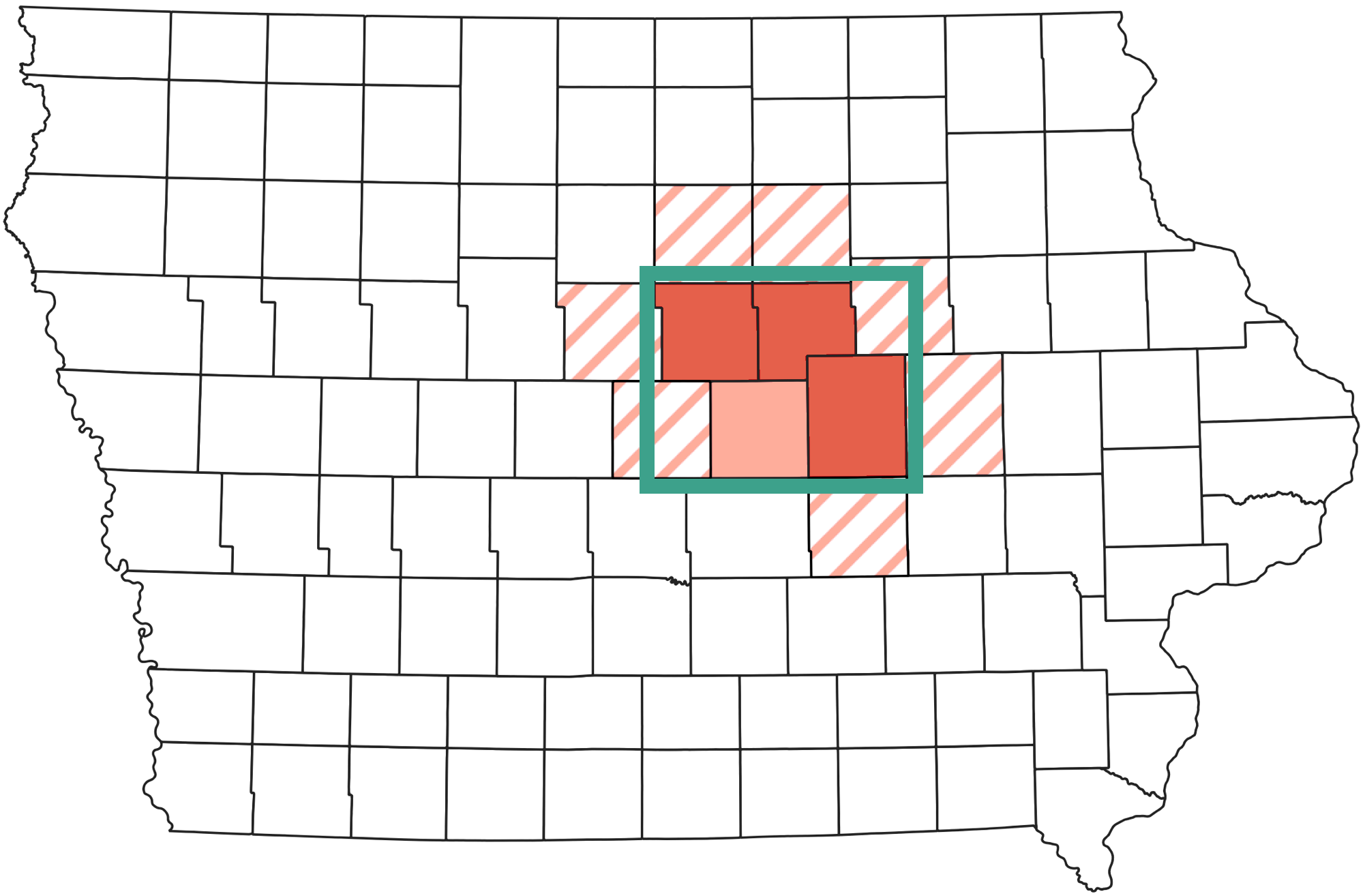}
      \caption{\label{fig:flood_fill_bb}Bounding-box variant}
    \end{subfigure}
    \quad
    \quad
    \begin{subfigure}[h]{0.4\textwidth}
      \includegraphics[width=\textwidth]{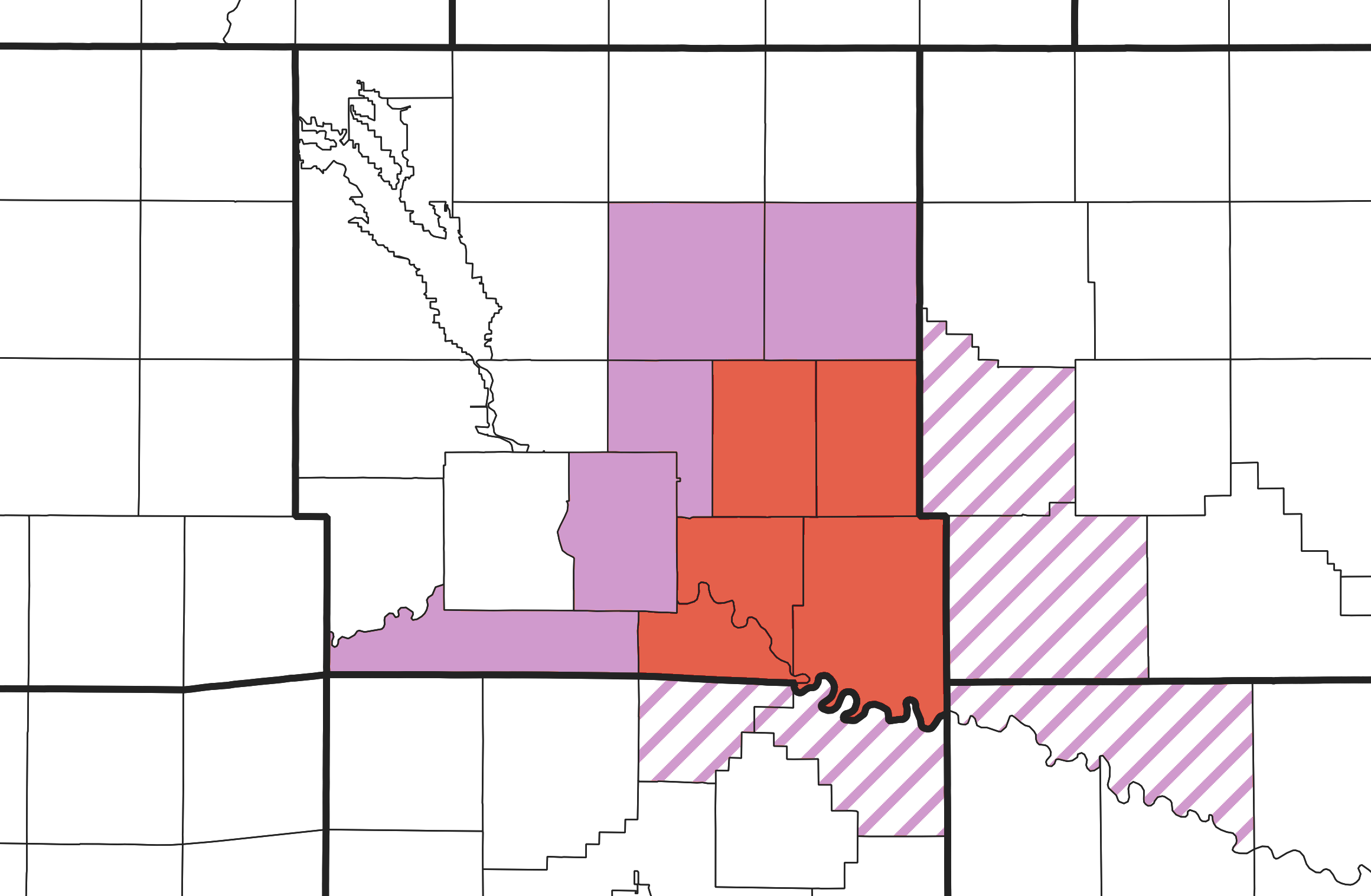}
      \caption{\label{fig:flood_fill_county}County-preserving variant}
    \end{subfigure}
\caption{\label{fig:flood_fill_variants}On the left is the bounding-box flood fill variant\cite{cirincione2000assessing}; the solid pink county lies entirely within the bounding box of the growing red district and are preferentially chosen over the striped pink neighboring counties. On the right is the county-respecting flood fill variant \cite{cirincione2000assessing}; the solid pink county subunits lie within the county of the growing red district and are preferentially chosen over the striped pink neighboring subunits in different counties.}  
\end{figure}



\begin{figure} 
\centering
    \begin{subfigure}[ht]{0.3\textwidth}
      \includegraphics[width=\textwidth, trim={2cm 2cm 2cm 2cm},clip]{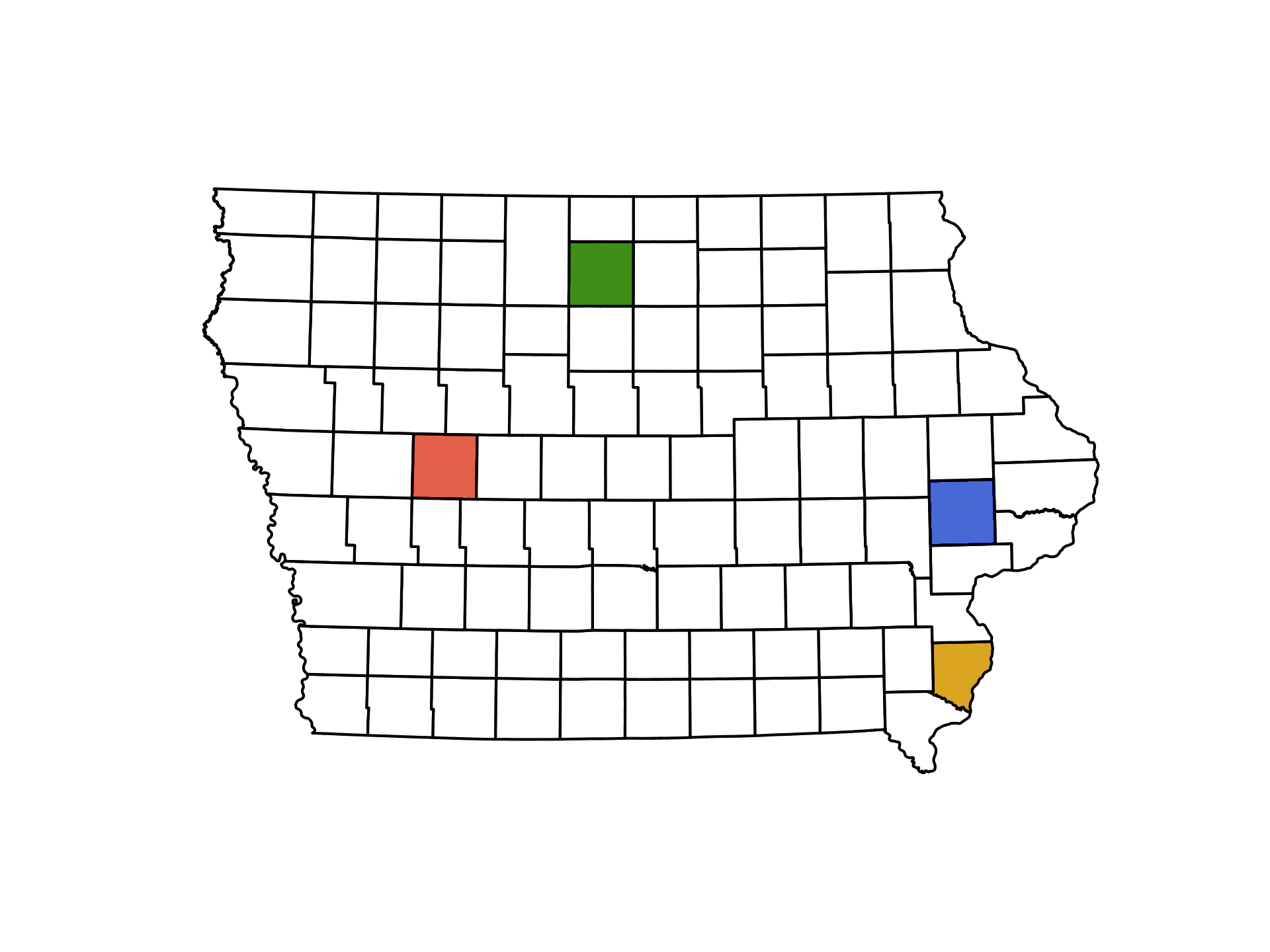}
    \end{subfigure}\ \ \ 
    \begin{subfigure}[ht]{0.3\textwidth}
      \includegraphics[width=\textwidth, trim={2cm 2cm 2cm 2cm}, clip]{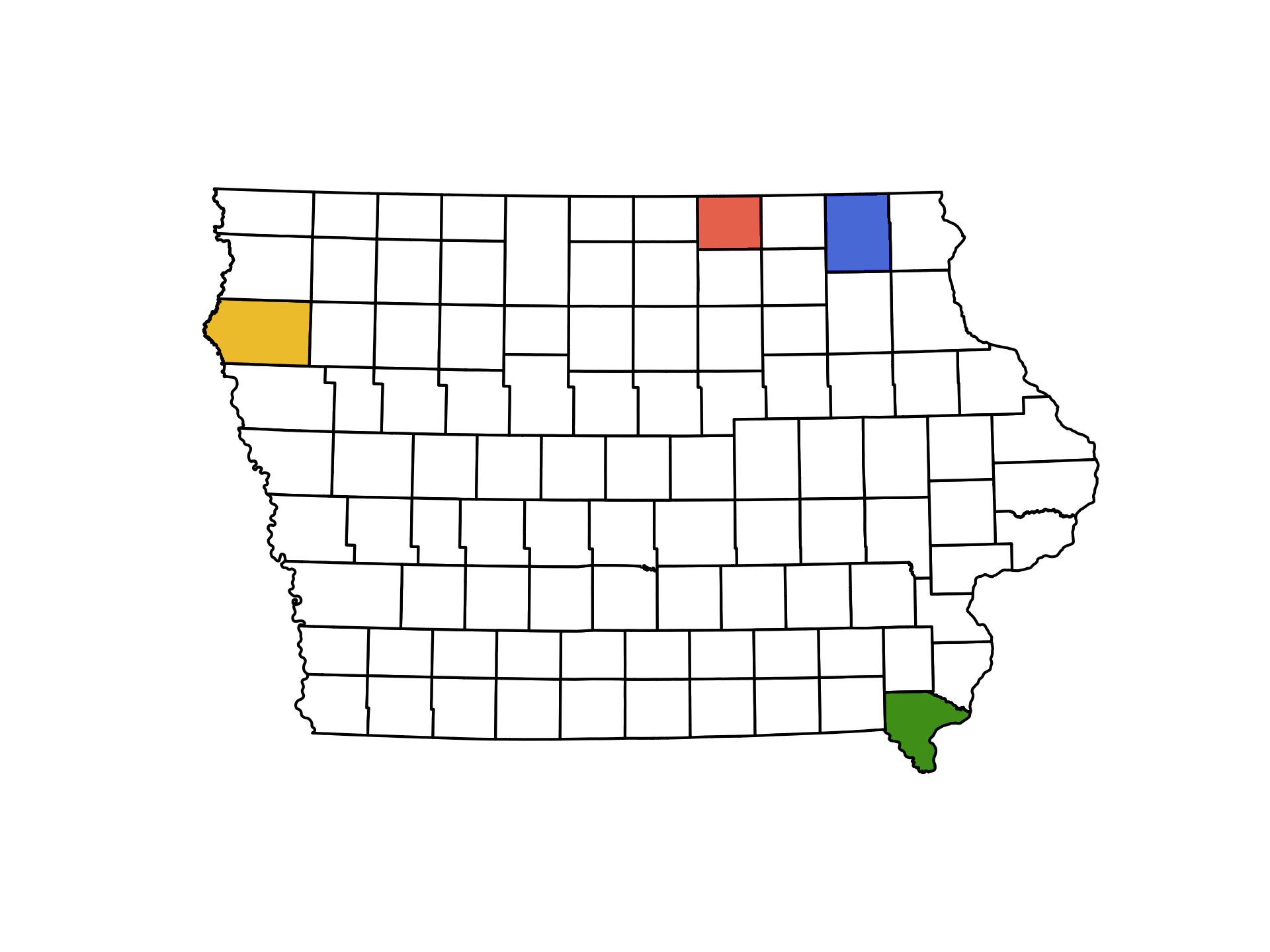}
    \end{subfigure}\ \ \ 
    \begin{subfigure}[ht]{0.3\textwidth}
      \includegraphics[width=\textwidth, trim={2cm 2cm 2cm 2cm}, clip]{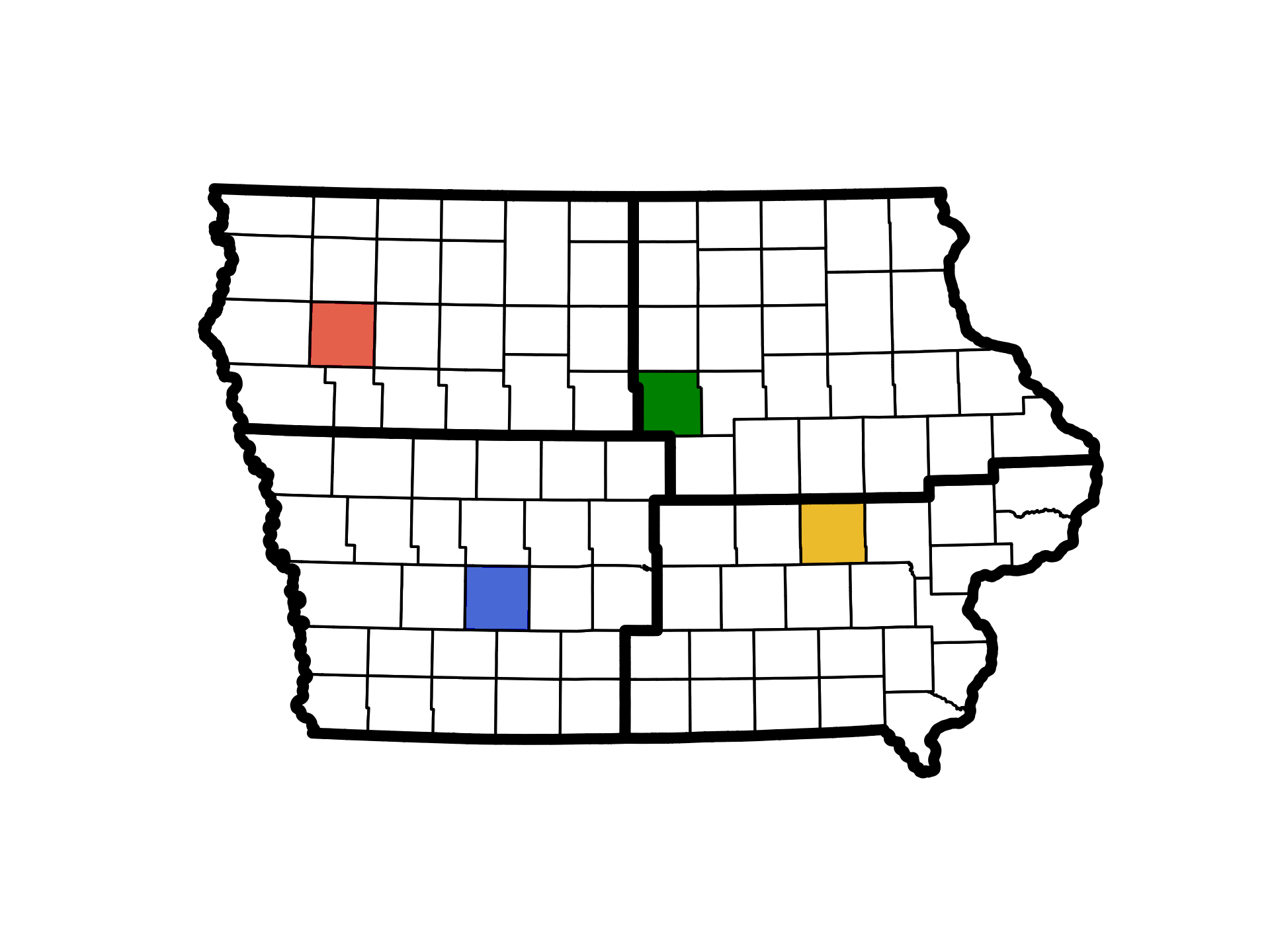}
    \end{subfigure}
\caption{\label{fig:flood_fill_seeds}Left: district seeds were chosen uniformly at random from all counties.  Center: district seeds were chosen uniformly at random from the boundary counties.  Right: one district seed was chosen uniformly at random from each predefined zone.}    
\end{figure}

\begin{figure} 
\centering
     \includegraphics[width=0.9\textwidth, trim={1cm 0cm 0cm 0cm},clip]{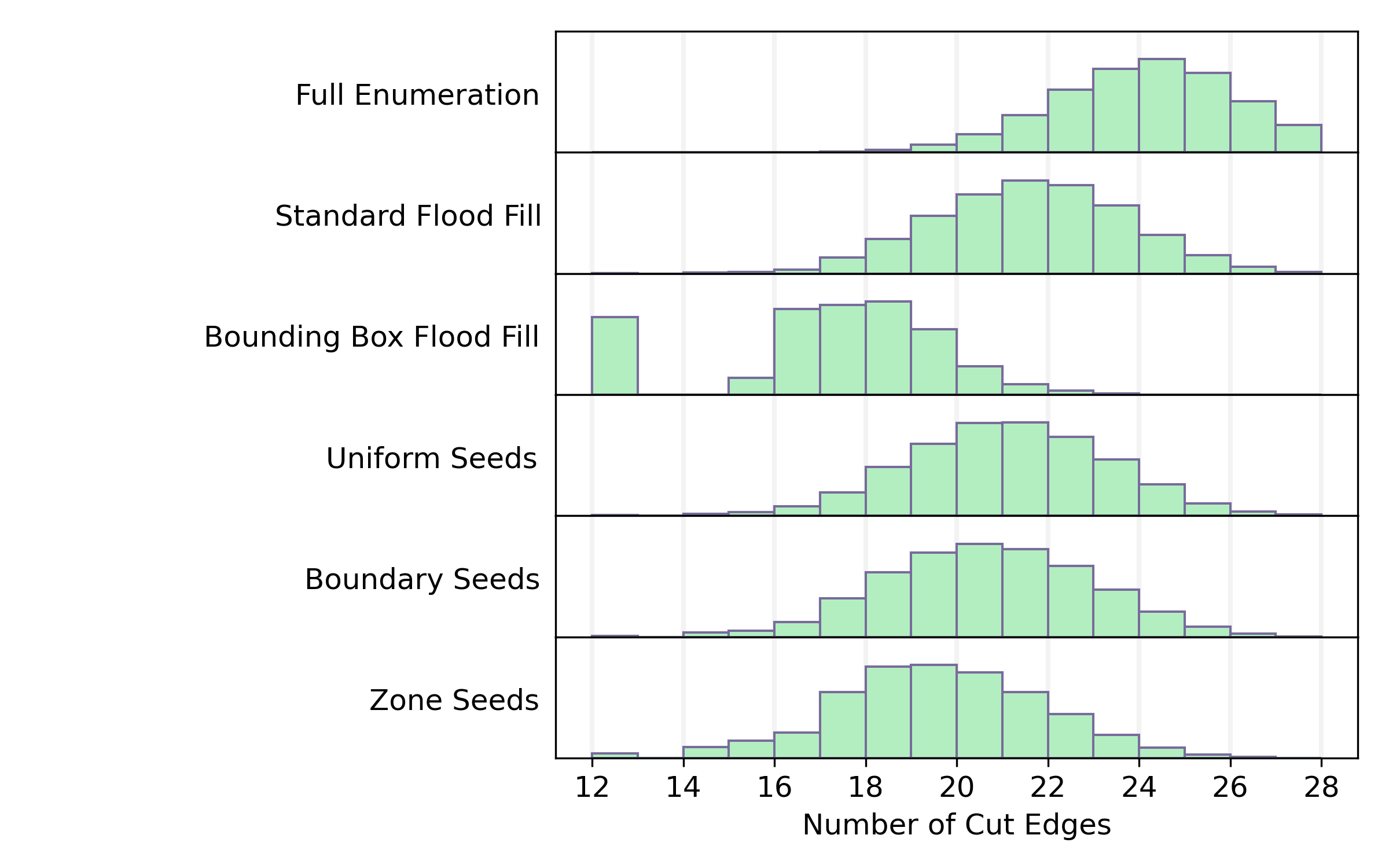}
    \caption{\label{fig:6_violin} Cut-edge comparison for flood fill methods that divide a $6 \times 6$ grid into four equal districts. The bounding box method succeeds at favoring more compact plans.}    
\end{figure}

\begin{figure} 
\centering
    \begin{subfigure}[ht]{0.3\textwidth}
     \includegraphics[width=\textwidth]{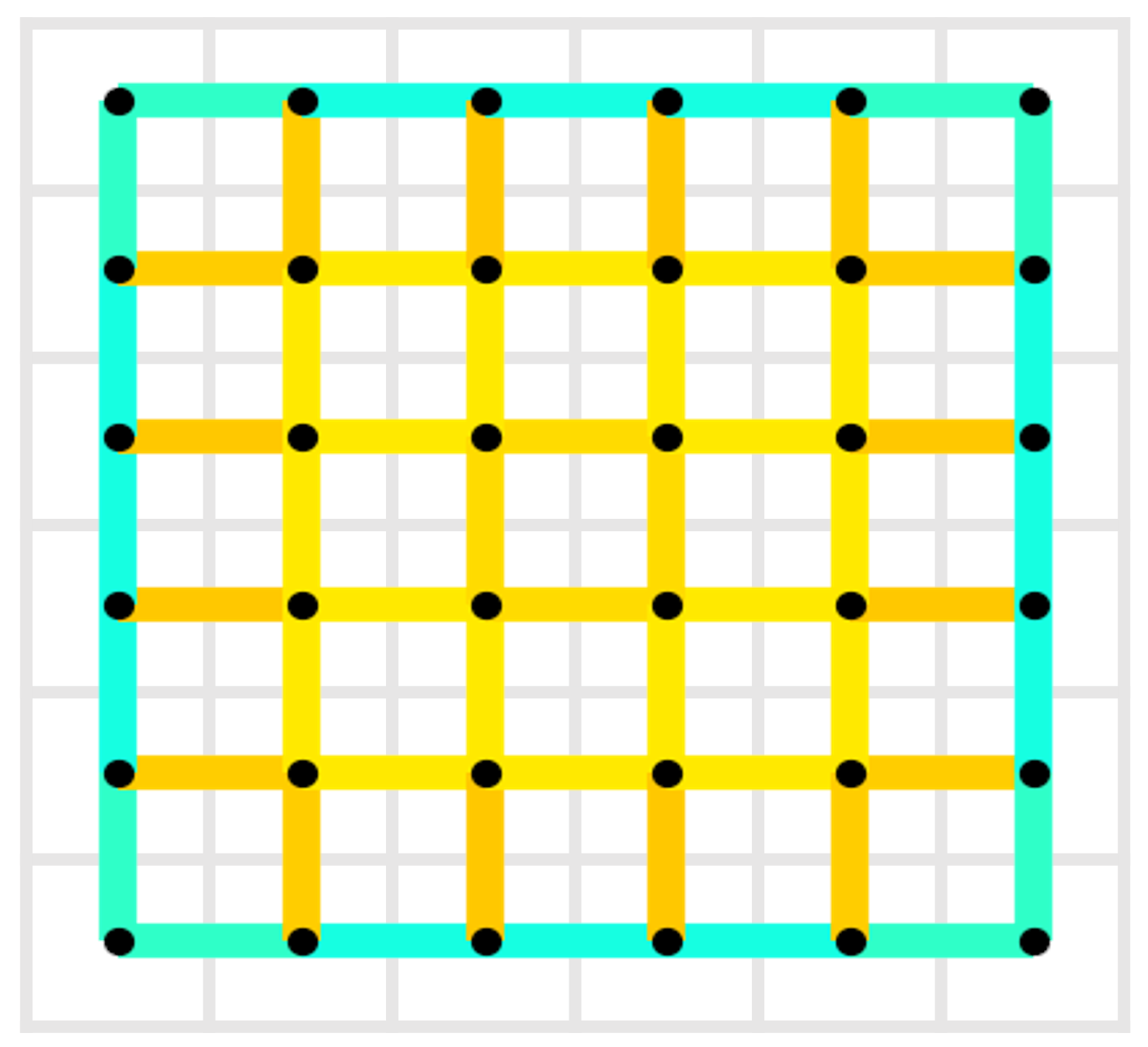}
     \caption{\label{fig:true_edge_freq} Full Enumeration}
     \end{subfigure}\ \ \ \ 
     \begin{subfigure}[ht]{0.3\textwidth}
     \includegraphics[width=\textwidth]{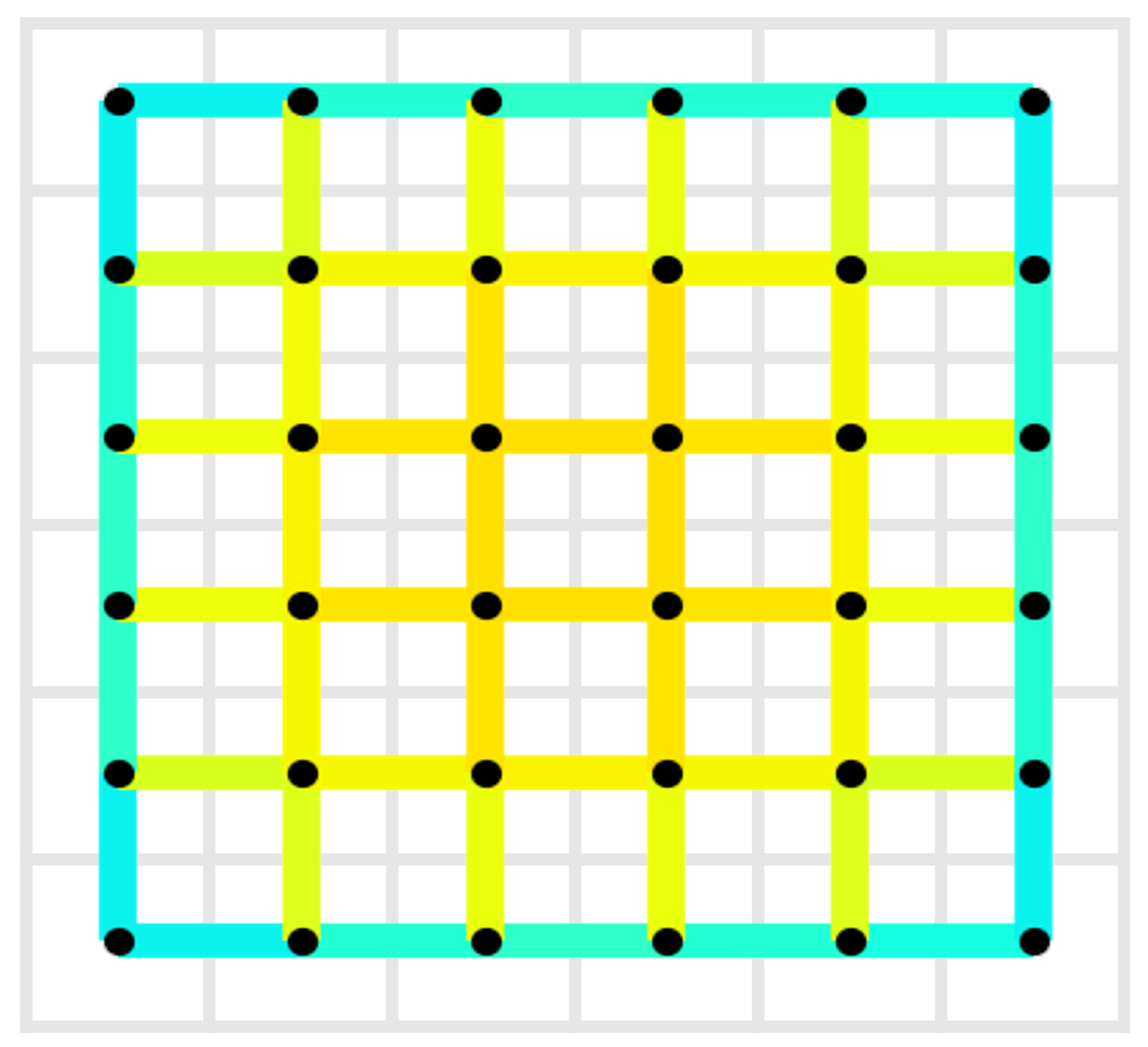}
     \caption{\label{fig:ff_neutral_edge_freq}Standard Flood Fill}
     \end{subfigure}\ \ \ \ 
    \begin{subfigure}[ht]{0.3\textwidth}
     \includegraphics[width=\textwidth]{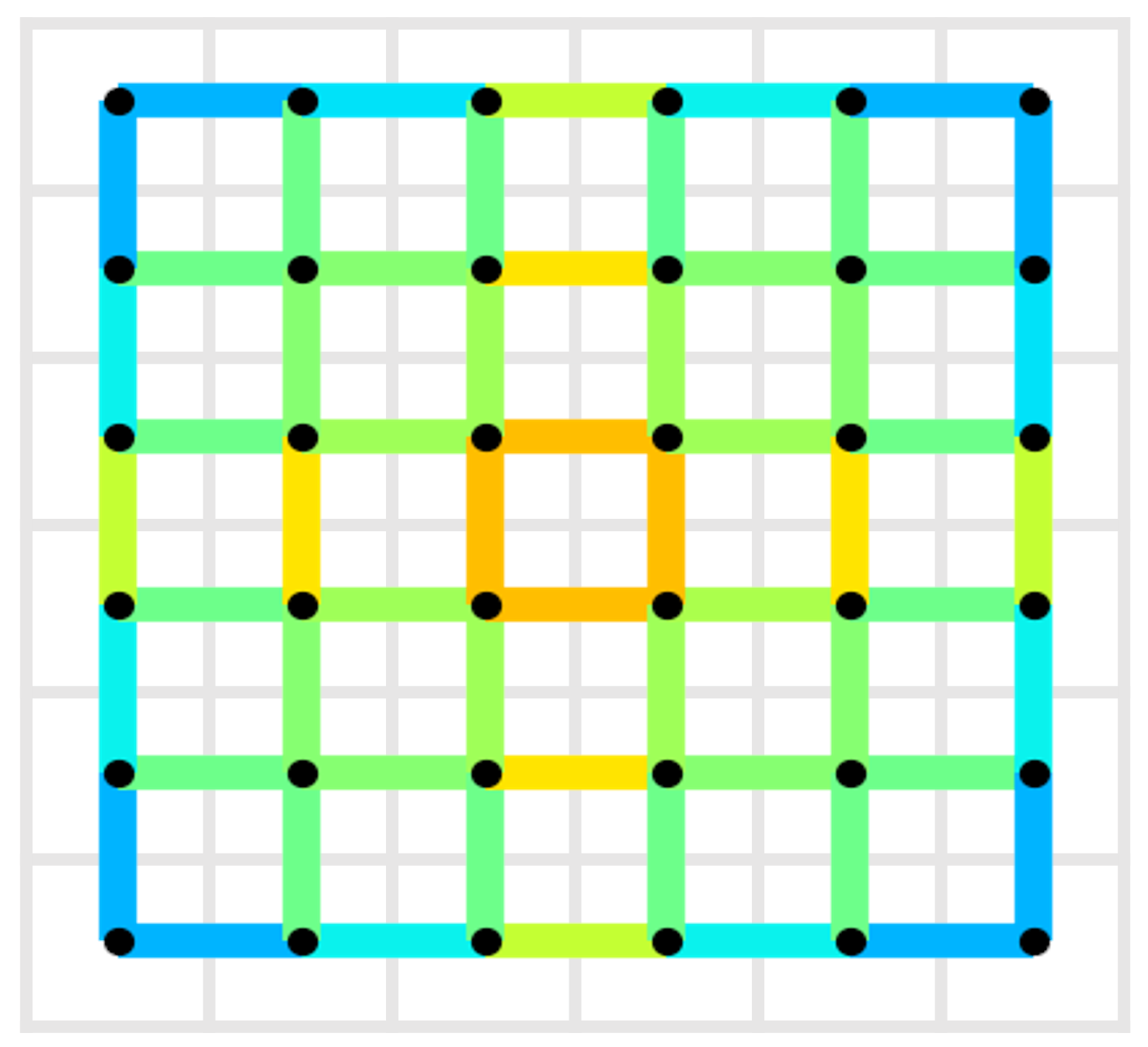}
    \caption{\label{fig:ff_bb_edge_freq}Bounding Box Flood Fill}
    \end{subfigure}\\
     \begin{subfigure}[ht]{0.3\textwidth}
     \includegraphics[width=\textwidth]{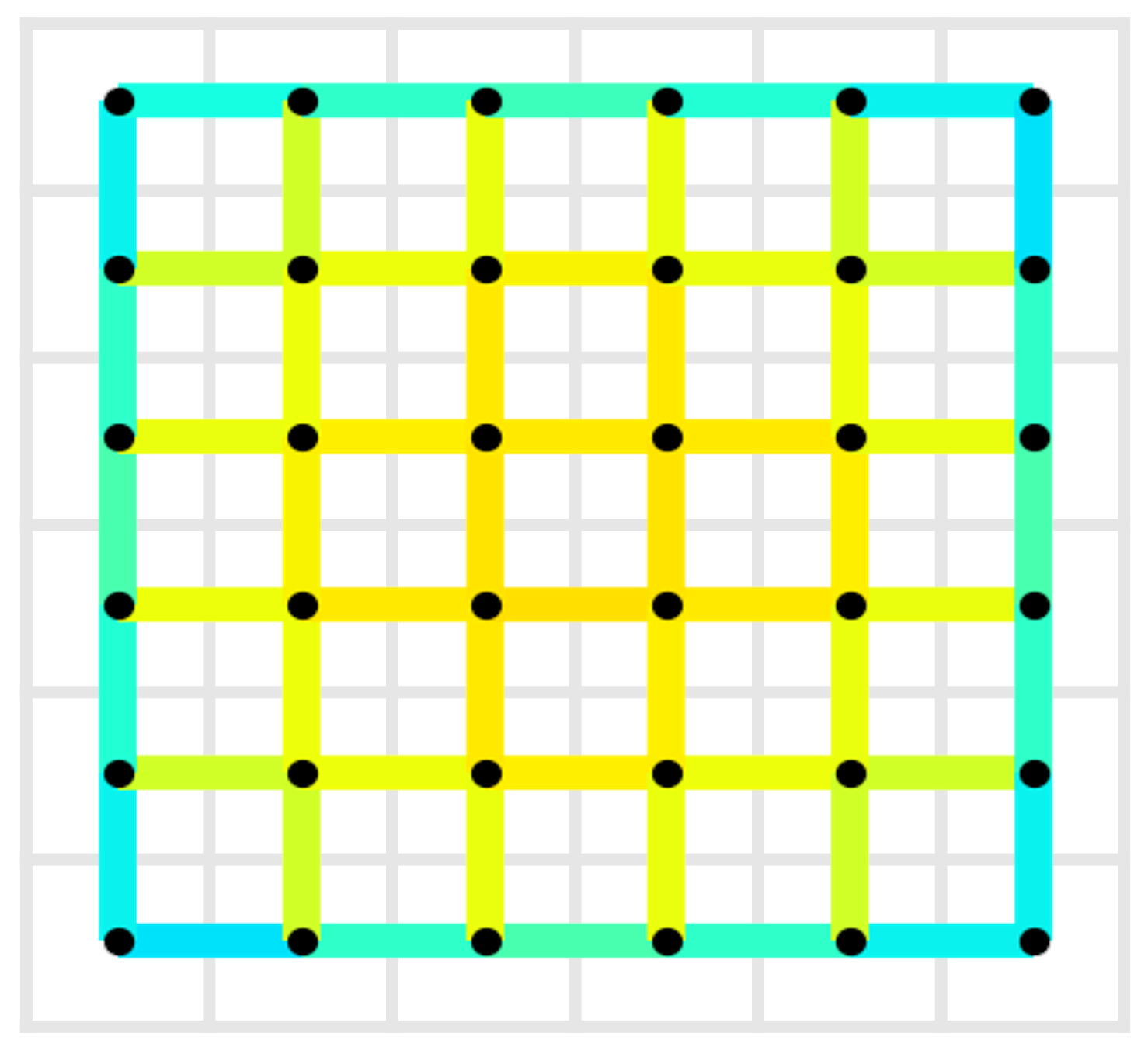}
     \caption{\label{fig:edge_freq_seeded}Whole-Plan Flood Fill (Uniform Seeds)}
     \end{subfigure}\ \ \ \ 
    \begin{subfigure}[ht]{0.3\textwidth}
     \includegraphics[width=\textwidth]{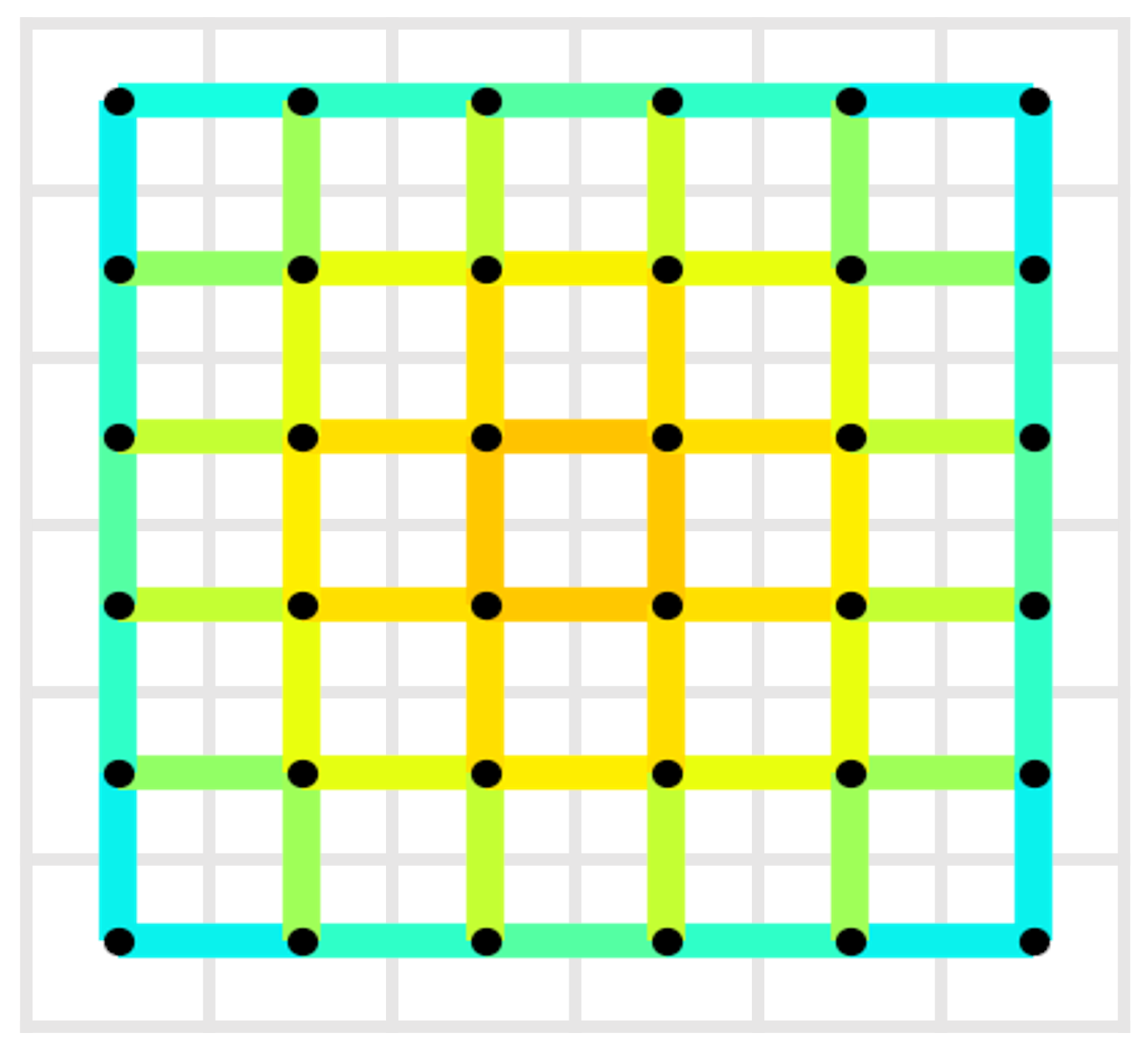}\caption{\label{fig:boundary_edge_freq_seeded} Whole-Plan Flood Fill (Boundary Seeds)}
     \end{subfigure}\ \ \ \ 
    \begin{subfigure}[ht]{0.3\textwidth}
    \includegraphics[width=\textwidth]{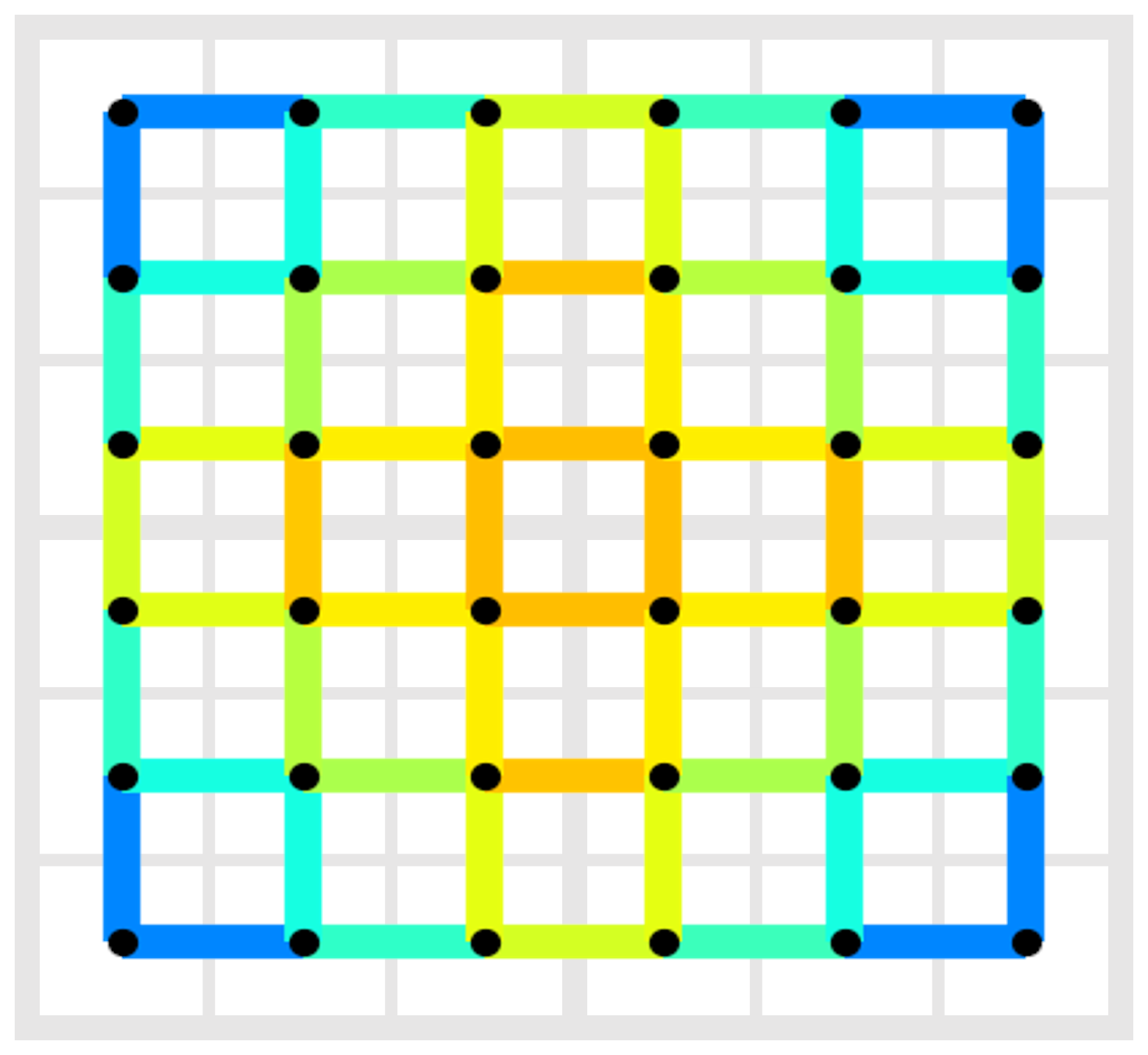}
    \caption{\label{fig:spaced_edge_freq_seeded} Whole-Plan Flood Fill (Zone Seeds)}
    \end{subfigure}\\
    \ \ \ \      \begin{subfigure}[ht]{0.5\textwidth}
     \includegraphics[width=\textwidth]{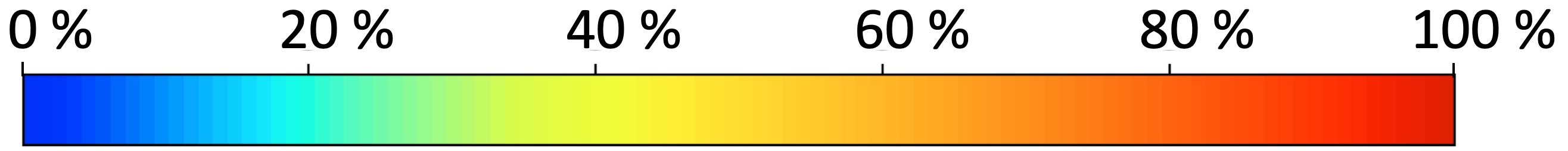}
     \end{subfigure}
\caption{\label{fig:flood_fill_6x6} Cut-edge heatmaps corresponding to samples in  Figure~\ref{fig:6_violin}. The color indicates the percent of plans in the sample in which that edge is a cut edge.
Figure~\ref{fig:6_violin} shows that all of the flood fill variants tend to have fewer cut edges than would be expected from a uniform sample (compare the top distribution to the bottom five). These heatmaps show that the relative distribution of \emph{where} the cut edges occur is similar for several of the variants (compare (b) to (d) and (e)). The plans made by the bounding box method (c) tend to have \emph{substantially} fewer cut edges than would be expected from a uniform sample (see Figure~\ref{fig:6_violin}), and we see here that edge cut frequency corresponds to proximity to the center of the grid. All three whole-plan variants tend to produce samples with a smaller number of cut edges than the full enumeration, where edge cut frequency increases closer to the center of the grid, and in the zone-seeded variant the edges close to the zone boundaries are cut substantially more frequently than the edges far from the zone boundaries.}    
\end{figure}

\begin{figure} 
\centering
    \begin{subfigure}[h]{0.3\textwidth}
      \includegraphics[width=\textwidth, trim={1cm 1cm 1cm 1cm},clip]{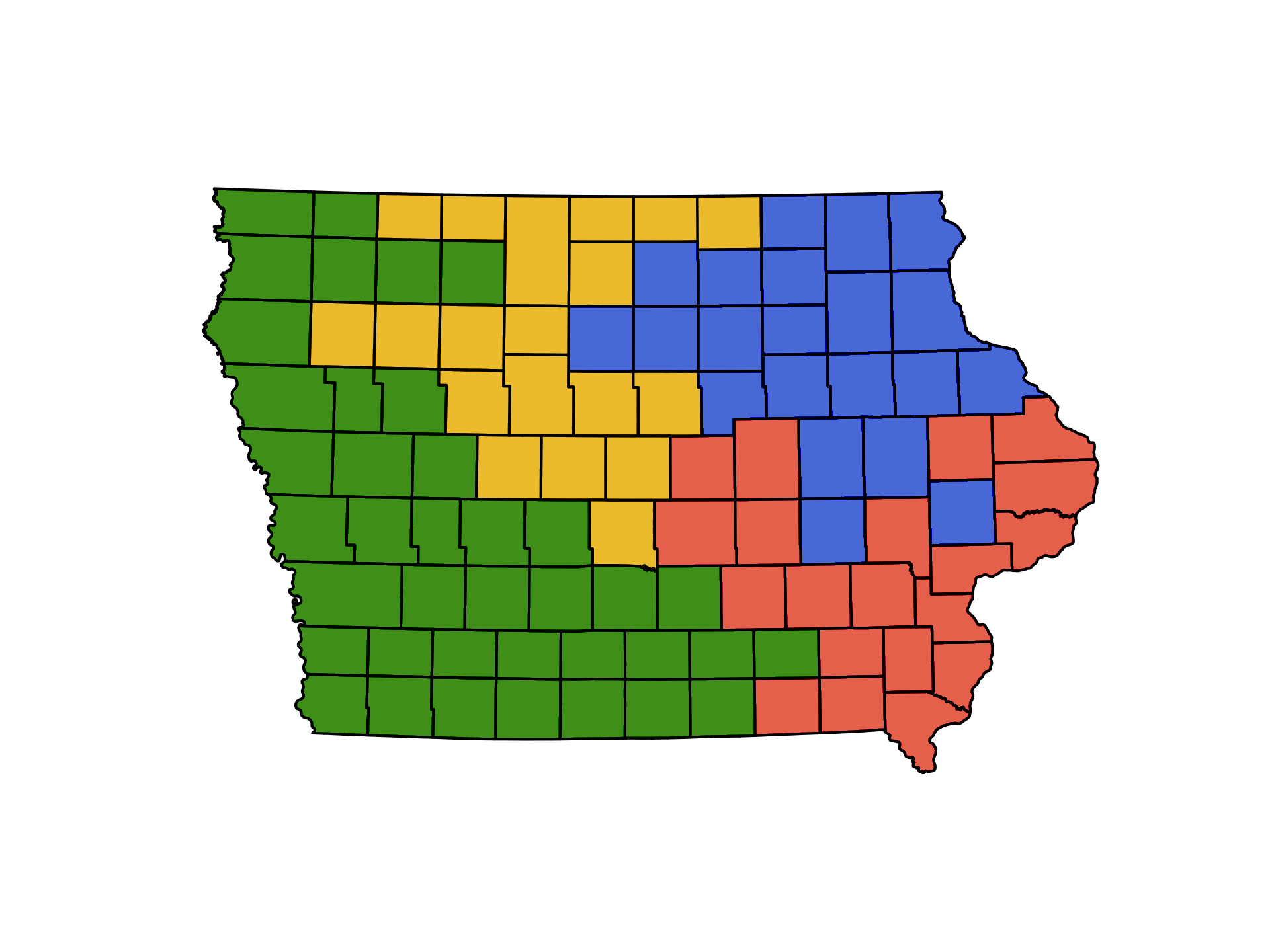}
    \end{subfigure}
    \begin{subfigure}[h]{0.3\textwidth}
      \includegraphics[width=\textwidth, trim={1cm 1cm 1cm 1cm},clip]{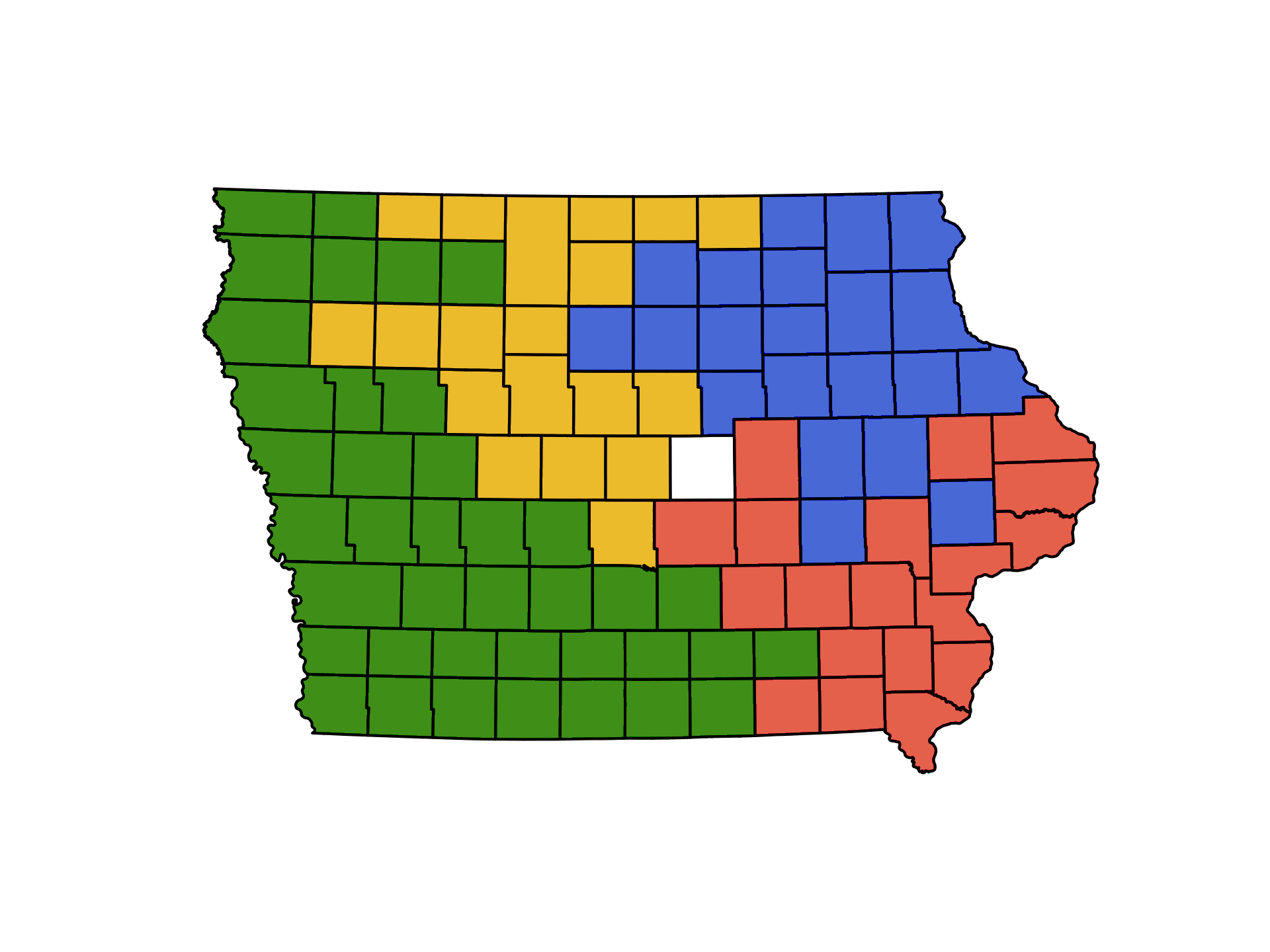}
    \end{subfigure}
    \begin{subfigure}[h]{0.3\textwidth}
      \includegraphics[width=\textwidth, trim={1cm 1cm 1cm 1cm},clip]{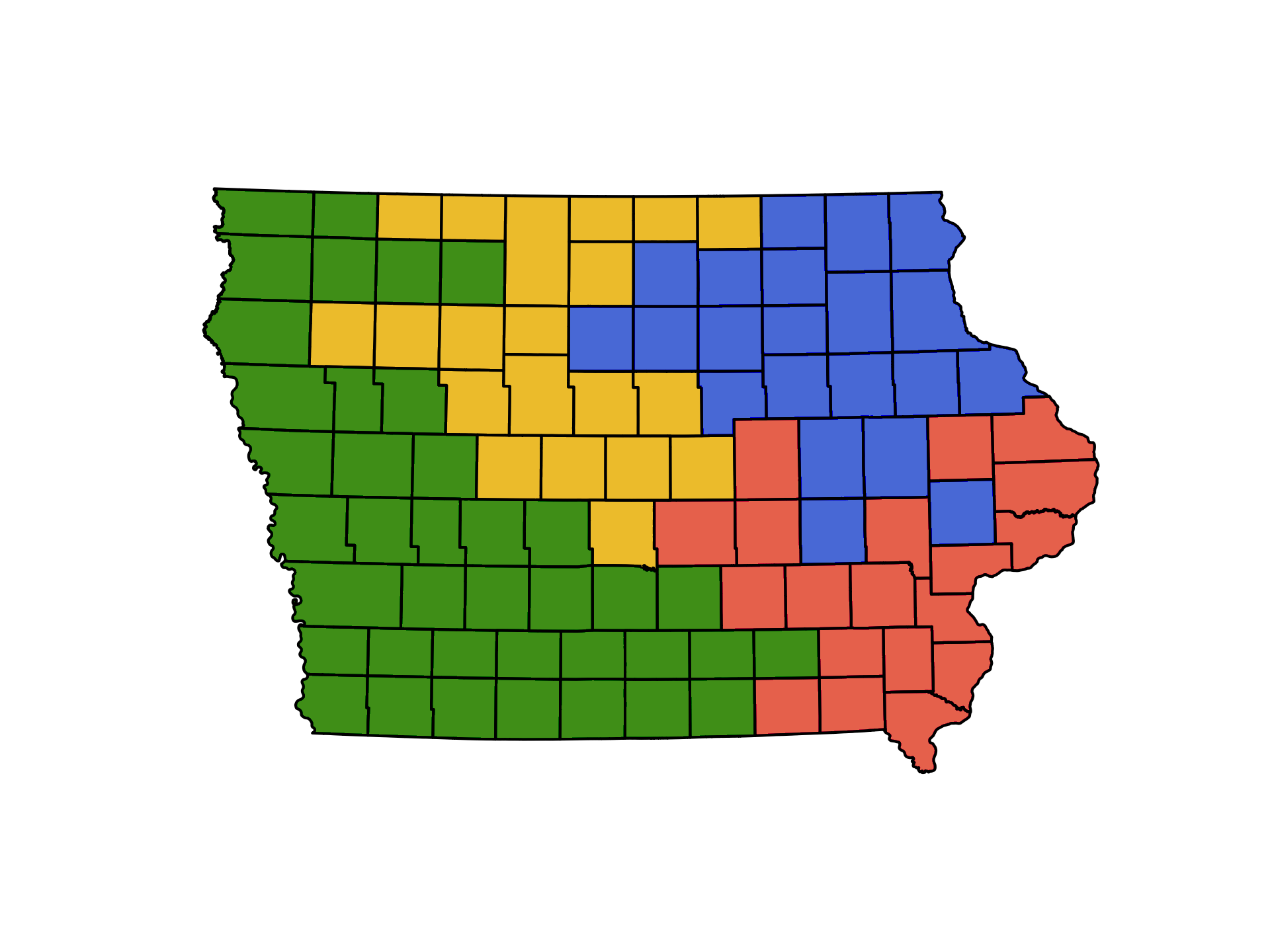}
    \end{subfigure}
\caption{\label{fig:flipstepexample} In this illustration of a {\bf flip step}, the white node in the middle figure is the randomly chosen district boundary unit.  The left figure shows the original plan and the right figure shows the plan after the white unit `flips' from red to yellow.}
\end{figure}

\begin{figure} 
\centering
    \begin{subfigure}[ht]{0.23\textwidth}
      \includegraphics[width=\textwidth, trim={2cm 2cm 2cm 2cm},clip]{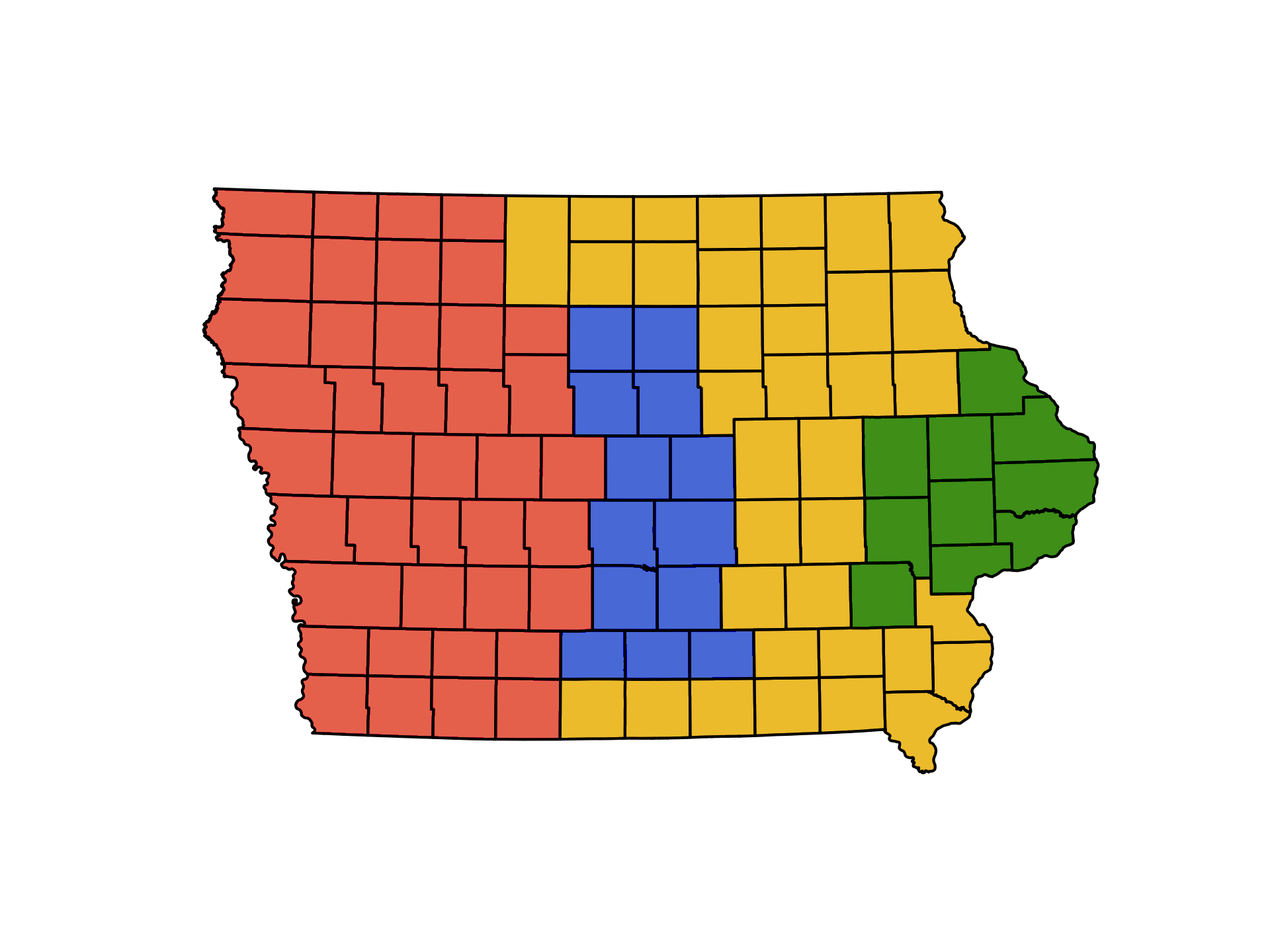}
    \end{subfigure}
    \vline\ 
    \begin{subfigure}[ht]{0.23\textwidth}
      \includegraphics[width=\textwidth, trim={2cm 2cm 2cm 2cm},clip]{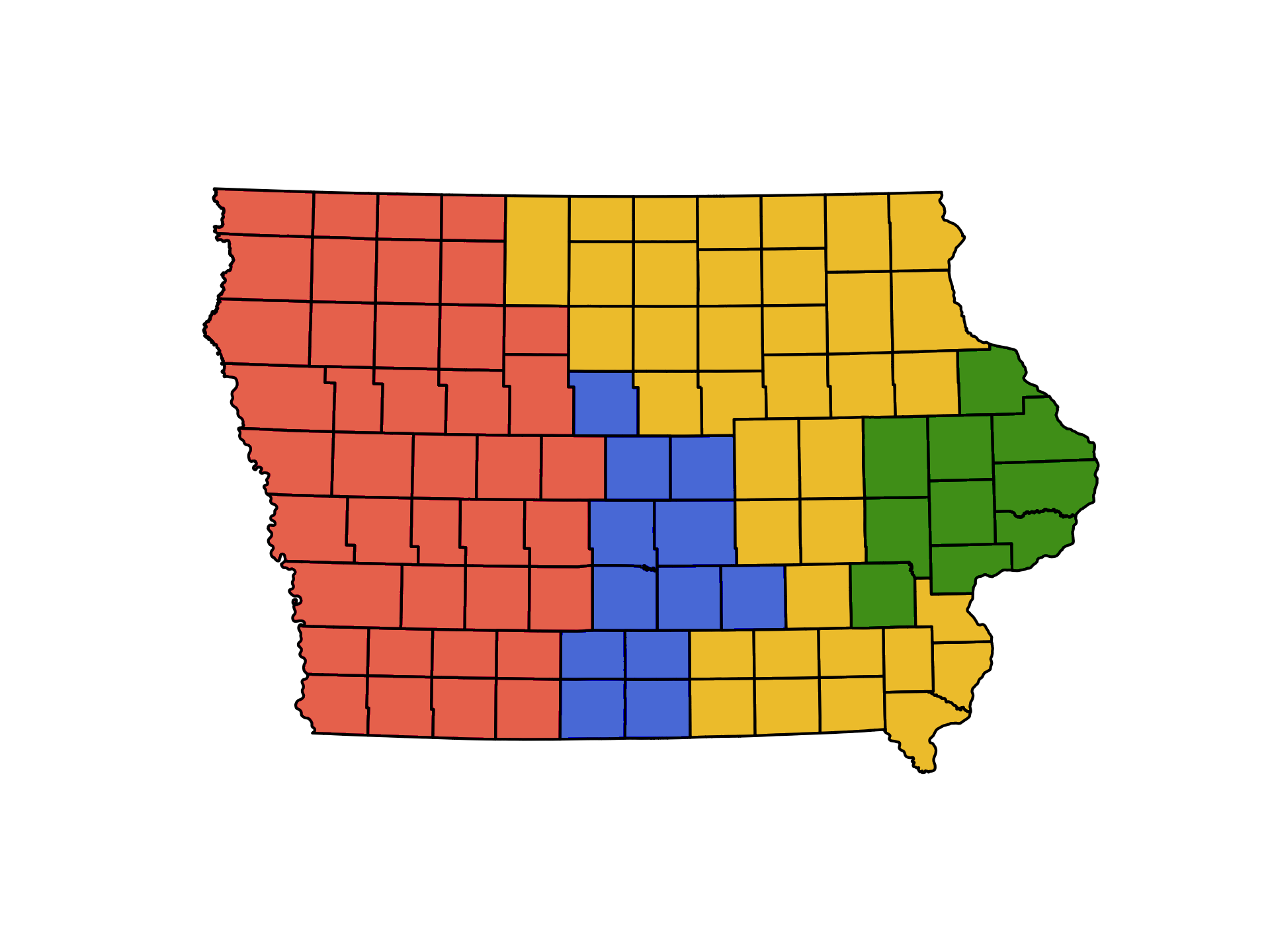}
    \end{subfigure}
    \begin{subfigure}[ht]{0.23\textwidth}
      \includegraphics[width=\textwidth, trim={2cm 2cm 2cm 2cm},clip]{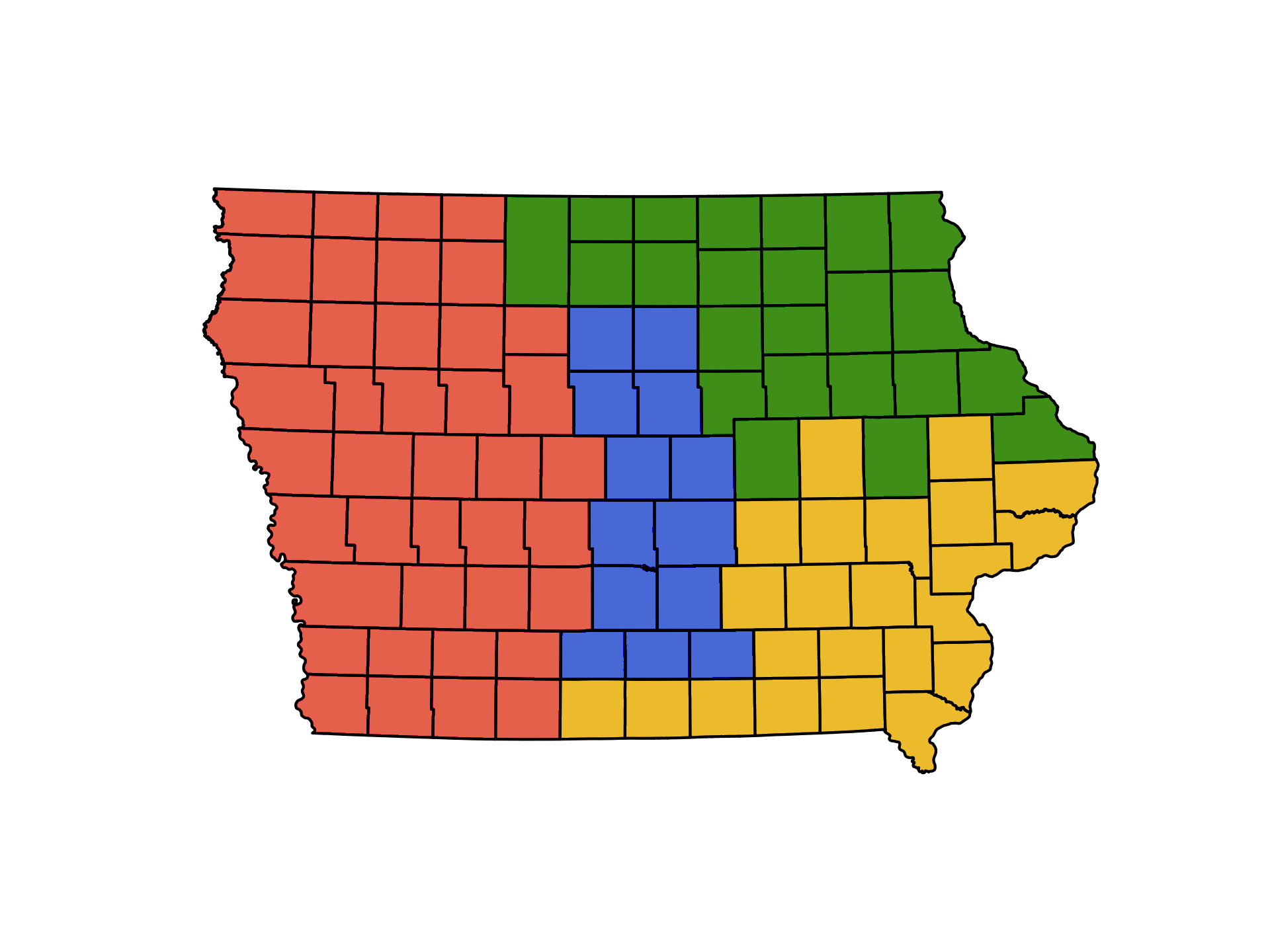}
    \end{subfigure}
       \begin{subfigure}[ht]{0.23\textwidth}
      \includegraphics[width=\textwidth, trim={2cm 2cm 2cm 2cm},clip]{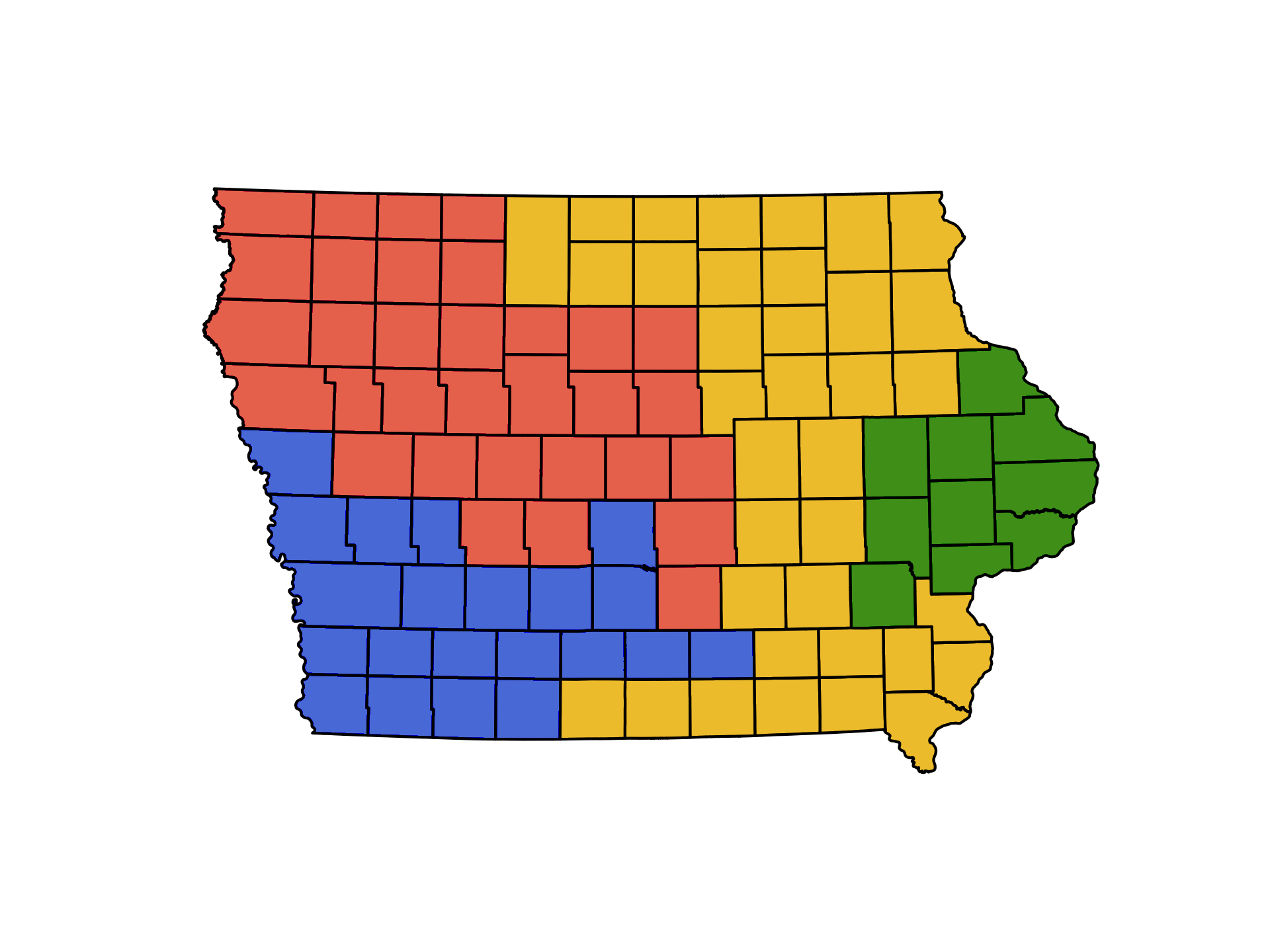}
    \end{subfigure}
\caption{\label{fig:recom_example}The figures on the right illustrate three different potential outcomes from taking a single {\bf recombination} step from the starting plan shown in the left figure.}
\end{figure}

\begin{figure}
\centering
     \includegraphics[width=0.9\textwidth, trim={1cm 0cm 0cm 0cm},clip]{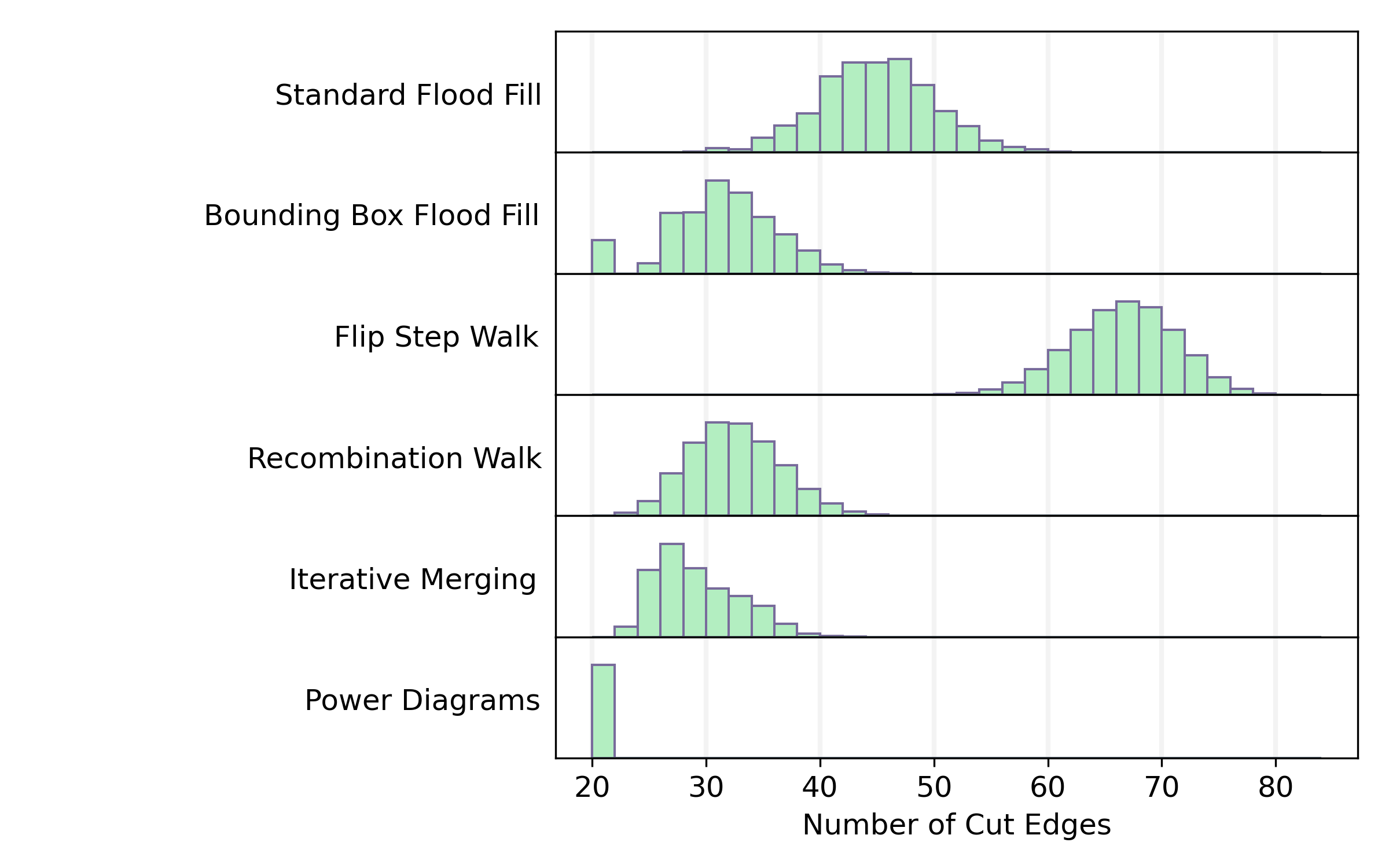}
    \caption{\label{fig:10_violin}Cut-edge distribution comparison for generating four-district plans with up to 5\% population deviation for a $10 \times 10$ grid.}    
\end{figure}

\begin{figure}
\centering
     \includegraphics[width=0.9\textwidth, trim={1cm 0cm 0cm 0cm},clip]{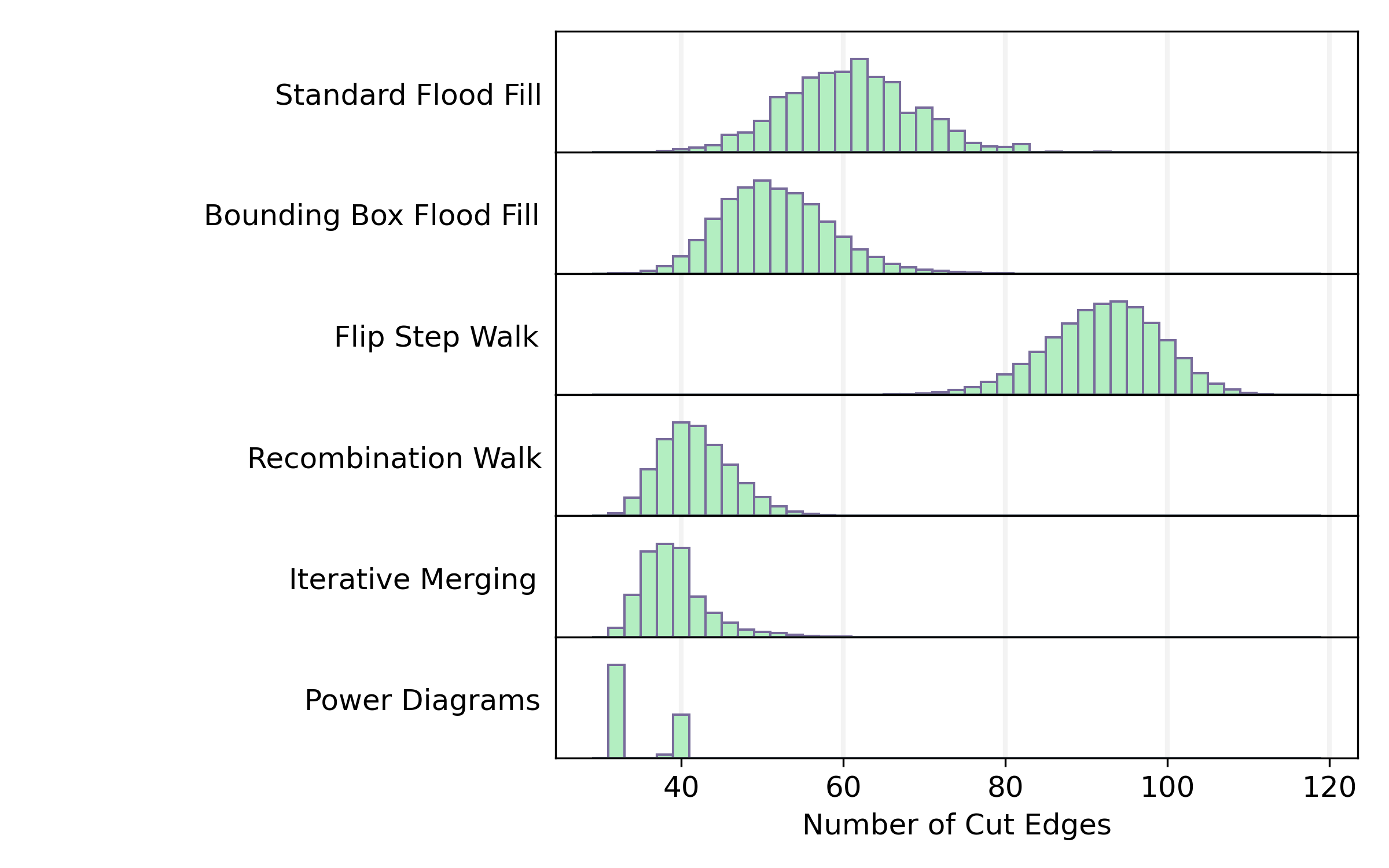}
    \caption{\label{fig:iowa_violin} Cut-edge distribution for generating four-district plans with up to 5\% population deviation using Iowa counties using various methods. The enacted plan (shown in Figure~\ref{fig:benchmarks}) 
    has 47 cut edges and 0.005 percent population deviation.}    
\end{figure}

\begin{figure}
\centering
        \begin{subfigure}[ht]{0.3\textwidth}
     \includegraphics[width=\textwidth]{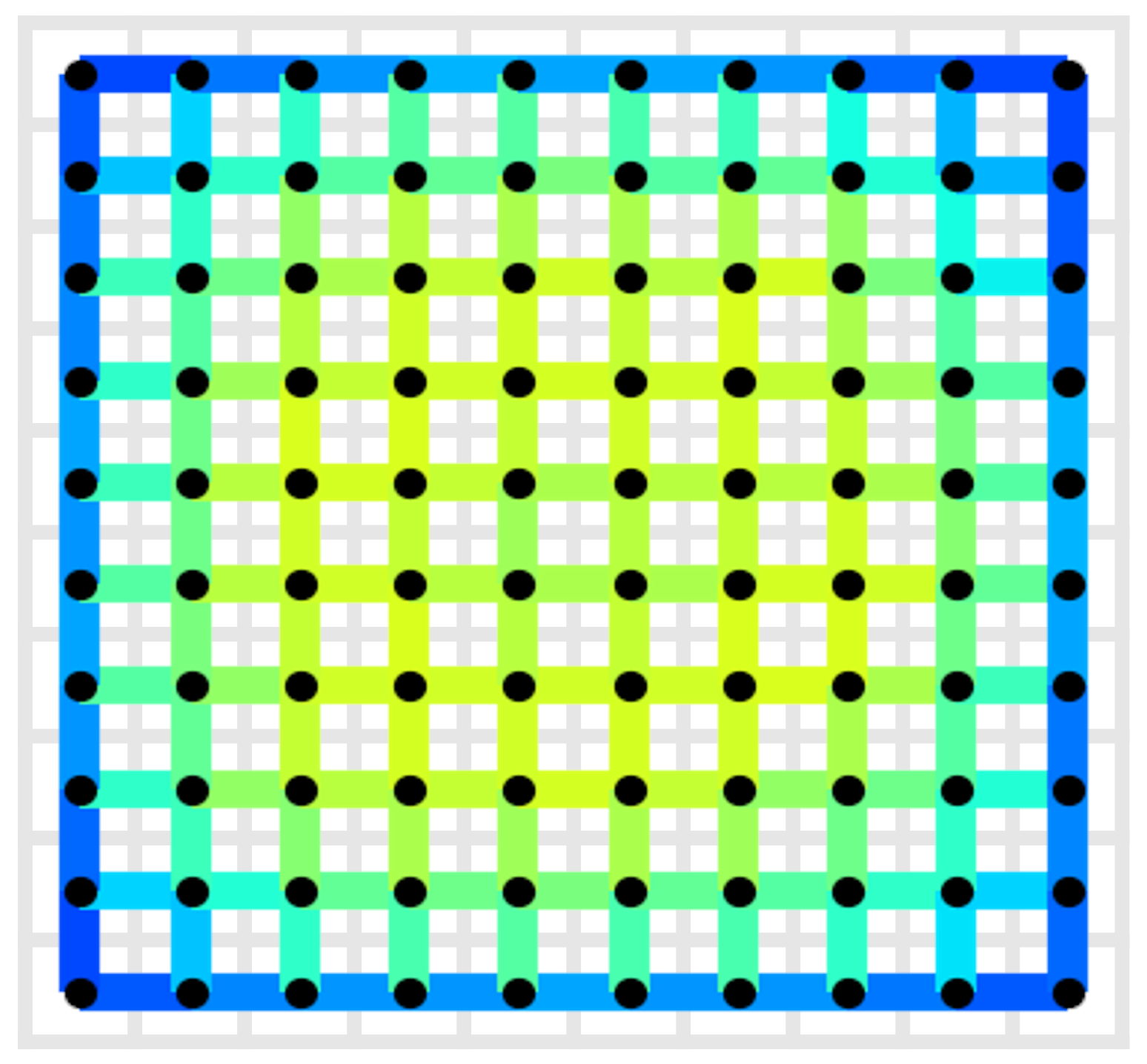}
      \caption{ Standard Flood Fill}
    \end{subfigure}\ \ \ \
    \begin{subfigure}[ht]{0.35\textwidth}
    \centering
     \includegraphics[width=.8571\textwidth]{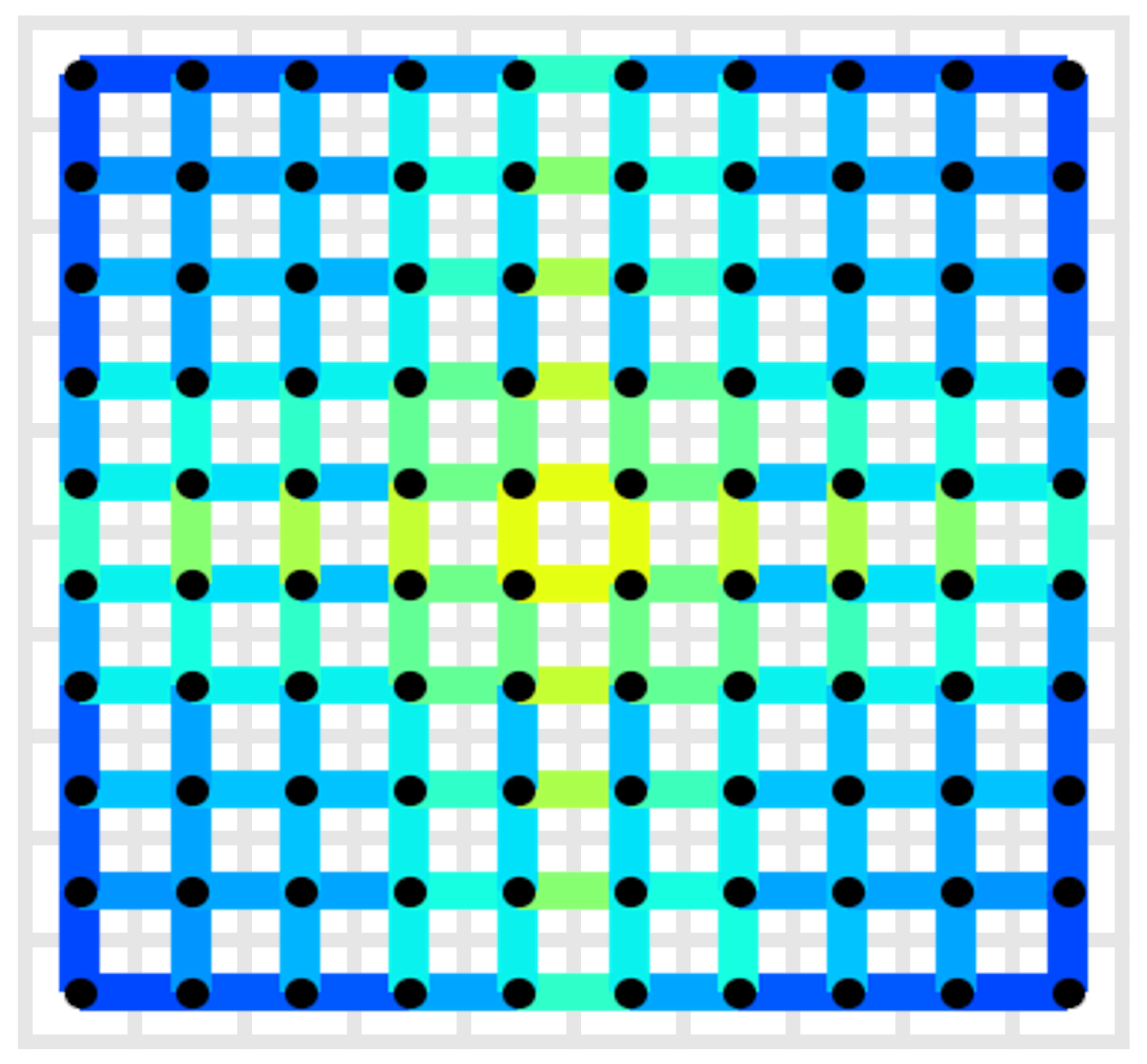}
      \caption{ Bounding Box Flood Fill}
    \end{subfigure}\ \ \ \
    \begin{subfigure}[ht]{0.3\textwidth}
     \includegraphics[width=\textwidth]{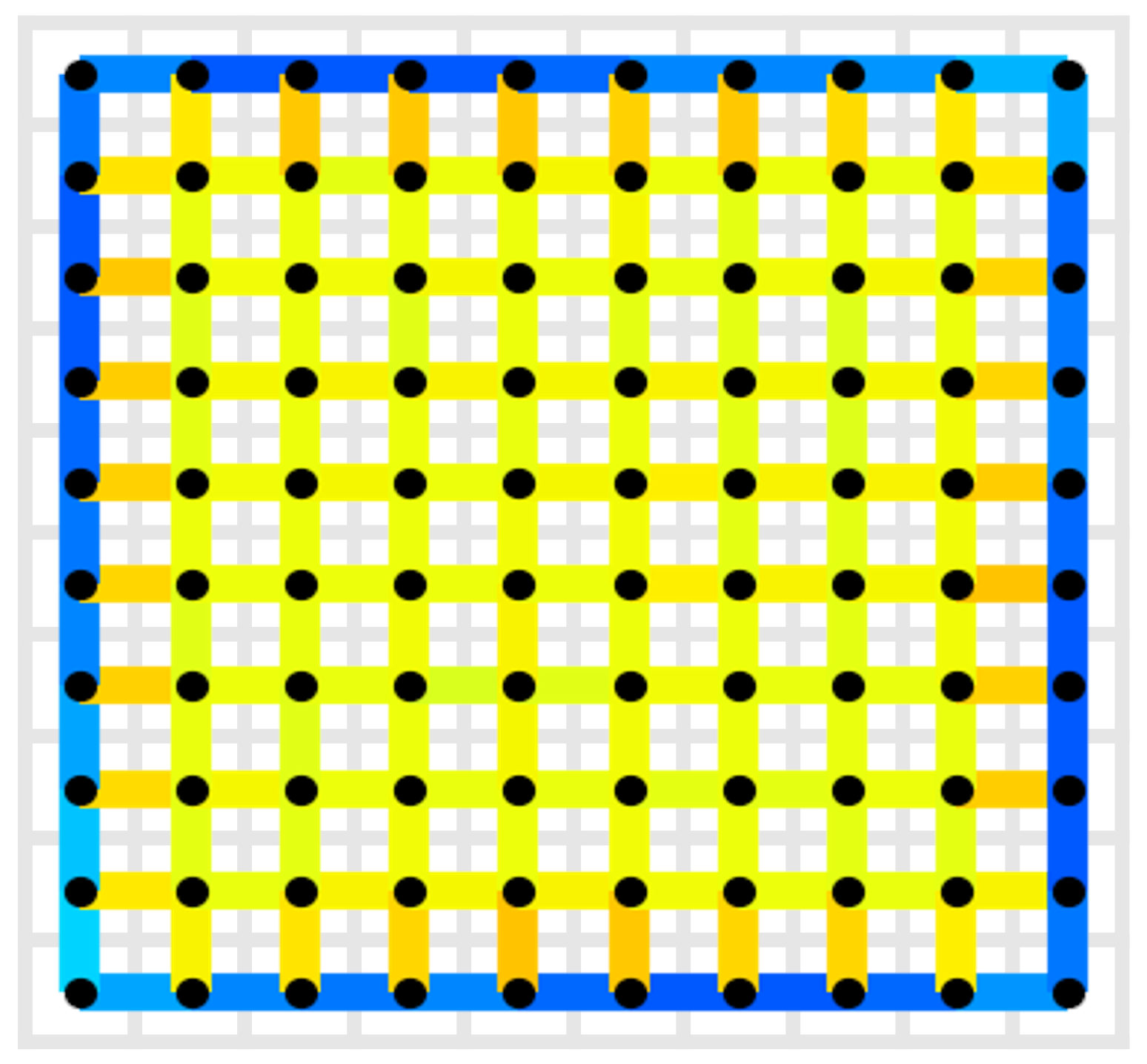}
      \caption{ Flip Step Walk}
    \end{subfigure}\\
    \begin{subfigure}[ht]{0.3\textwidth}
     \includegraphics[width=\textwidth]{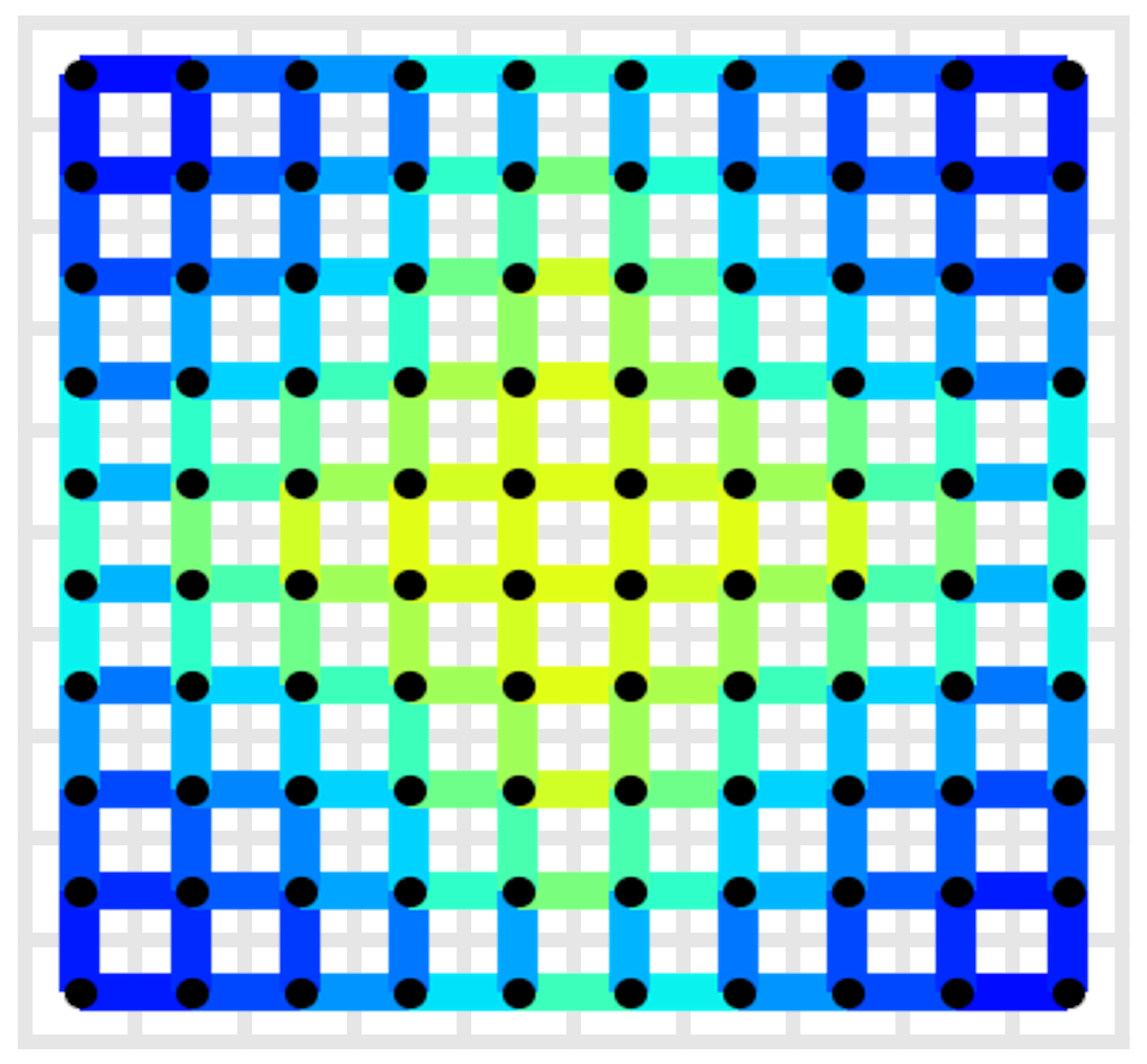}
      \caption{ Recombination Walk}
    \end{subfigure}\ \ \ \ \ \ \ \ 
    \begin{subfigure}[ht]{0.3\textwidth}
     \includegraphics[width=\textwidth]{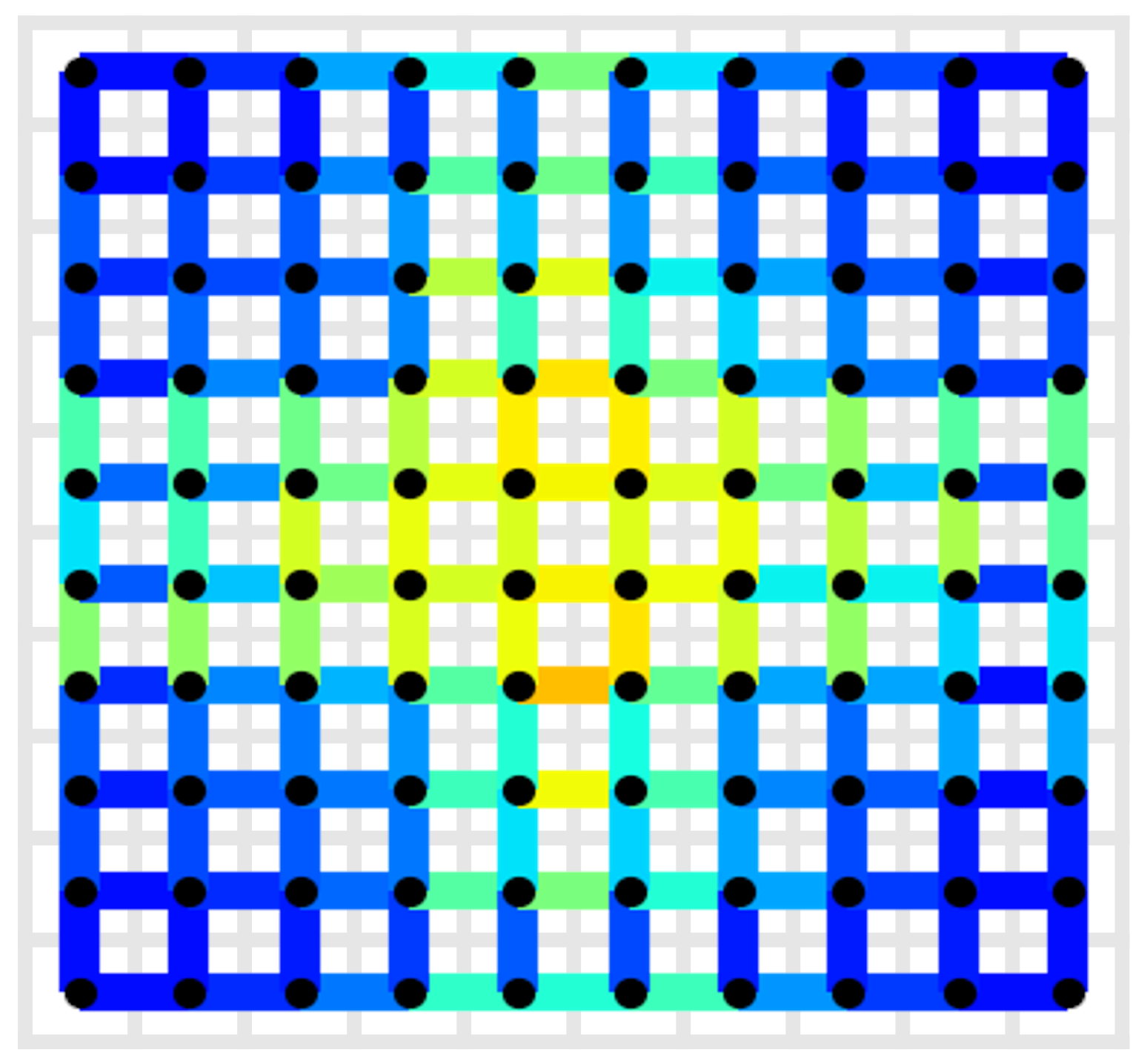}
      \caption{ Iterative Merging}
    \end{subfigure}\ \ \ \ \ \ \ \ 
     \begin{subfigure}[ht]{0.3\textwidth}
     \includegraphics[width=\textwidth]{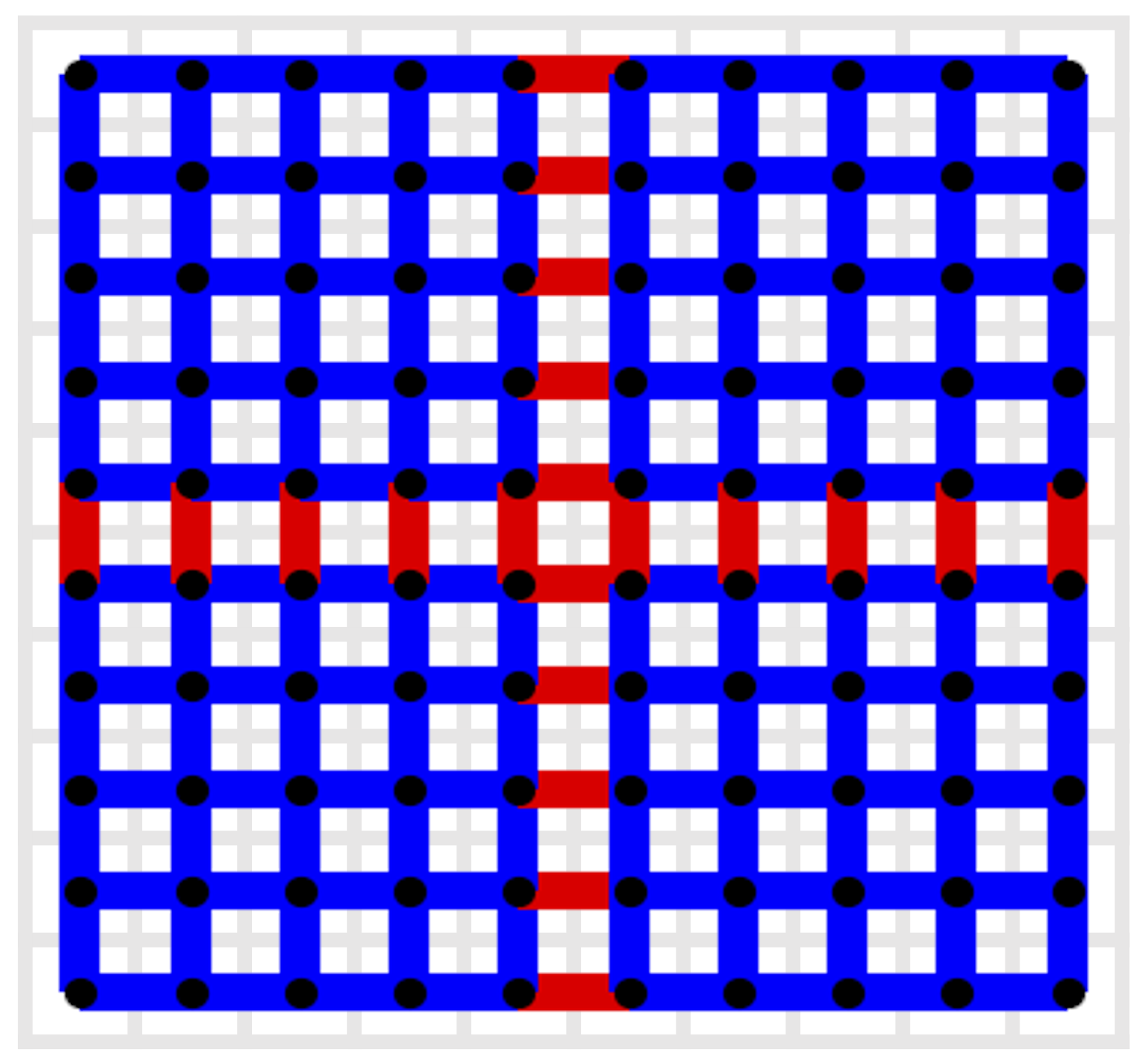}
      \caption{ Power Diagrams}
    \end{subfigure}\\
     \ \ \ \
     \begin{subfigure}[h]{0.5\textwidth}
     \includegraphics[width=\textwidth]{./BeckerSolomon/figs/heat_map_colorbar.png}
    \end{subfigure}
    \caption{\label{fig:flood_fill_10grid_exp} Cut-edge heatmap corresponding to samples in  Figure~\ref{fig:10_violin}. The color indicates the percent of plans in the sample in which that edge is a cut edge.}
\end{figure}

\begin{figure} 
\centering
        \begin{subfigure}[ht]{0.3\textwidth}
     \includegraphics[width=\textwidth]{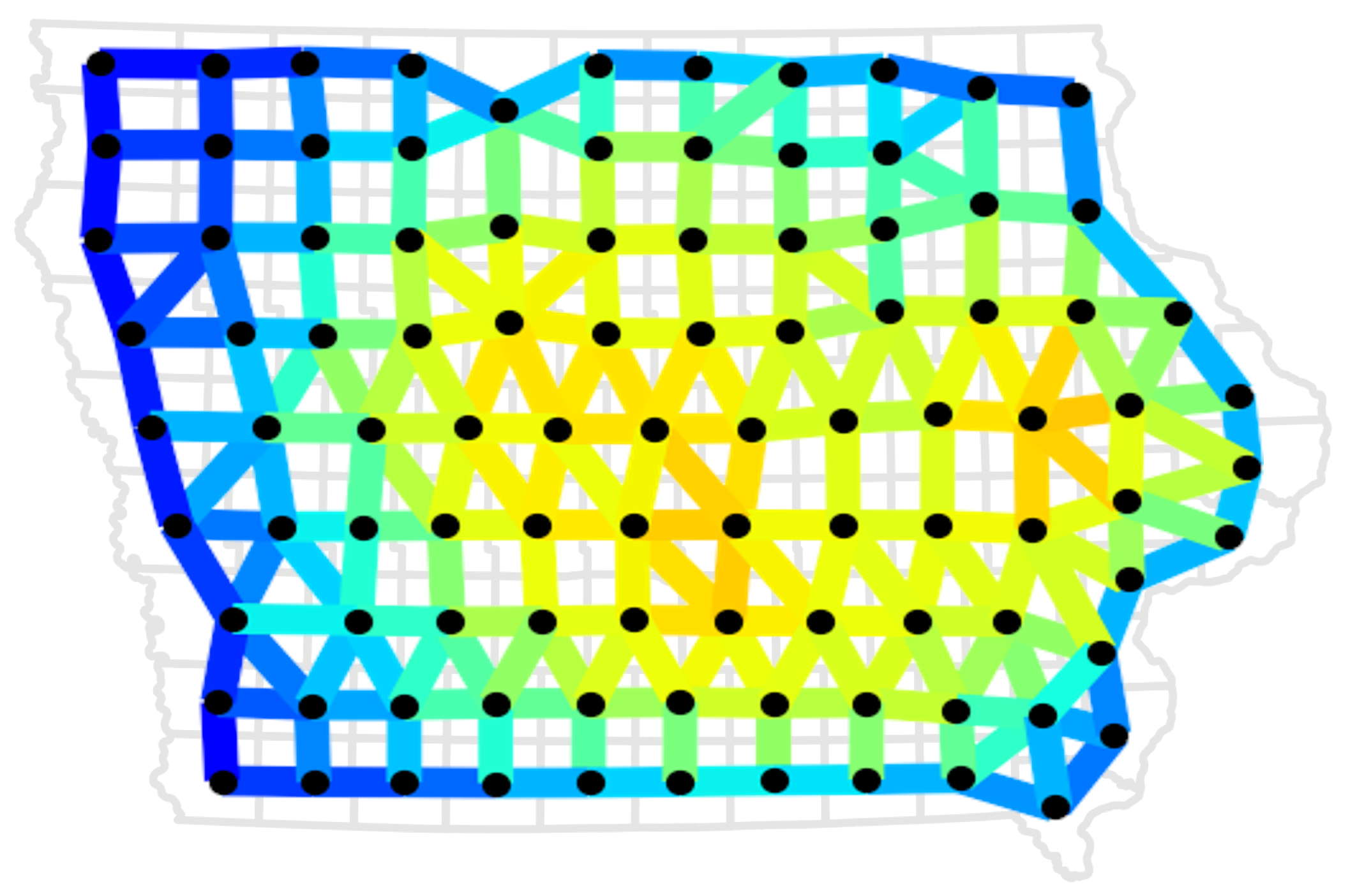}
      \caption{ Standard Flood Fill}
    \end{subfigure}\ \ \ \
    \begin{subfigure}[ht]{0.35\textwidth}
    \centering
     \includegraphics[width=.8571\textwidth]{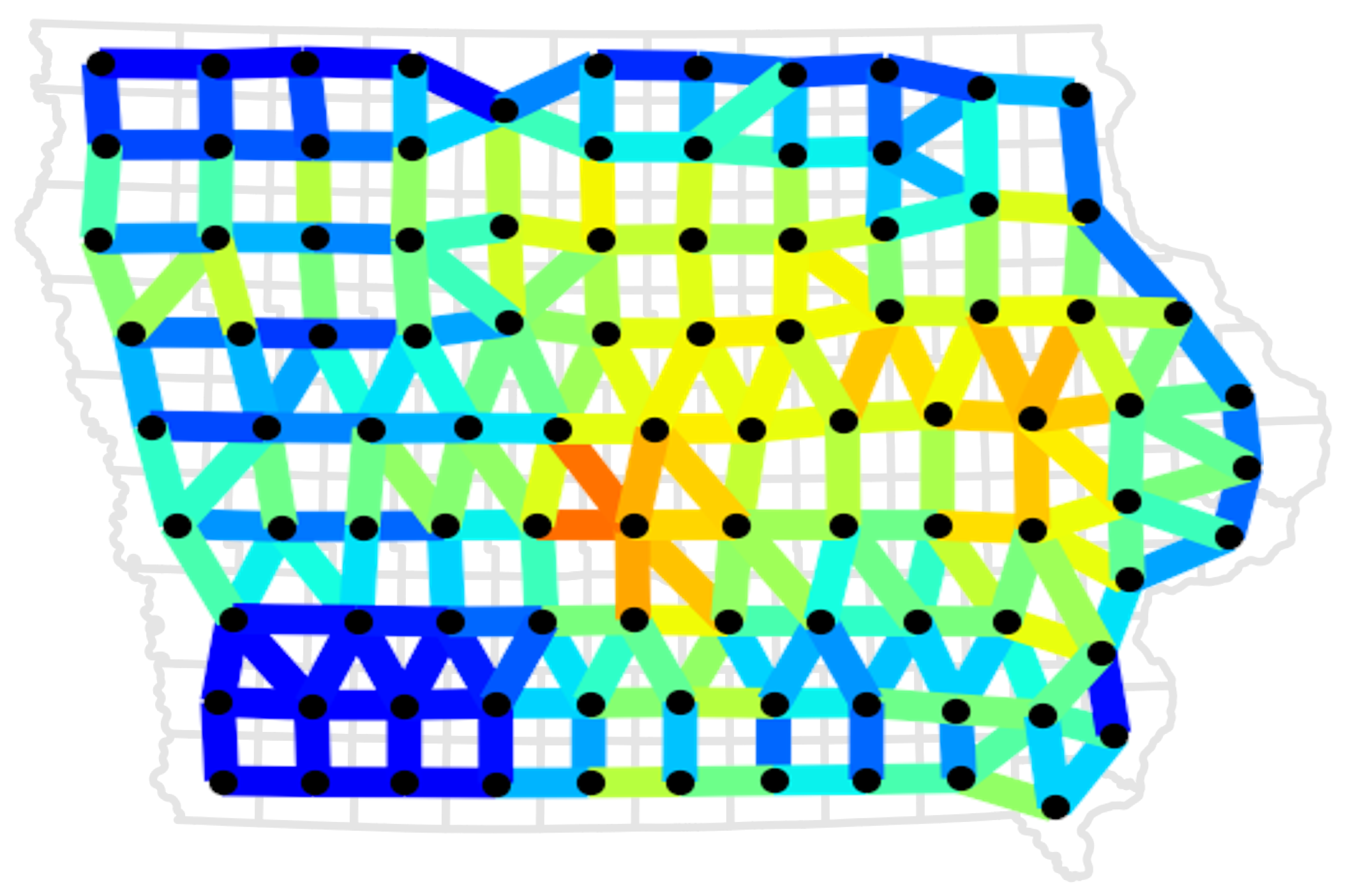}
      \caption{ Bounding Box Flood Fill}
    \end{subfigure}\ \ \ \
    \begin{subfigure}[ht]{0.3\textwidth}
     \includegraphics[width=\textwidth]{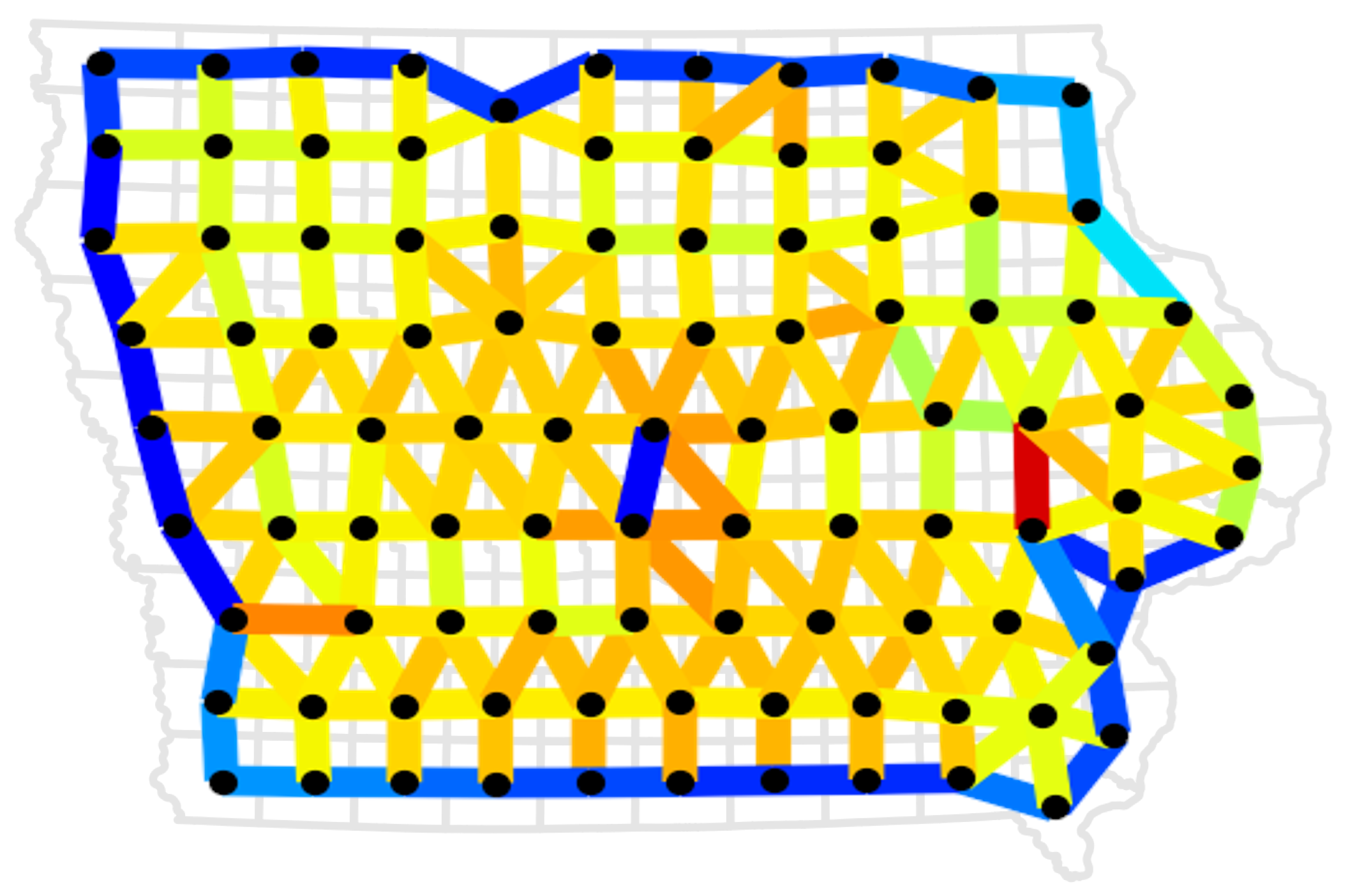}
      \caption{ Flip Step Walk}
    \end{subfigure}\\
    \begin{subfigure}[ht]{0.3\textwidth}
     \includegraphics[width=\textwidth]{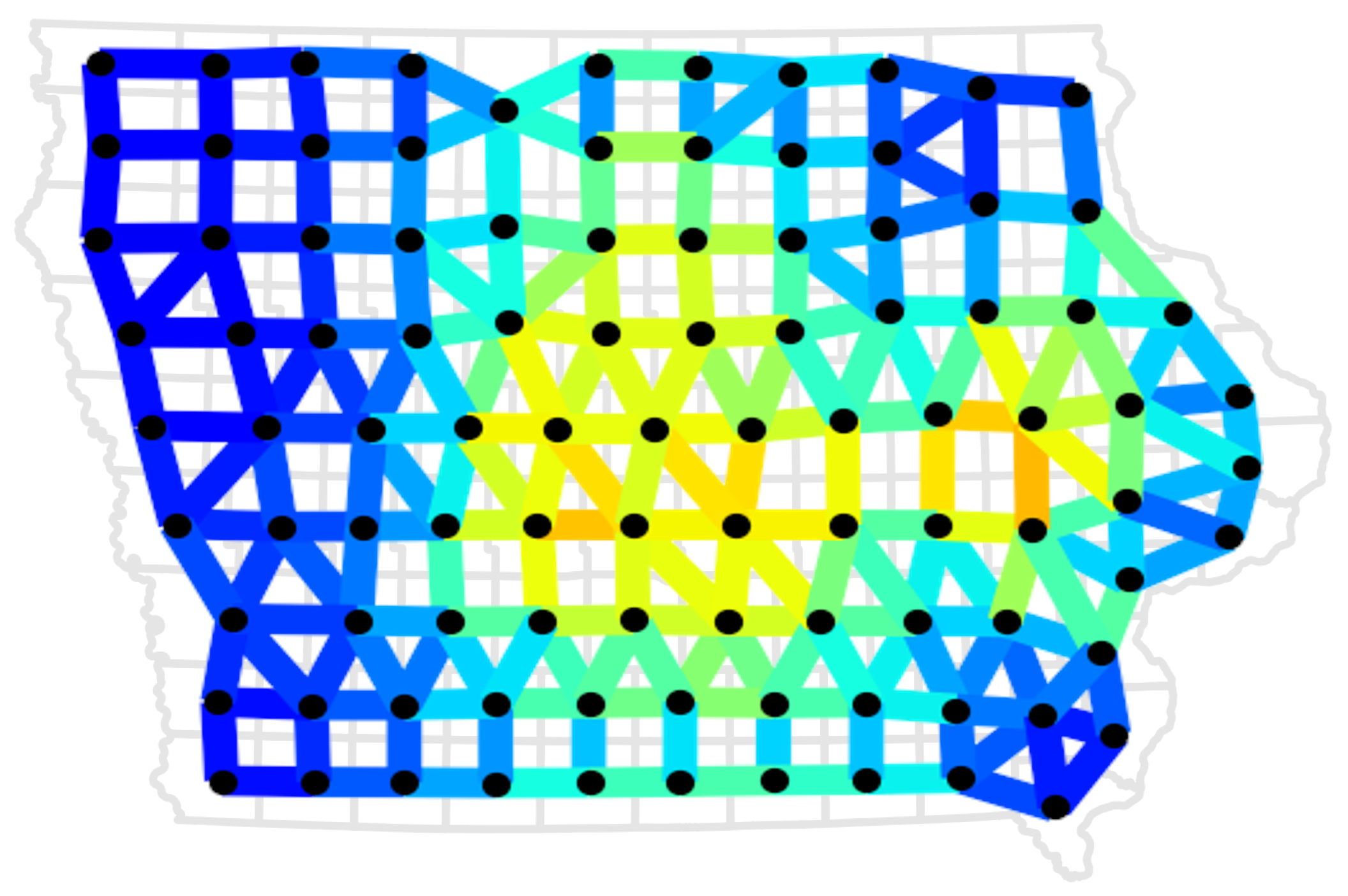}
      \caption{ Recombination Walk}
    \end{subfigure}\ \ \ \ \ \ \ \ 
    \begin{subfigure}[ht]{0.3\textwidth}
     \includegraphics[width=\textwidth]{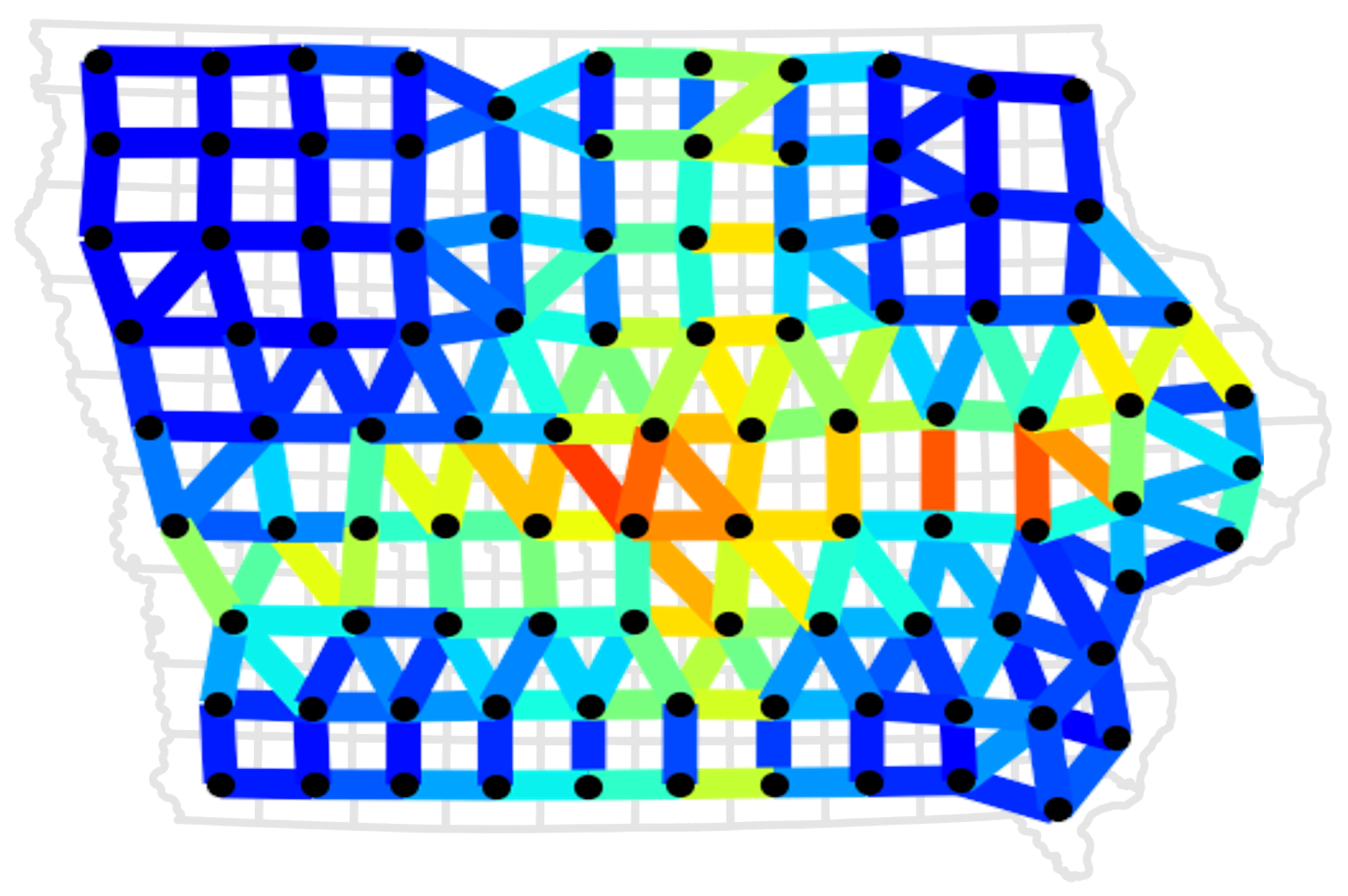}
      \caption{ Iterative Merging}
    \end{subfigure}\ \ \ \ \ \ \ \ 
    \begin{subfigure}[ht]{0.3\textwidth}
     \includegraphics[width=\textwidth]{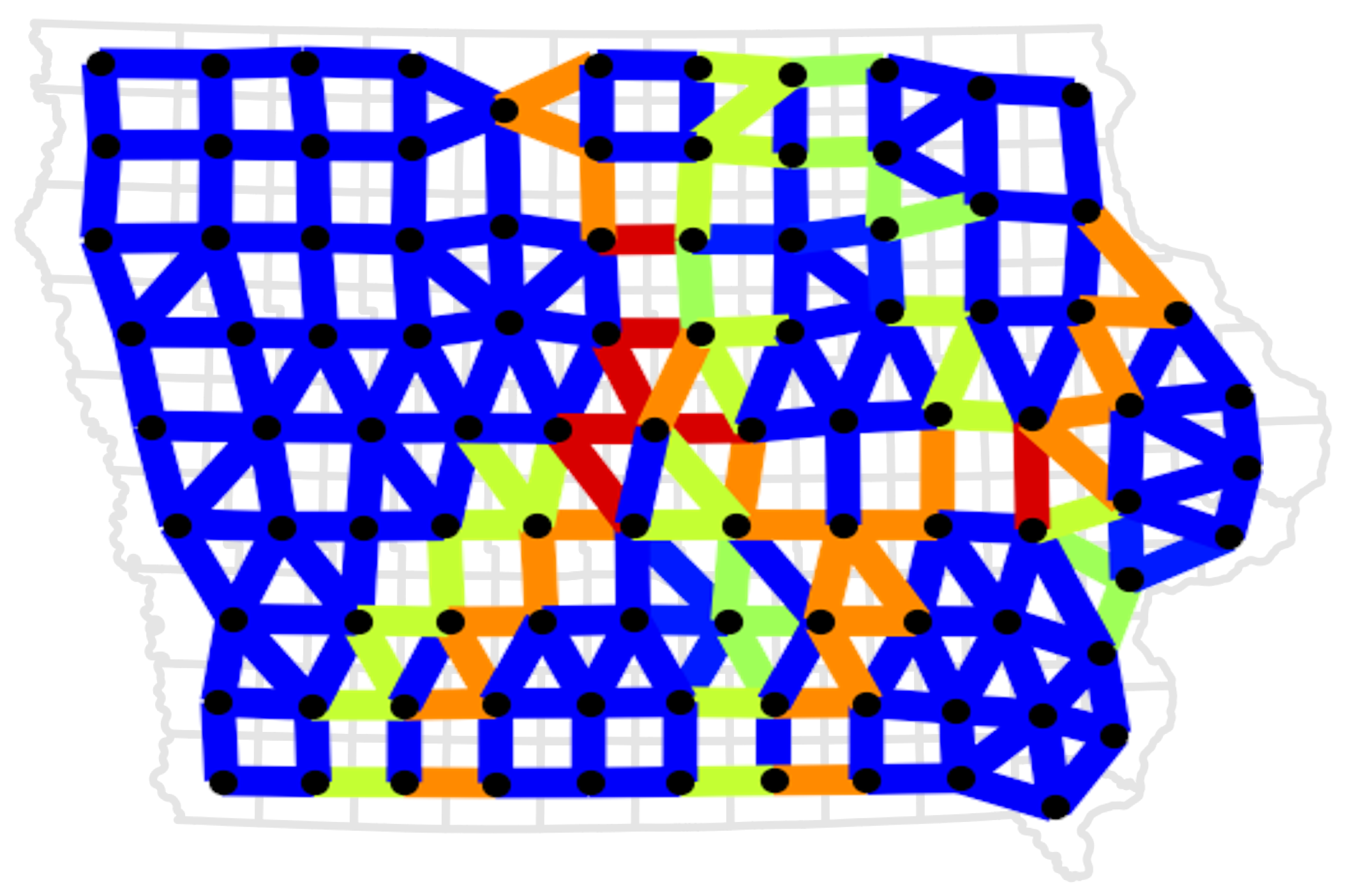}
      \caption{ Power Diagrams}
    \end{subfigure}\\
     \ \ \ 
     \begin{subfigure}[h]{0.5\textwidth}
     \includegraphics[width=\textwidth]{./BeckerSolomon/figs/heat_map_colorbar.png}
    \end{subfigure}
    \caption{\label{fig:flood_fill_iowa_exp} Cut-edge frequency comparison corresponding to samples in  Figure~\ref{fig:iowa_violin}. The color indicates the percent of plans in the sample in which that edge is a cut edge.}
\end{figure}

\begin{figure} 
\centering
    \begin{subfigure}[h]{0.23\textwidth}
      \includegraphics[width=\textwidth]{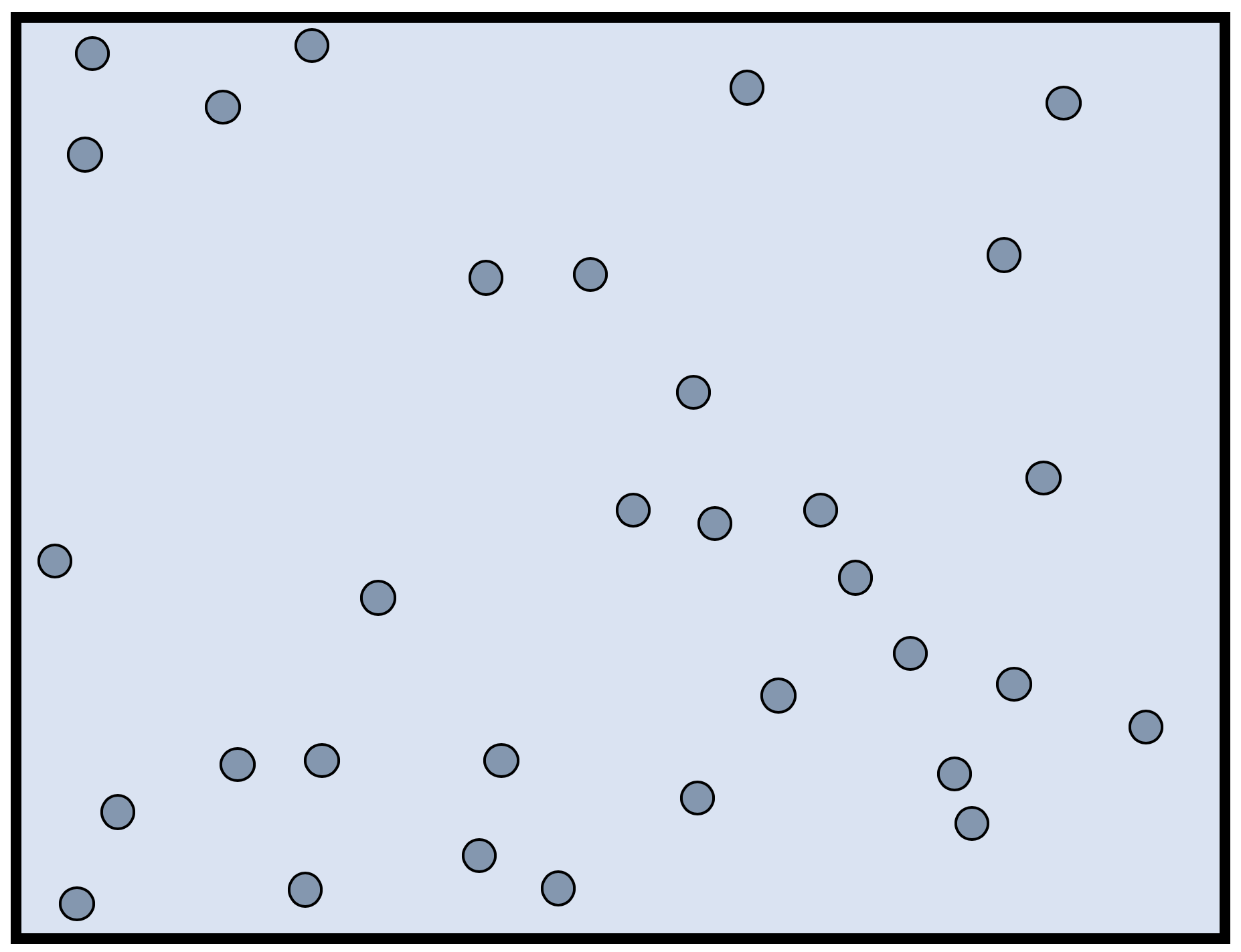}
    \end{subfigure}\ \ 
    \begin{subfigure}[h]{0.23\textwidth}
      \includegraphics[width=\textwidth]{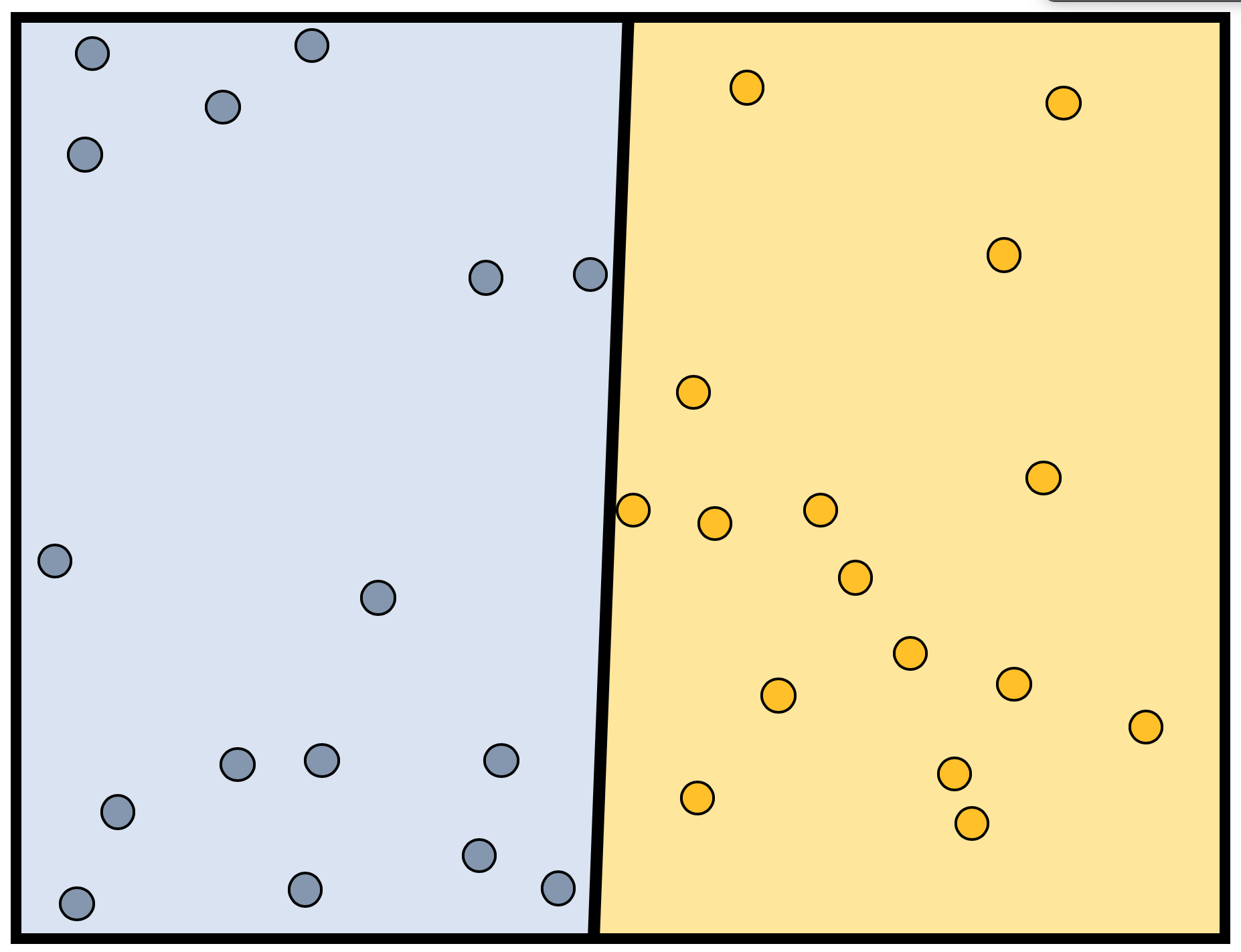}
    \end{subfigure}\ \ 
    \begin{subfigure}[h]{0.23\textwidth}
      \includegraphics[width=\textwidth]{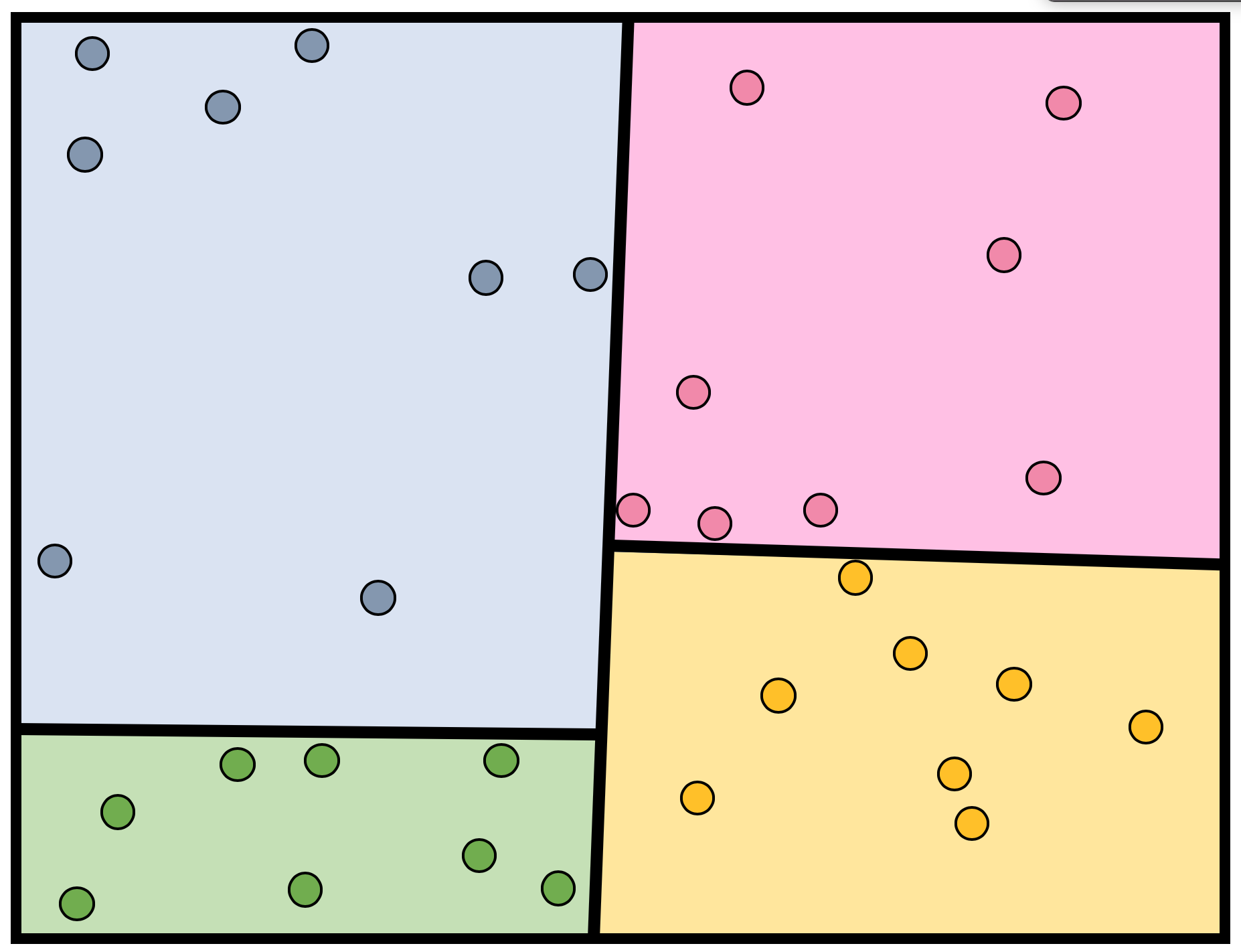}
    \end{subfigure}\ \ 
    \begin{subfigure}[h]{0.23\textwidth}
      \includegraphics[width=\textwidth]{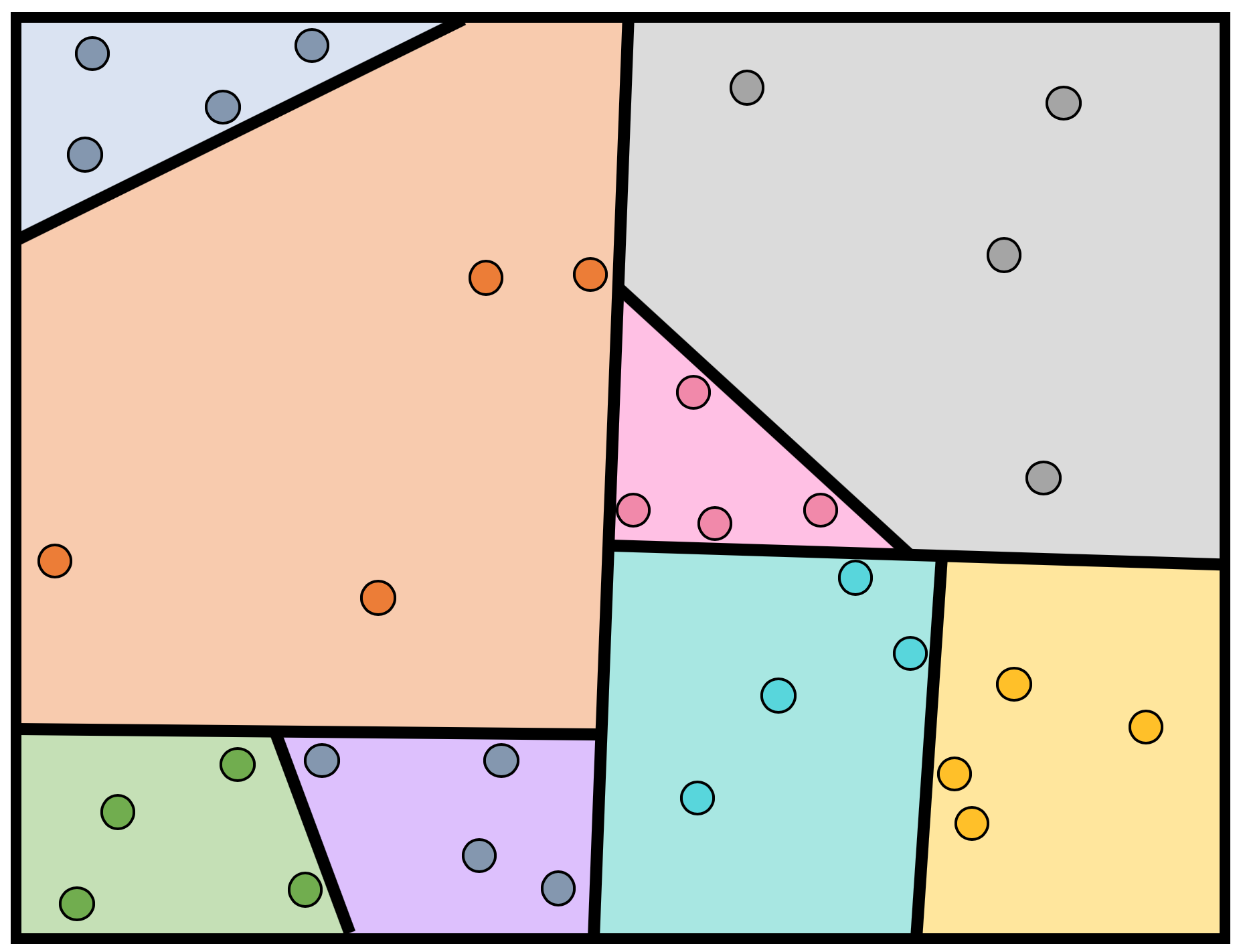}
    \end{subfigure}
\caption{\label{fig:splitline} In this illustration of the shortest-splitline algorithm, the left figure shows the population distribution of a fictional region.  The subsequent figures show the regions being bisected using the shortest line that evenly divides the region's population.}
\end{figure}

\begin{figure} 
\centering
    \begin{subfigure}[h]{0.21\textwidth}
      \includegraphics[width=\textwidth]{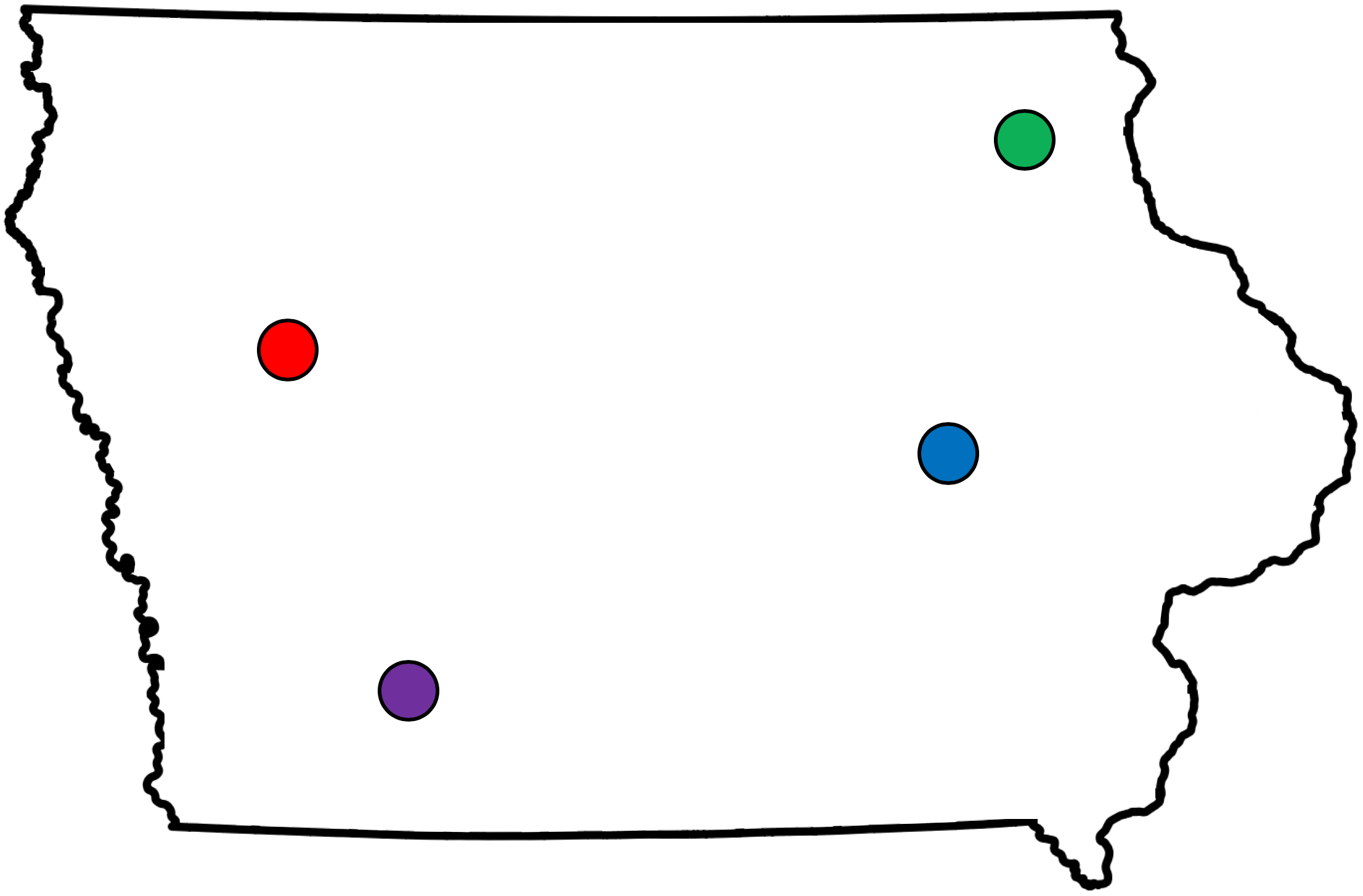}
      \label{fig:centers1}
    \end{subfigure}
    \begin{subfigure}[h]{0.21\textwidth}
      \includegraphics[width=\textwidth]{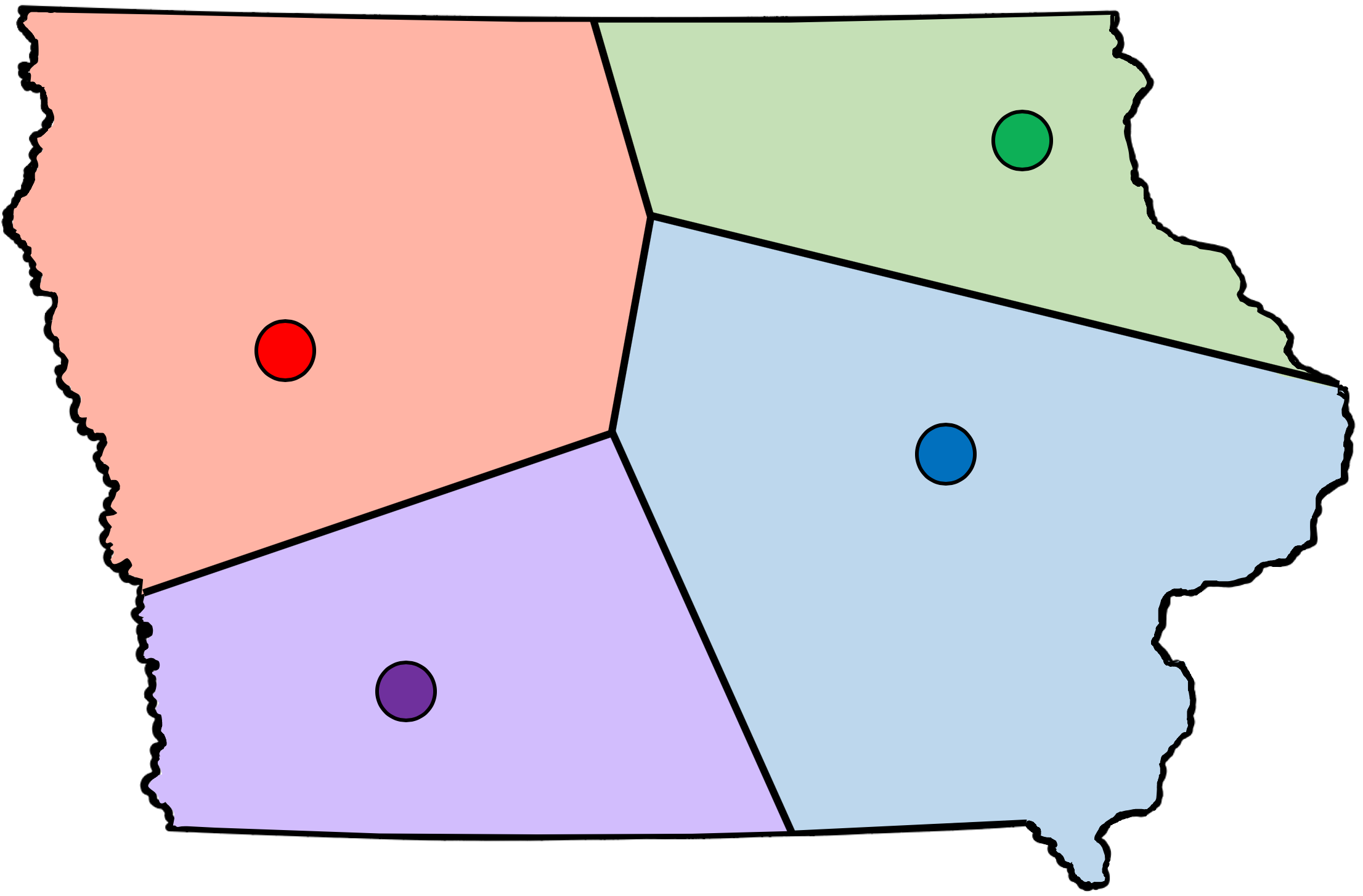}
      \label{fig:voronoi1}
    \end{subfigure}\ \ \ \ \ \ \ \ \ \ \ \ \ 
    \begin{subfigure}[h]{0.21\textwidth}
      \includegraphics[width=\textwidth]{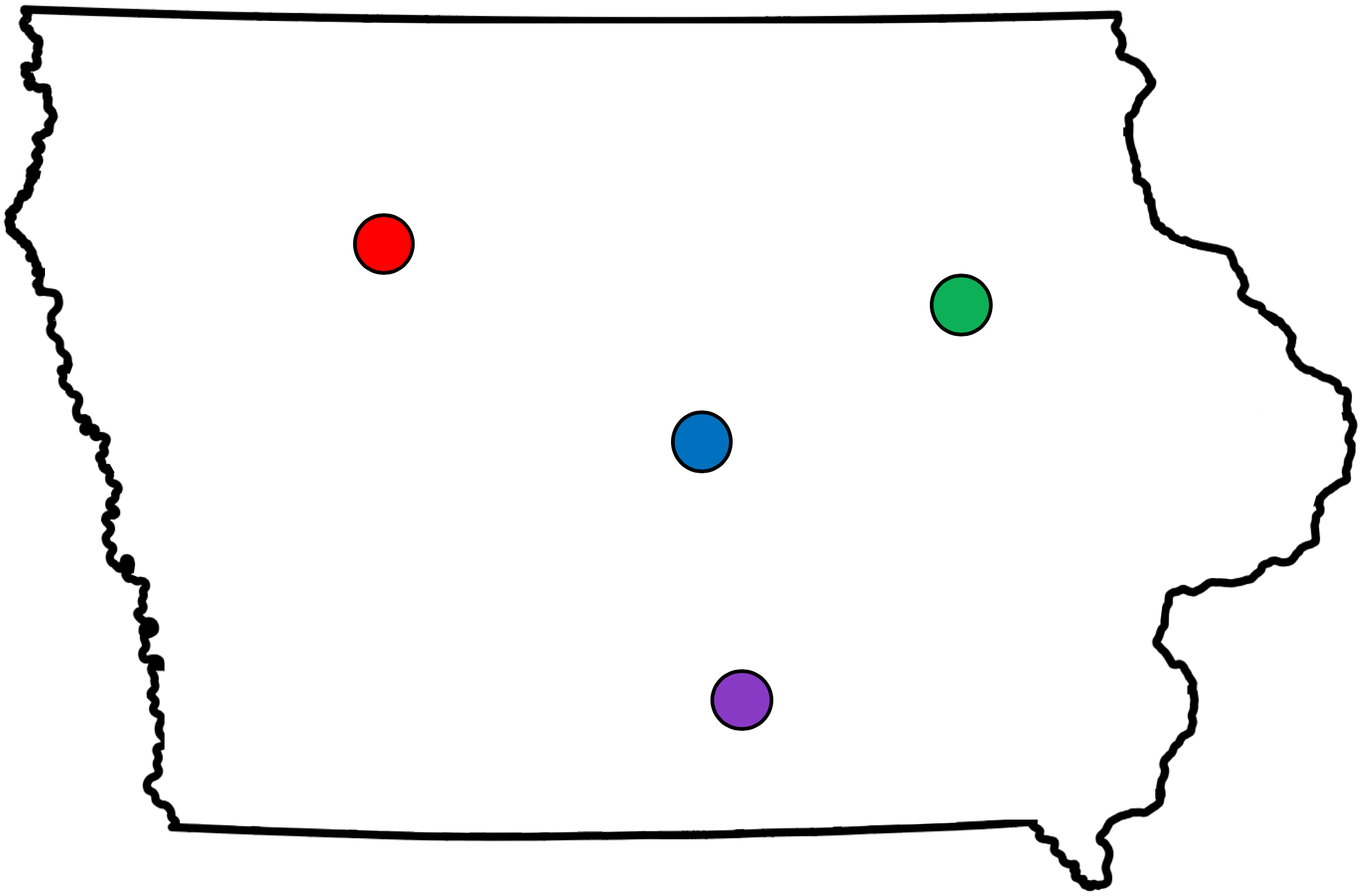}
      \label{fig:centers2}
    \end{subfigure}
    \begin{subfigure}[h]{0.21\textwidth}
      \includegraphics[width=\textwidth]{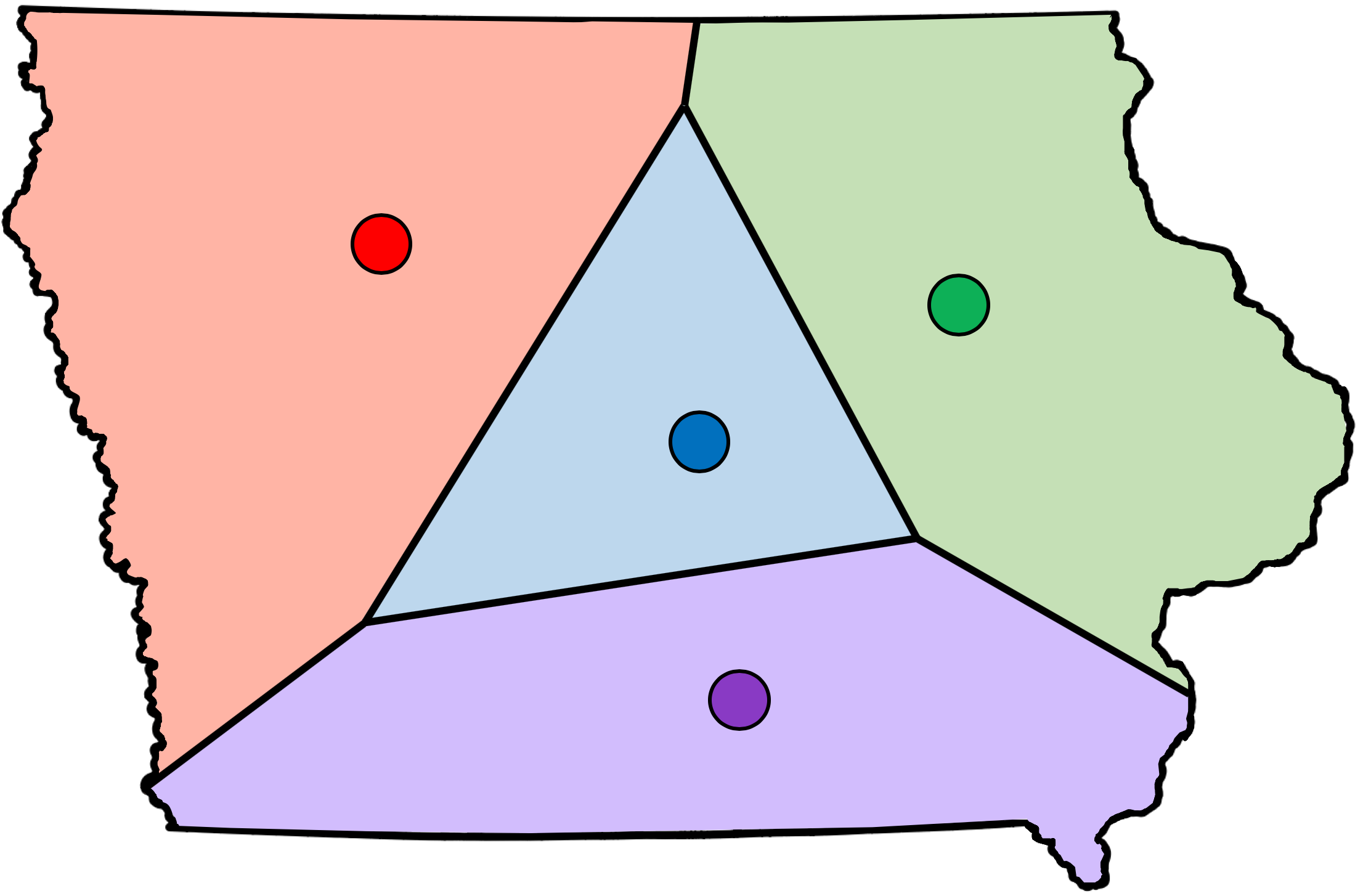}
      \label{fig:voronoi2}
    \end{subfigure}\\
    \begin{subfigure}[h]{0.21\textwidth}
    \label{fig:centers3}
      \includegraphics[width=\textwidth]{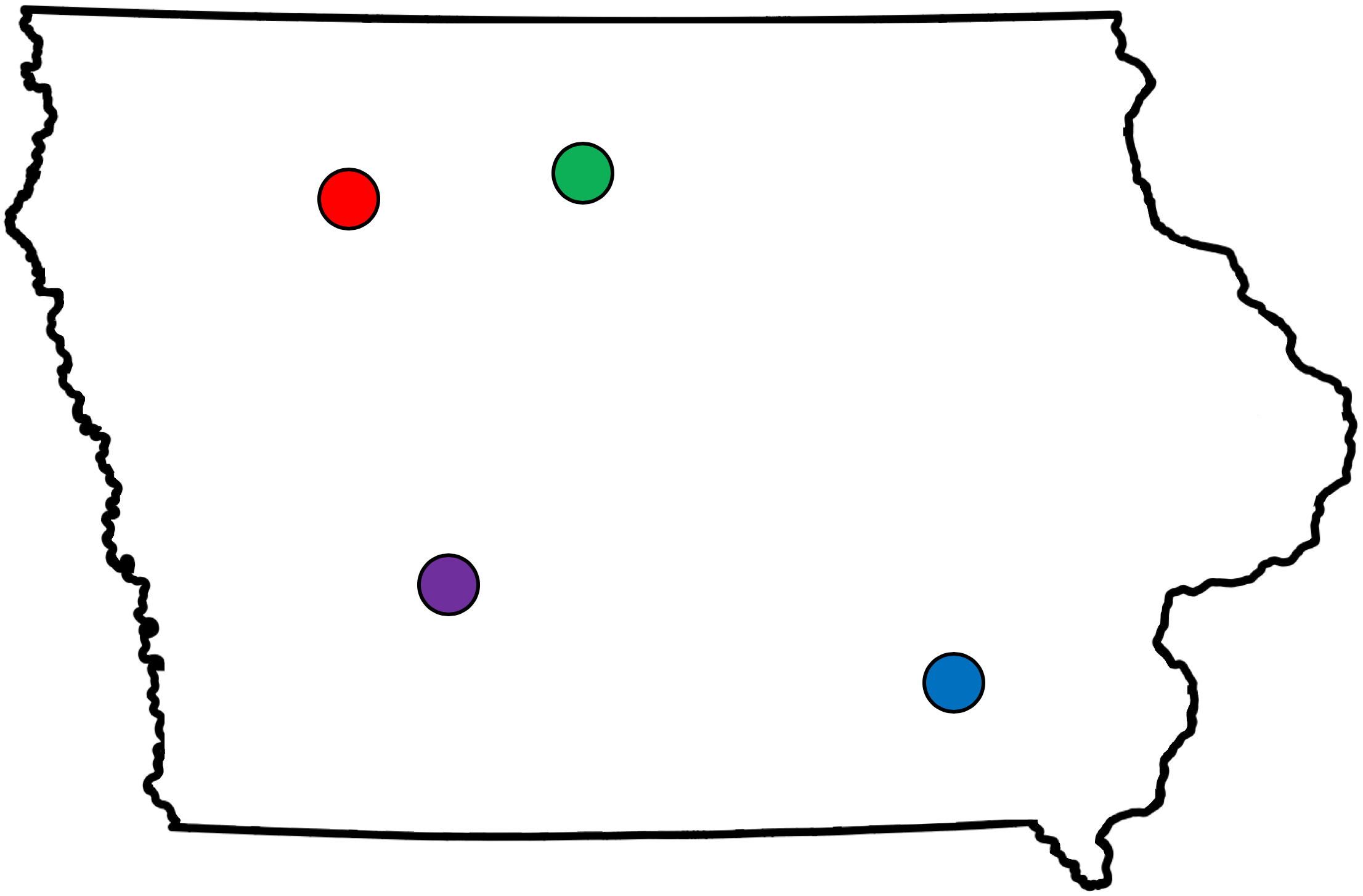}
    \end{subfigure}
    \begin{subfigure}[h]{0.21\textwidth}
    \label{fig:voronoi3}
      \includegraphics[width=\textwidth]{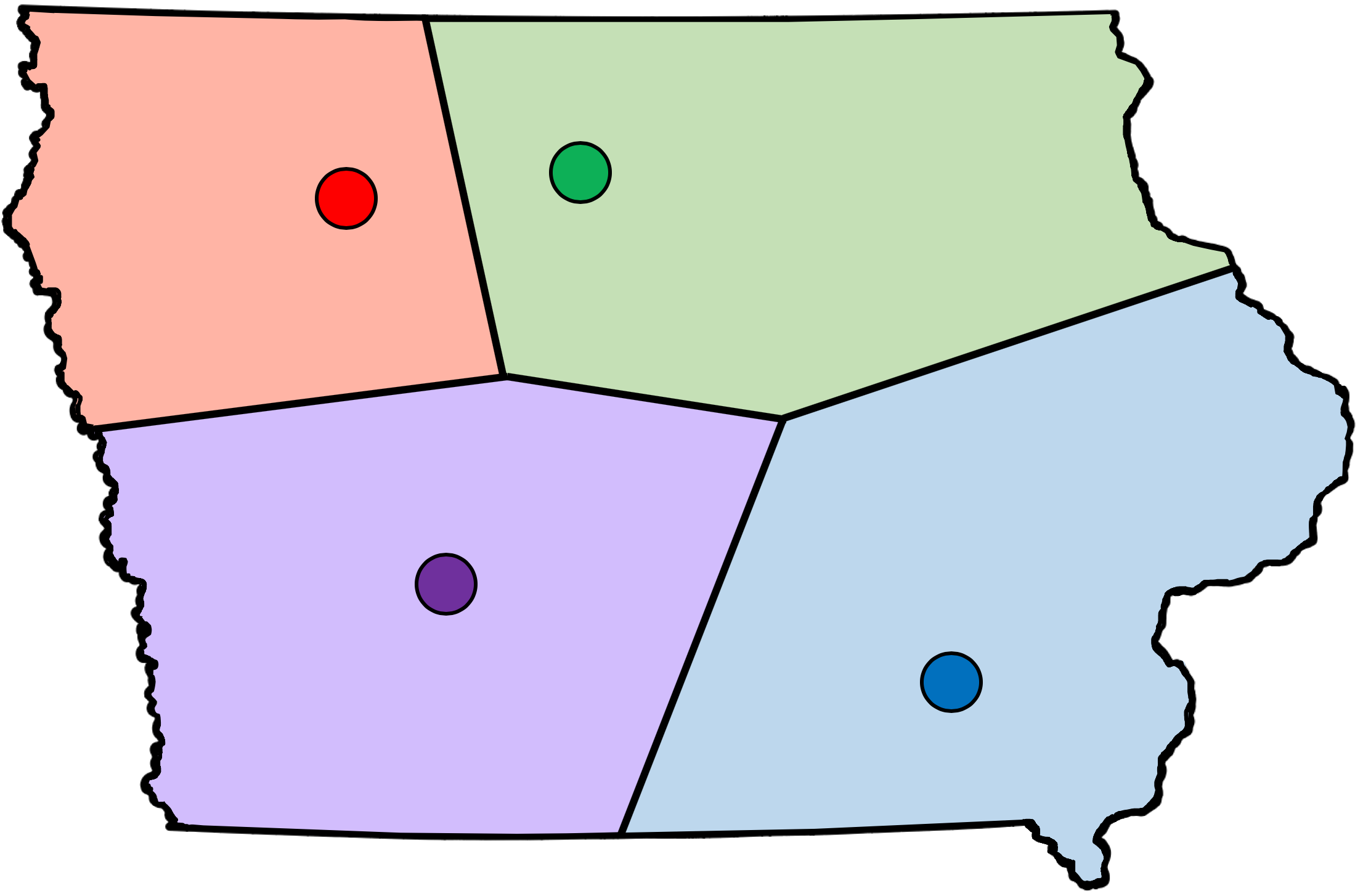}
    \end{subfigure}
\caption{\label{fig:Voronoi_diagrams} Three different ways to draw four district hubs in Iowa and the corresponding Voronoi diagrams.}    
\end{figure}

\begin{figure} 
\centering
    \begin{subfigure}[ht]{0.4\textwidth}
      \includegraphics[width=\textwidth]{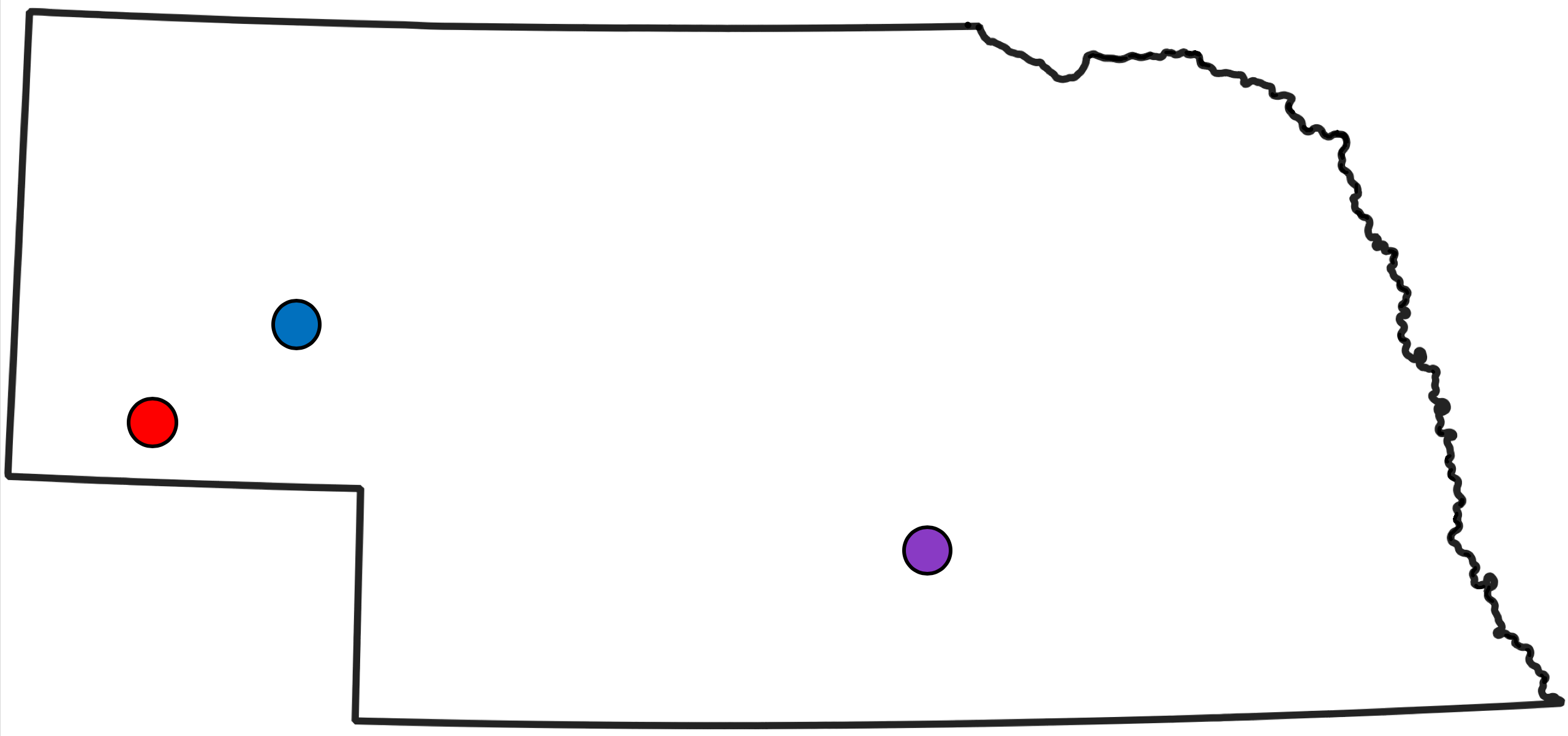}
    \end{subfigure}
    \quad
    \begin{subfigure}[ht]{0.4\textwidth}
      \includegraphics[width=\textwidth]{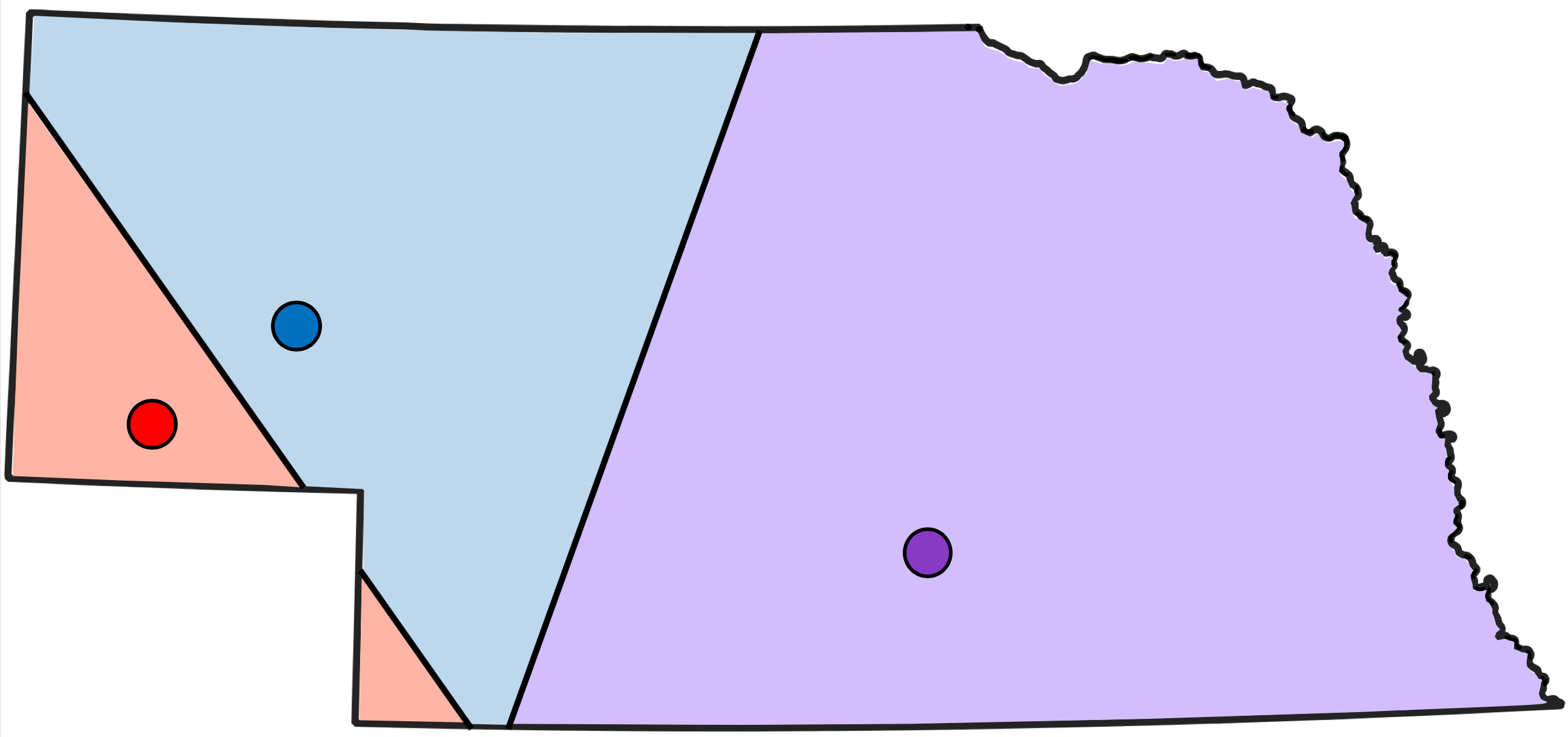}
    \end{subfigure}
\caption{\label{fig:Nebraska_Voronoi} The three centroids on the left correspond to the Voronoi diagram on the left. Because Nebraska's geography is nonconvex, the red district is disconnected.}    
\end{figure}

\begin{figure} 
\centering
    \begin{subfigure}[h]{0.4\textwidth}
      \includegraphics[width=\textwidth, trim={0cm 3cm 0cm 3cm},clip]{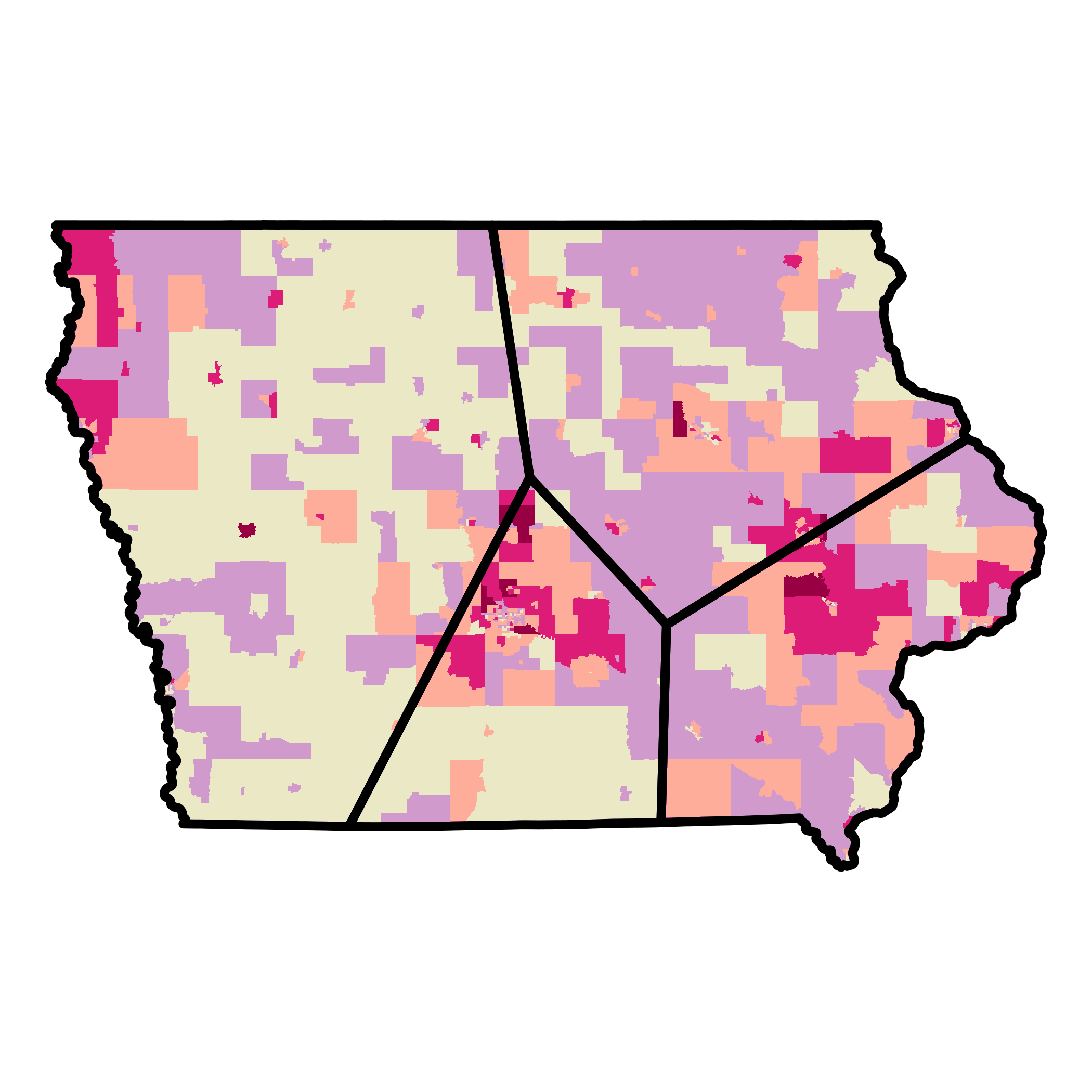}
    \end{subfigure}
    \quad
    \vline\ 
    \quad
    \begin{subfigure}[h]{0.3\textwidth}
      \includegraphics[width=\textwidth, trim={0cm 1cm 0cm 1cm},clip]{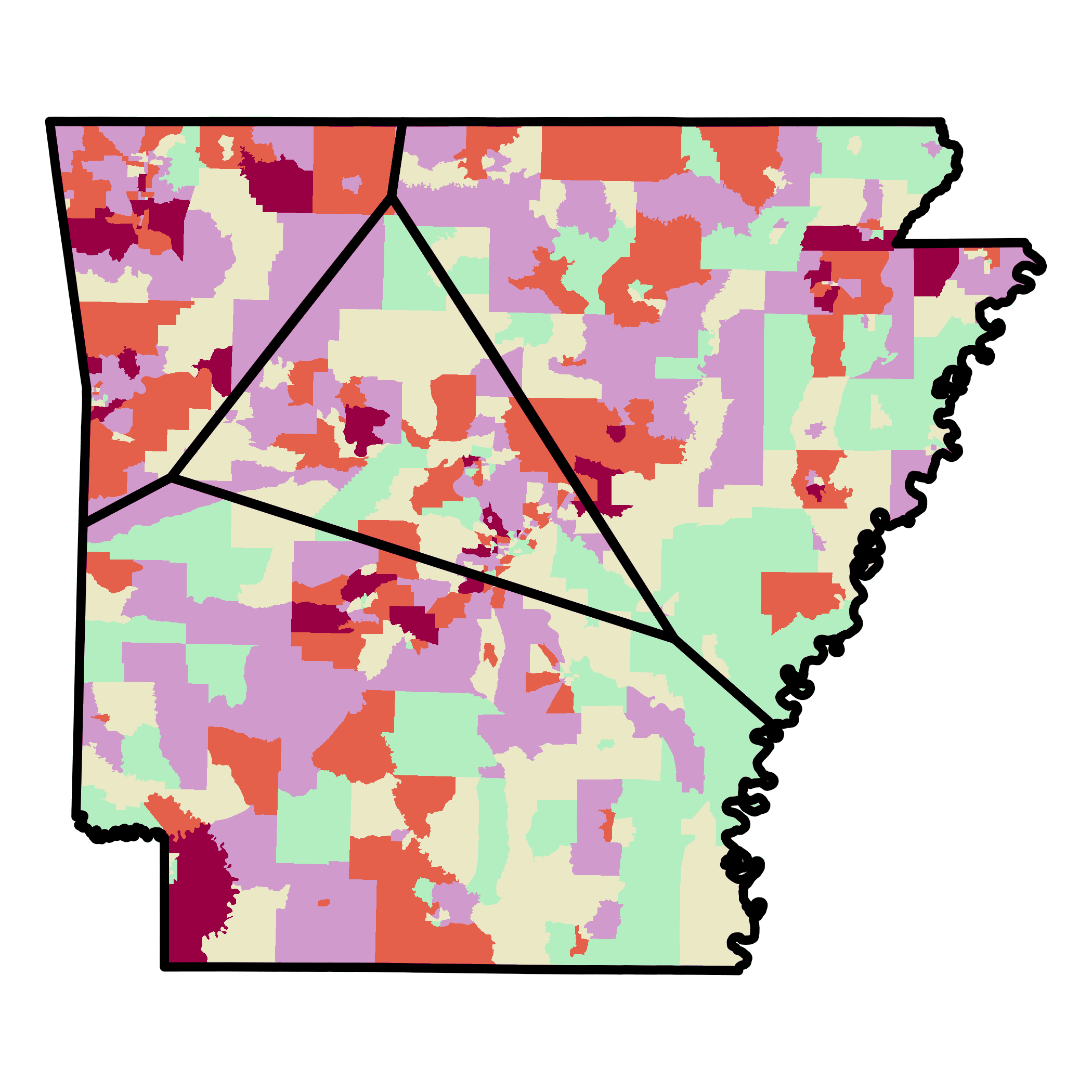}
    \end{subfigure}
\caption{\label{fig:power_diagrams} Power diagrams are shown for Iowa and Arkansas. The shading of each map represents population density.  Figures provided by Richard Barnes. 
}    
\end{figure}

\begin{figure} 
  \centering
    \includegraphics[width=0.5\textwidth]{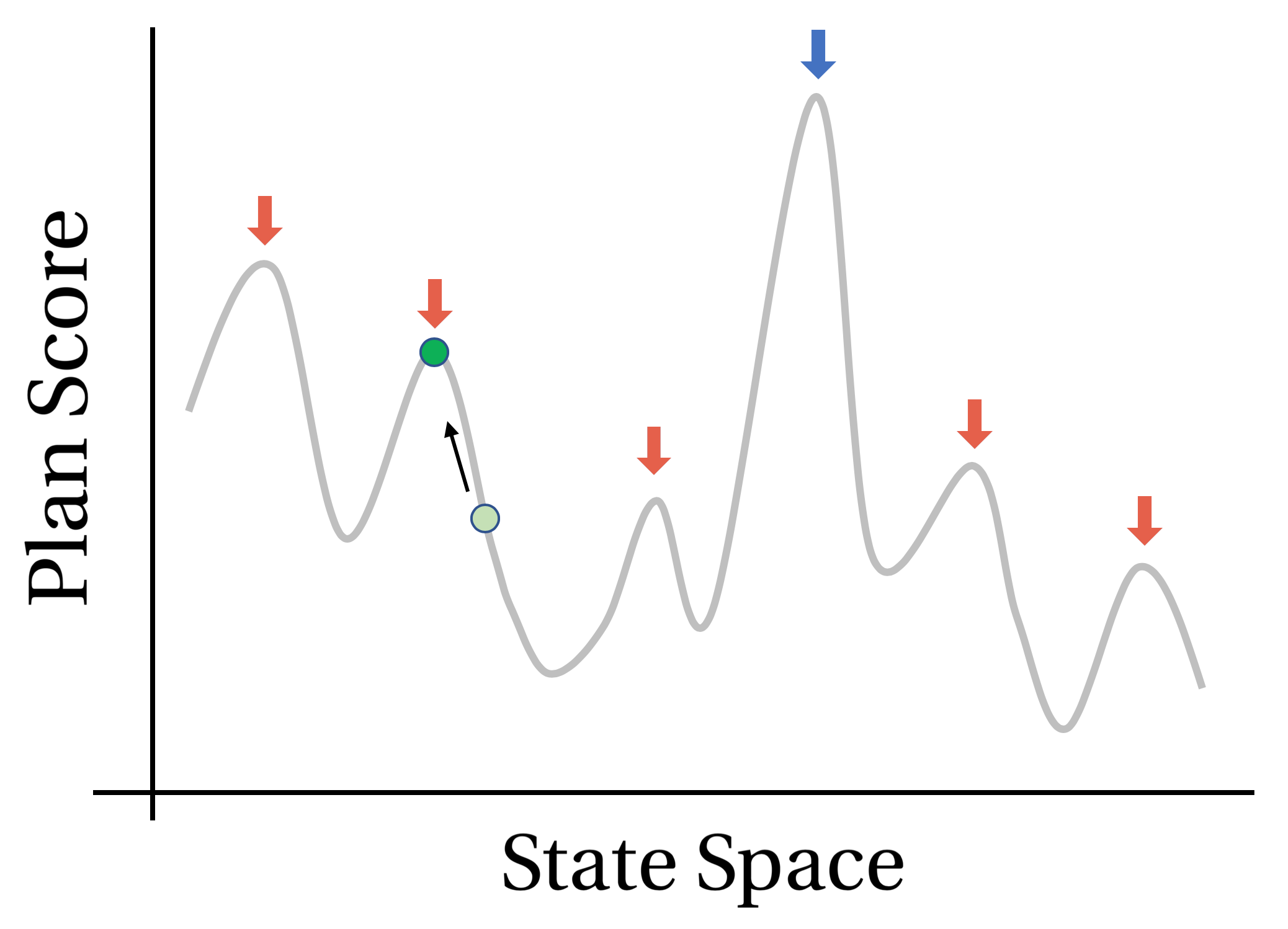}
    \caption{In this visualization of the state space, the global maximum is indicated with a blue arrow and local maxima are indicated with red arrows.  If hill-climbing optimization begins at the light green circle the algorithm will identify the local maxima at the dark green circle, but will not find the global maximum.}\label{fig:optima}
\end{figure}

\begin{figure}[h]
\centering
\begin{subfigure}[h]{0.48\textwidth}
\includegraphics[width=\textwidth, trim={1cm 0cm 1.5cm 1.4cm},clip]{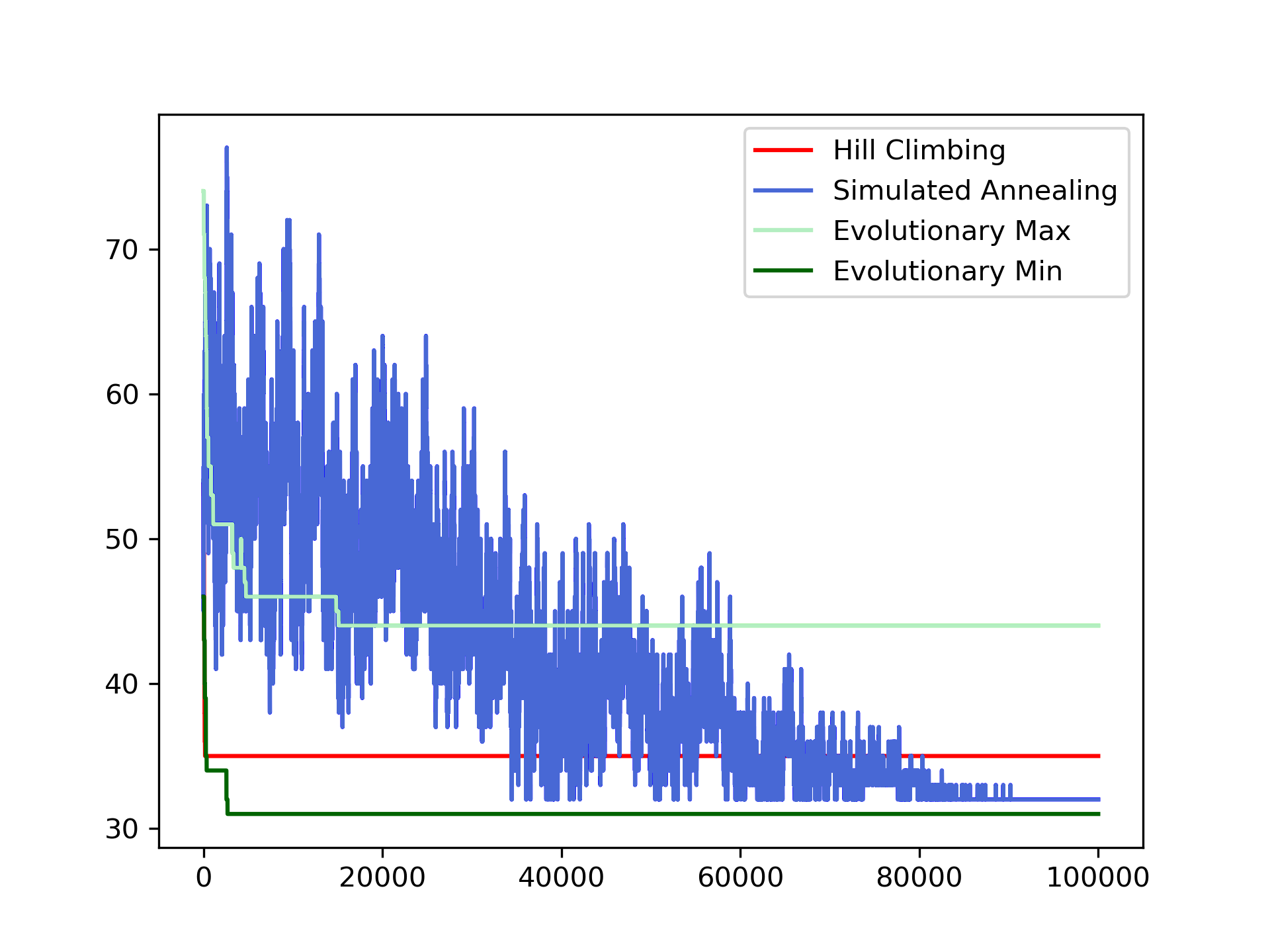}
\end{subfigure}
\begin{subfigure}[h]{0.48\textwidth}
\includegraphics[width=\textwidth, trim={1cm 0cm 1.5cm 1.4cm},clip]{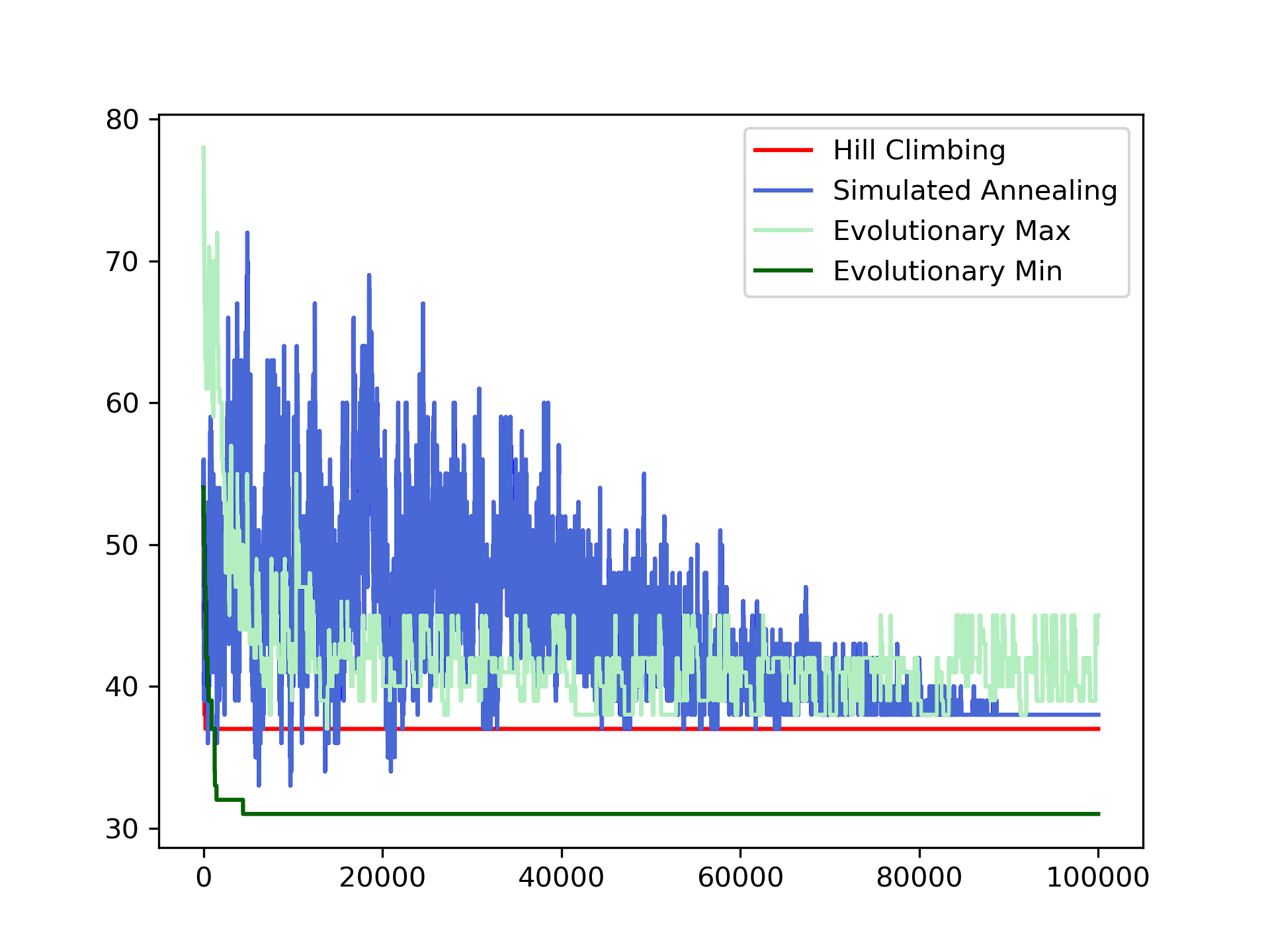}
\end{subfigure}
\caption{\label{fig:opt_combined} These plots compare two different runs of three metaheuristic strategies: hill climbing (red), simulated annealing (blue), and an evolutionary heuristic (dark green is the population minimum and light green is the population maximum number of cut edges)} 
\end{figure}

\begin{figure}[h]\label{fig:evol_steps}
\centering
    \begin{subfigure}[h]{0.3\textwidth}
      \includegraphics[width=\textwidth, trim={1cm 1cm 1cm 1cm},clip]{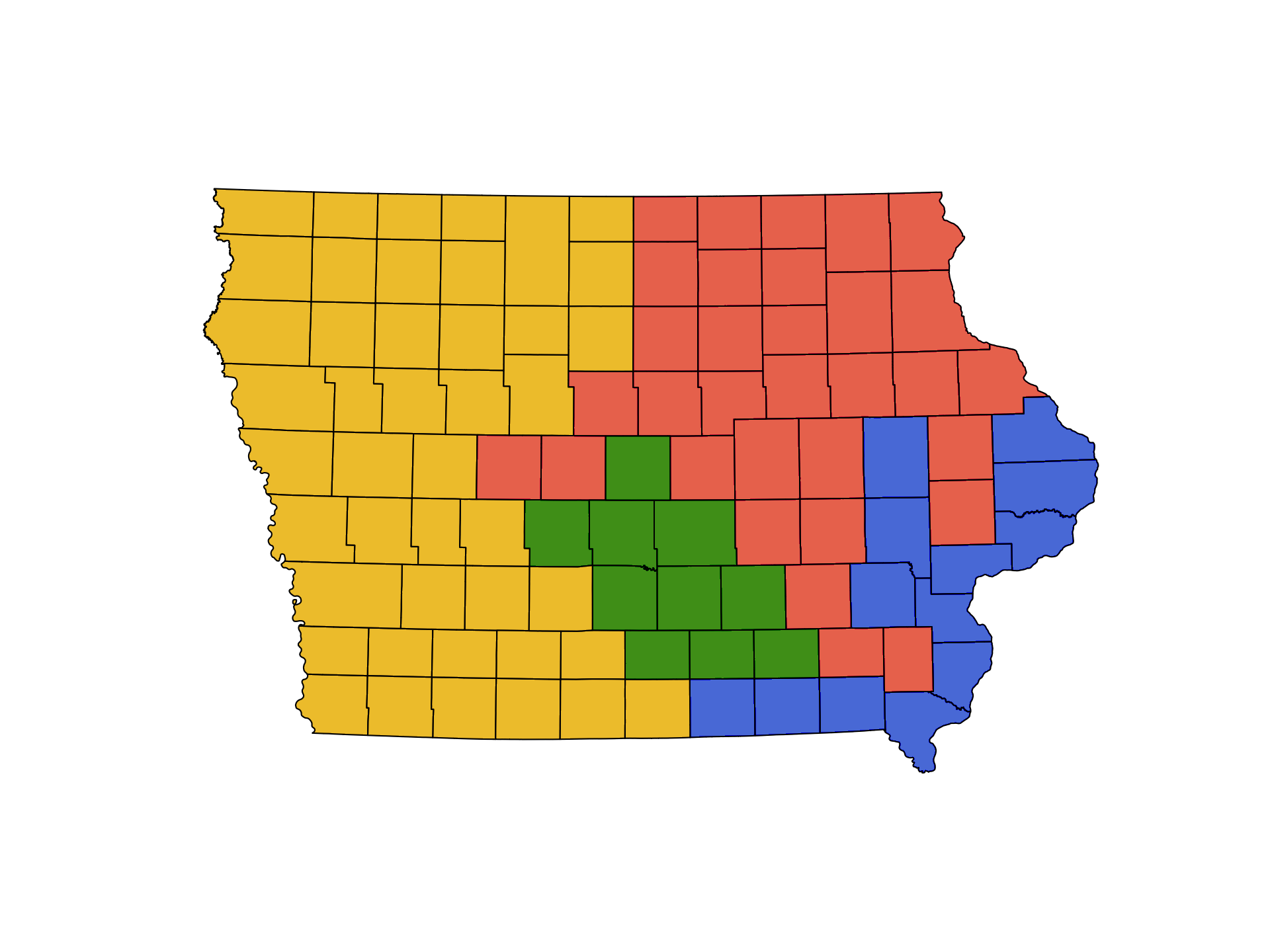}
      \caption{\label{fig:evol_parent1} Parent Plan 1}
    \end{subfigure}
    \begin{subfigure}[h]{0.3\textwidth}
      \includegraphics[width=\textwidth, trim={1cm 1cm 1cm 1cm},clip]{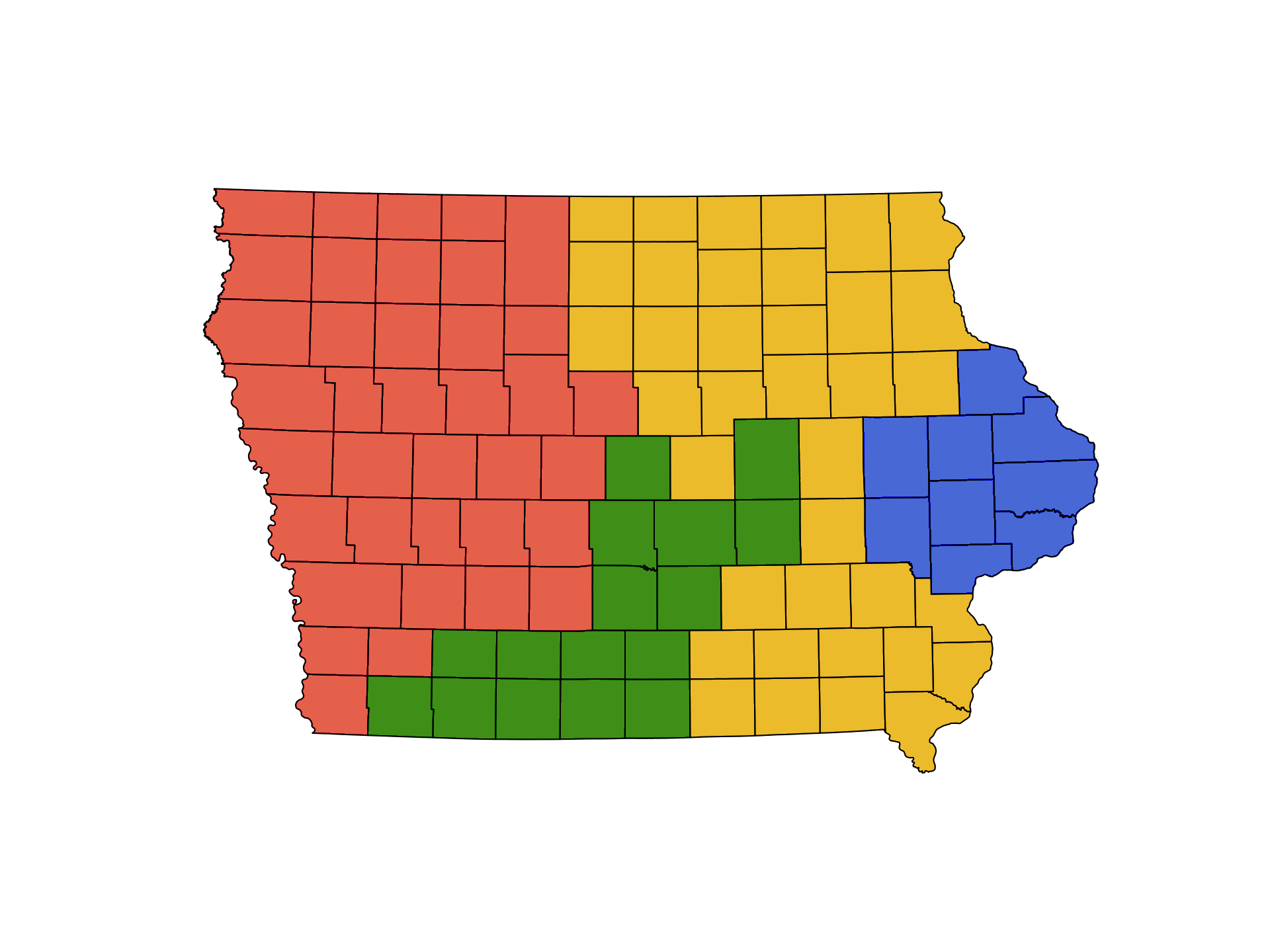}
      \caption{\label{fig:evol_parent2} Parent Plan 2}
    \end{subfigure}\\
    \begin{subfigure}[ht]{0.32\textwidth}
      \includegraphics[width=\textwidth, trim={1cm 1cm 1cm 1cm},clip]{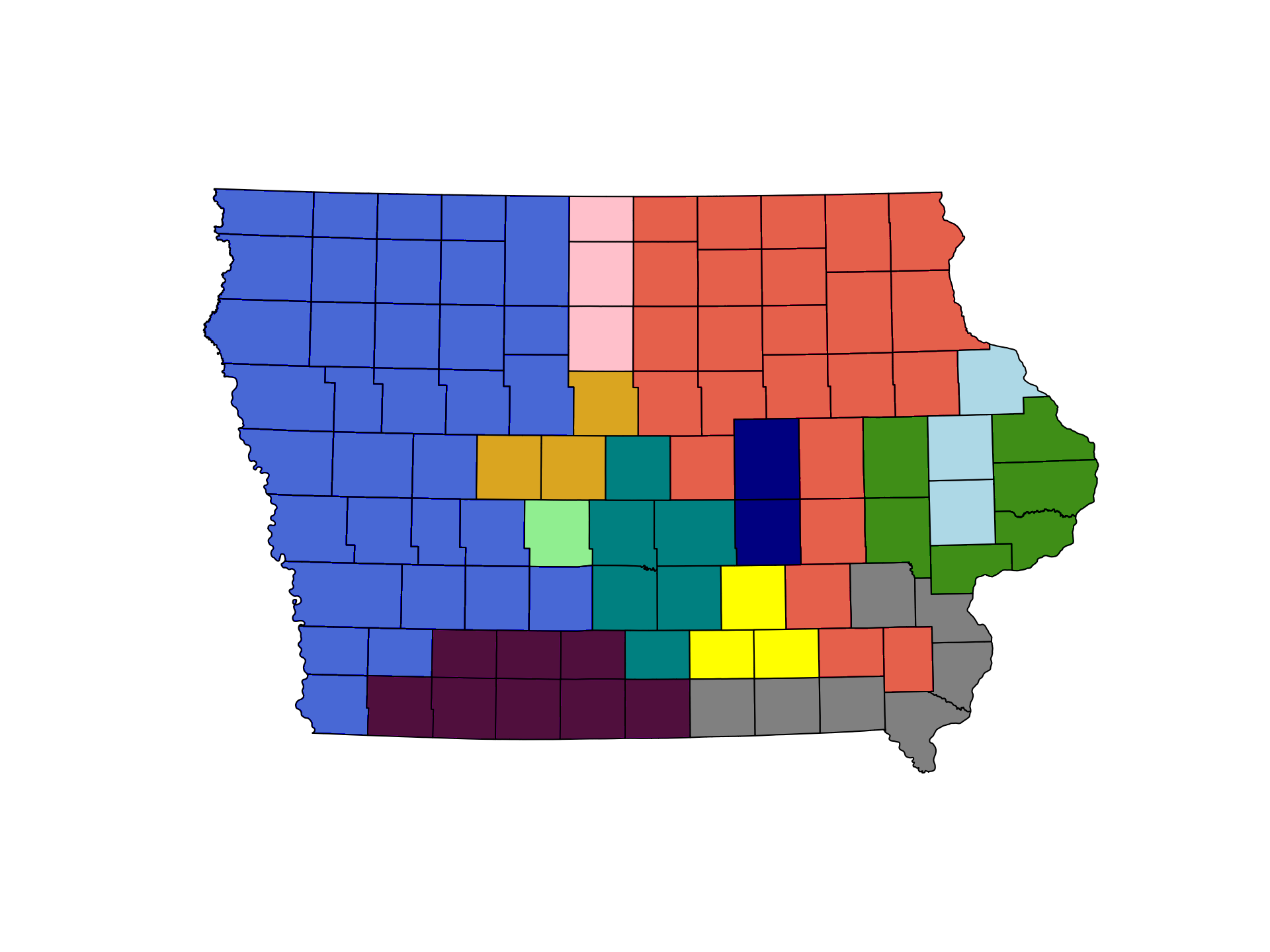}
      \caption{\label{fig:evol_refine} Common Refinement}
    \end{subfigure}
    \begin{subfigure}[ht]{0.32\textwidth}
      \includegraphics[width=\textwidth, trim={1cm 1cm 1cm 1cm},clip]{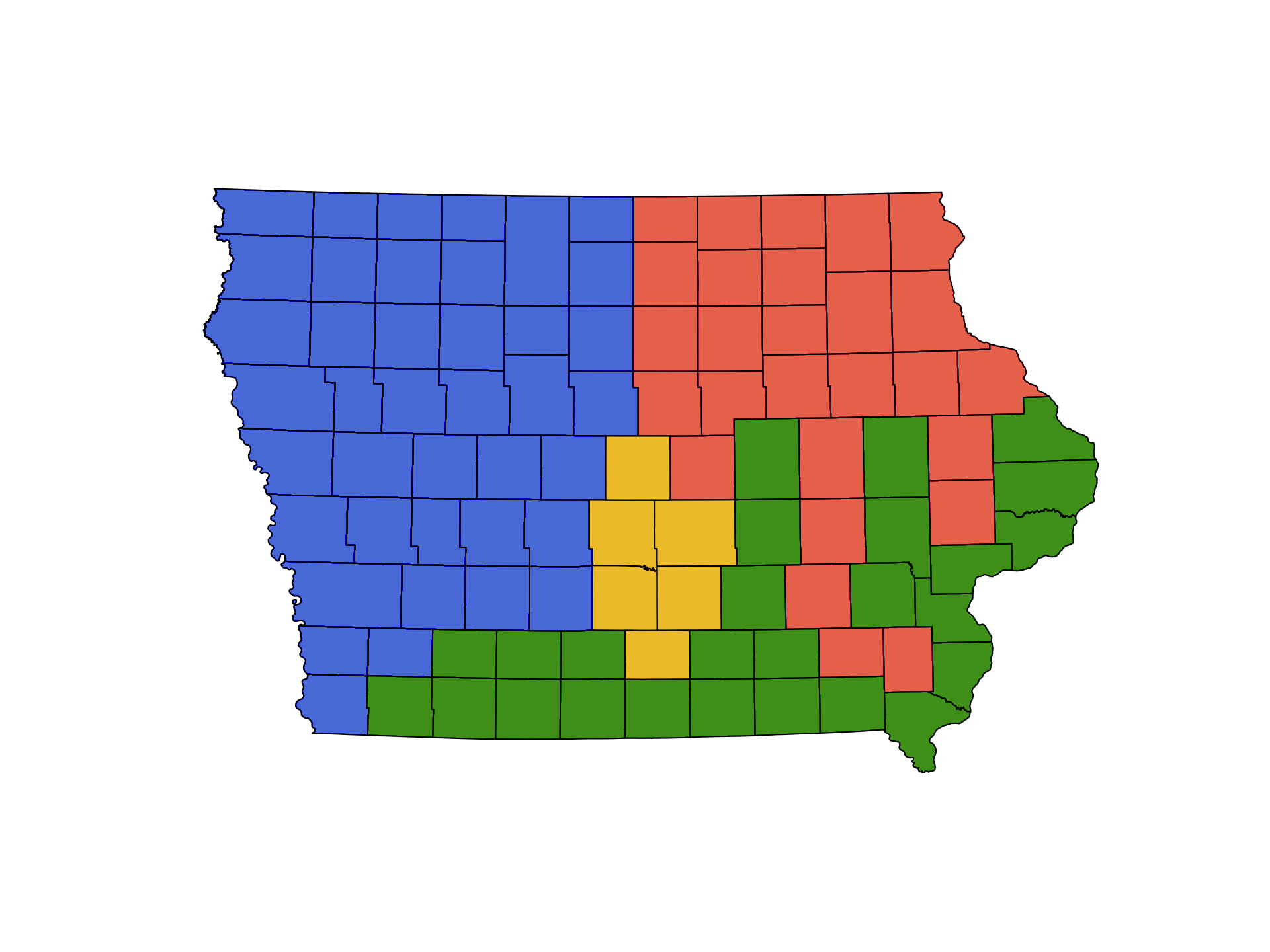}
      \caption{\label{fig:evol_merge} Merged Districts}
    \end{subfigure}
    \begin{subfigure}[ht]{0.32\textwidth}
      \includegraphics[width=\textwidth, trim={1cm 1cm 1cm 1cm},clip]{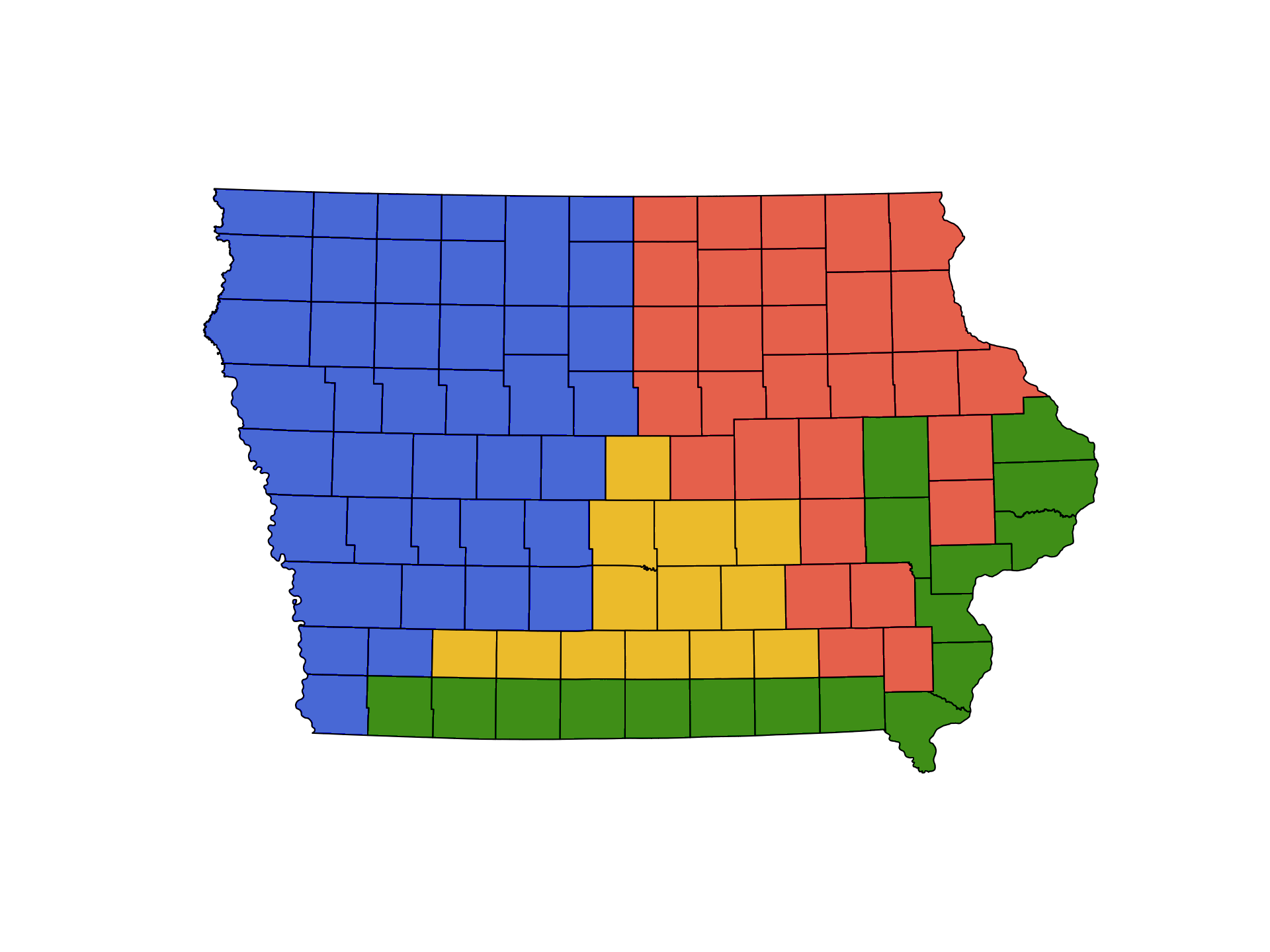}
      \caption{\label{fig:evol_adjust} Child Plan}
    \end{subfigure}
\caption{ \label{fig:evol_crossover} These figures show an example of a crossover step similar to the one used in \cite{liu2016pear}.  Two \emph{parent} plans are chosen from the population.  Their common refinement is computed and the resulting regions are merged until there are four districts.  The merged districts are adjusted to achieve population balance, and the resulting child plan is added to the population.}    
\end{figure}

\begin{figure}[h]
\centering
\includegraphics[width=0.3\textwidth, trim={1cm 1cm 1cm 2cm},clip]{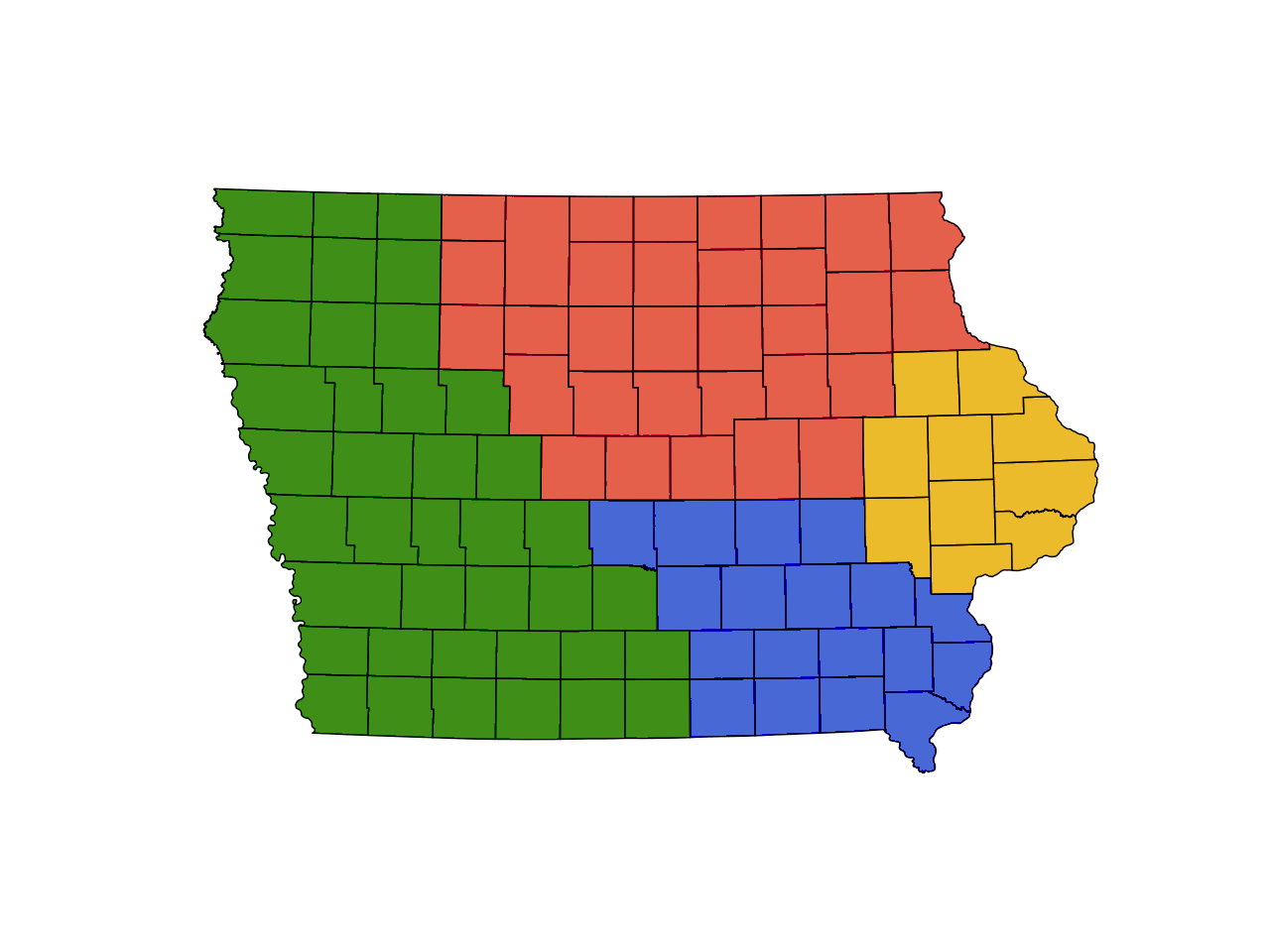}
\caption{\label{fig:five_pct_iowa_opt} This plan has the fewest number of cut edges (29) for partitioning Iowa into four districts with at most 5\% population deviation from ideal (this plan has a population deviation of less than 3.5\%).}
\end{figure}


\begin{figure} 
\centering
    \begin{subfigure}[ht]{0.4\textwidth}
      \includegraphics[width=\textwidth, trim={1cm 1.5cm 1cm 1cm},clip]{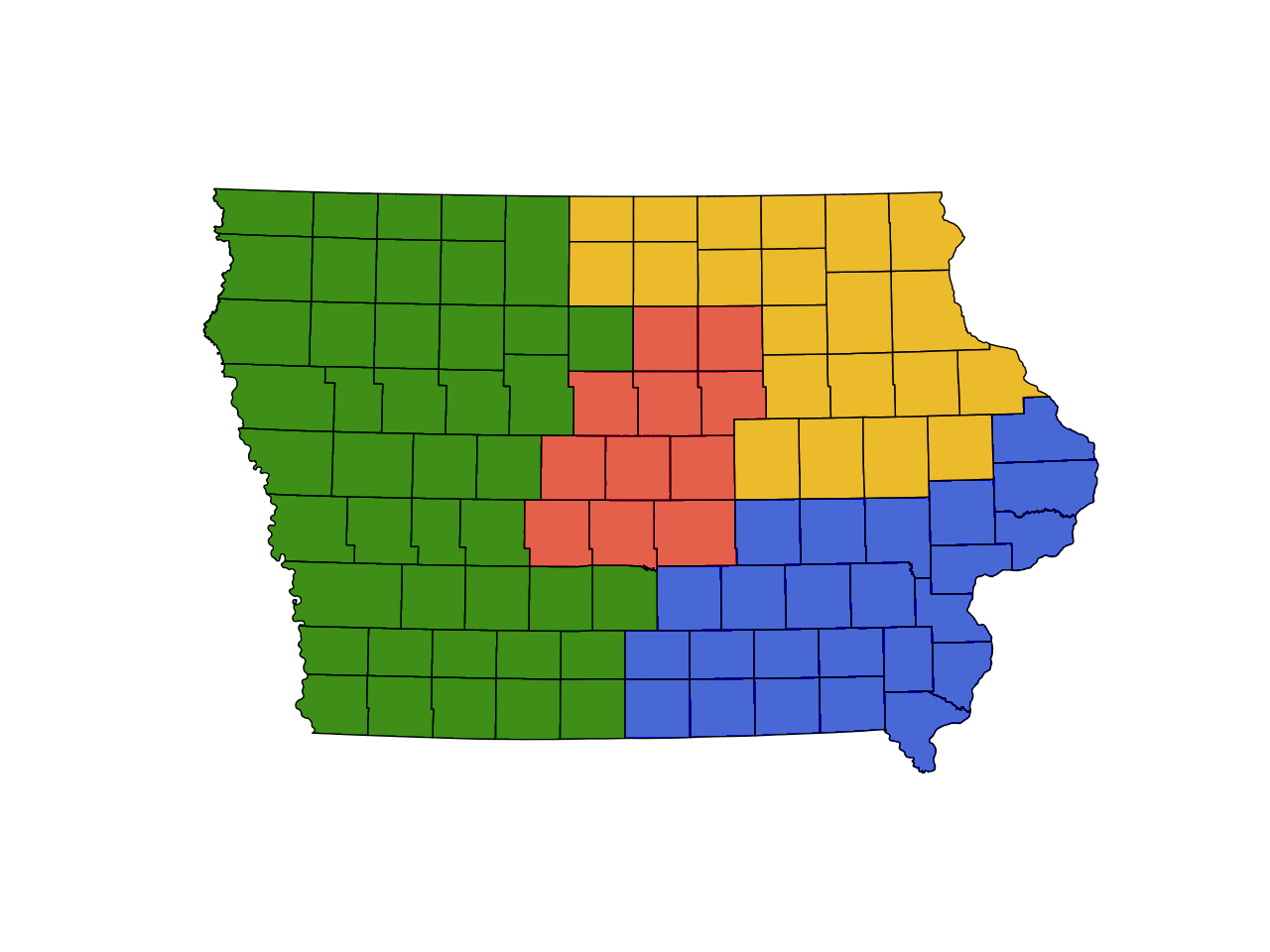}
      \caption{\label{fig:iowa_opt_0.001} 0.1\% Deviation, 35 cut edges}
    \end{subfigure}
    \vline\ 
    \begin{subfigure}[ht]{0.4\textwidth}
      \includegraphics[width=\textwidth, trim={1cm 1.5cm 1cm 1cm},clip]{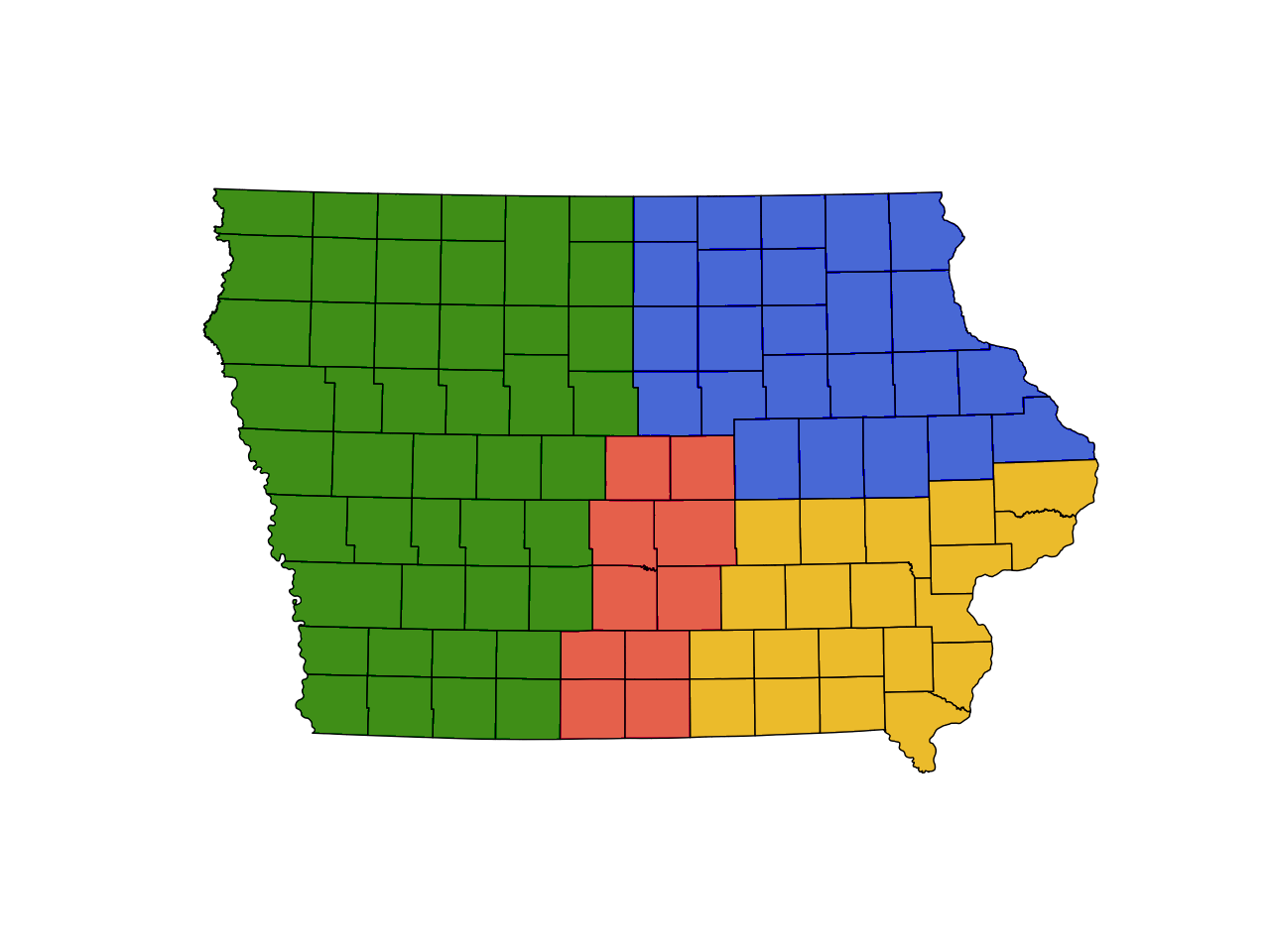}
      \caption{\label{fig:iowa_opt_0.1} 10\% Deviation, 29 cut edges}
    \end{subfigure}
    \begin{subfigure}[ht]{0.4\textwidth}
      \includegraphics[width=\textwidth, trim={1cm 1.5cm 1cm 1cm},clip]{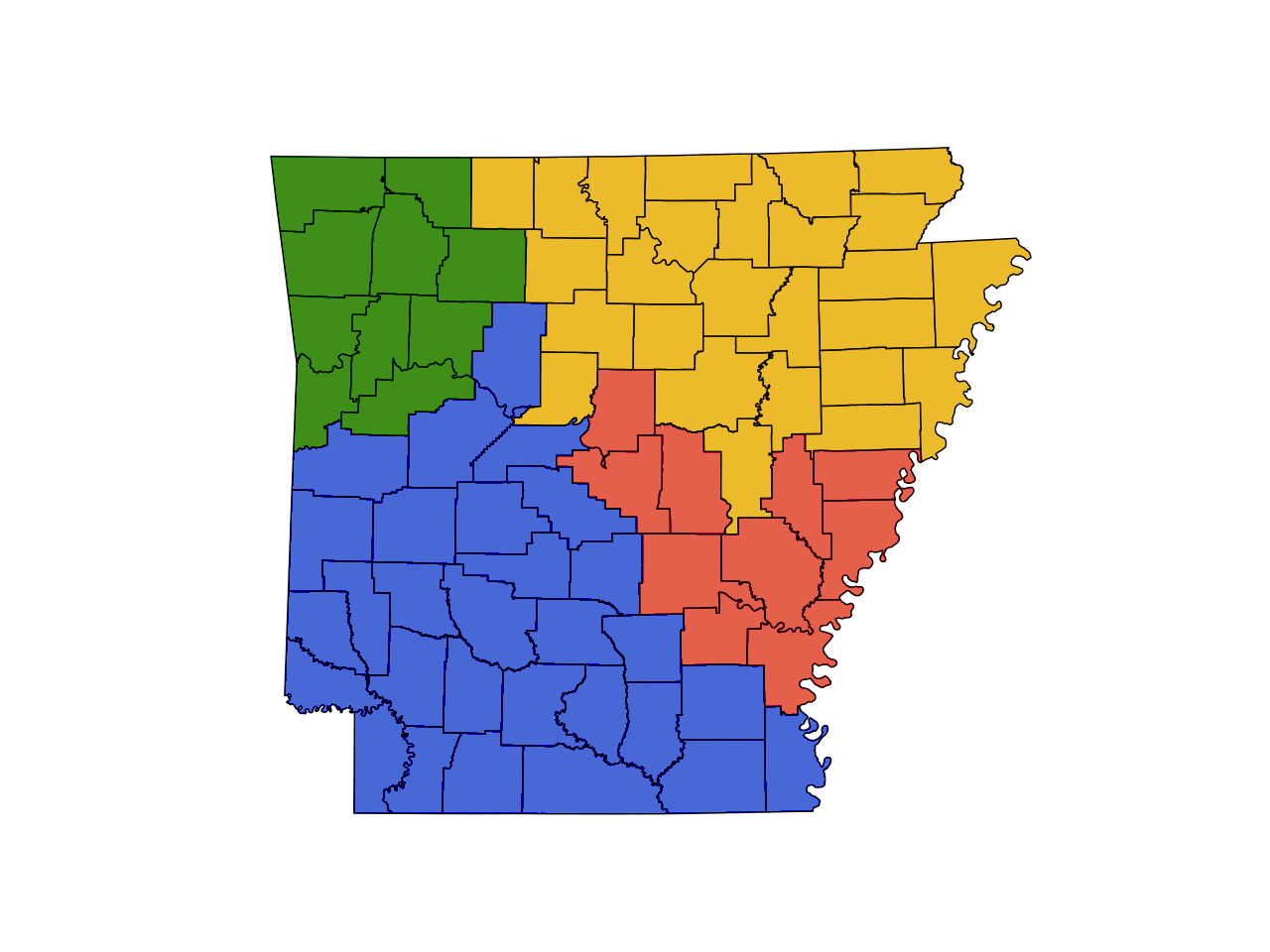}
      \caption{\label{fig:arkansas_opt_001} 0.1\% Deviation, 36 cut edges}
    \end{subfigure}
    \vline\ 
    \begin{subfigure}[ht]{0.4\textwidth}
      \includegraphics[width=\textwidth, trim={1cm 1.5cm 1cm 1cm},clip]{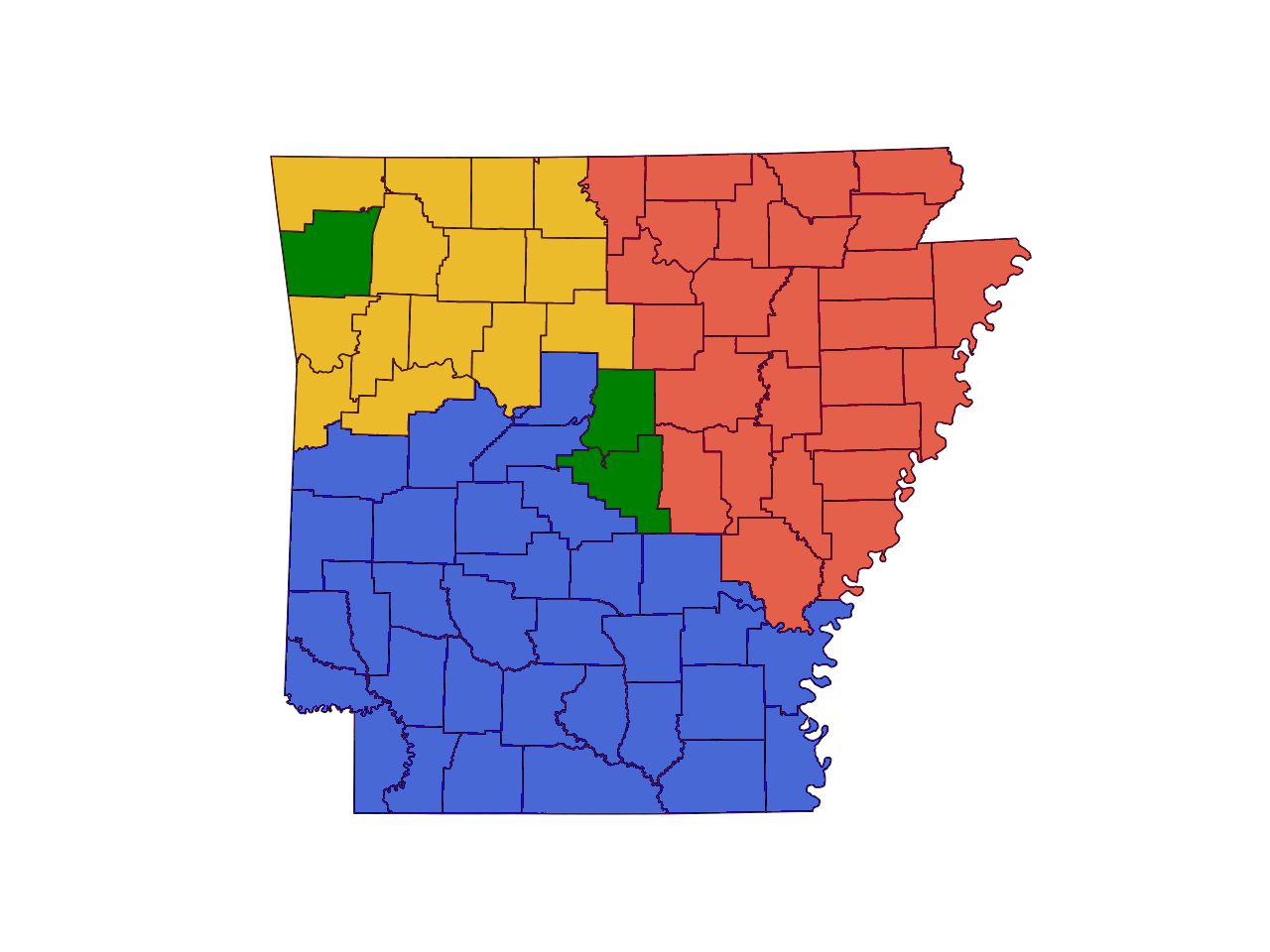}
      \caption{\label{fig:arkansas_opt_01} 10\% Deviation, 30 cut edges}
    \end{subfigure}
\caption{ \label{fig:integer_program_cut_edges} Plans built out of Iowa counties with minimum cut edges for allowed population deviations of 0.1\% and 10\% (top), and analogous plans built out of Arkansas counties (bottom).}    
\end{figure}

\begin{figure} 
\centering
\begin{subfigure}[ht]{0.45\textwidth}
      \includegraphics[width=\textwidth]{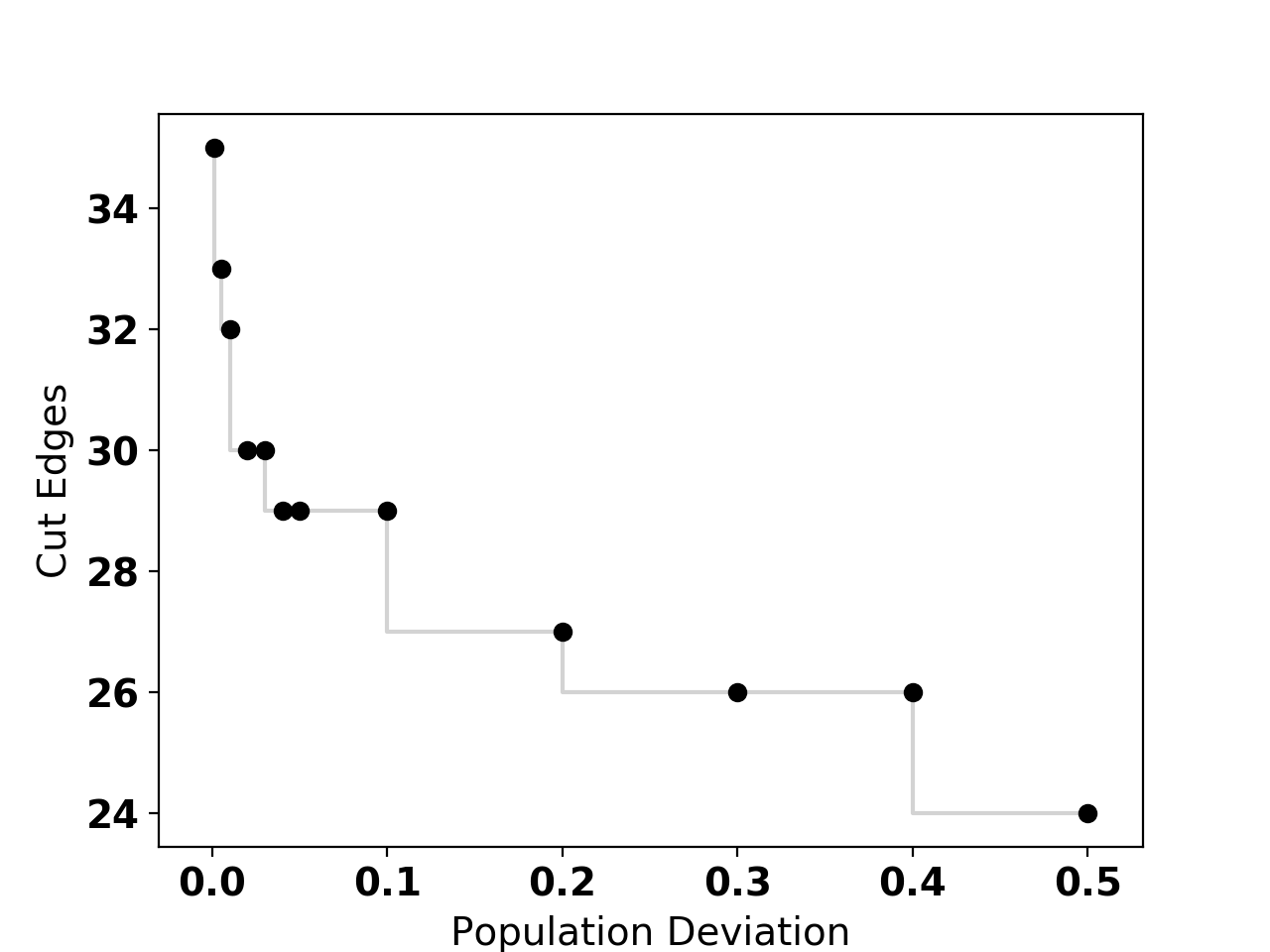}
      \caption{\label{fig:iowa_opt_edges_chart} Iowa}
    \end{subfigure}\ \ \ 
\begin{subfigure}[ht]{0.45\textwidth}
      \includegraphics[width=\textwidth]{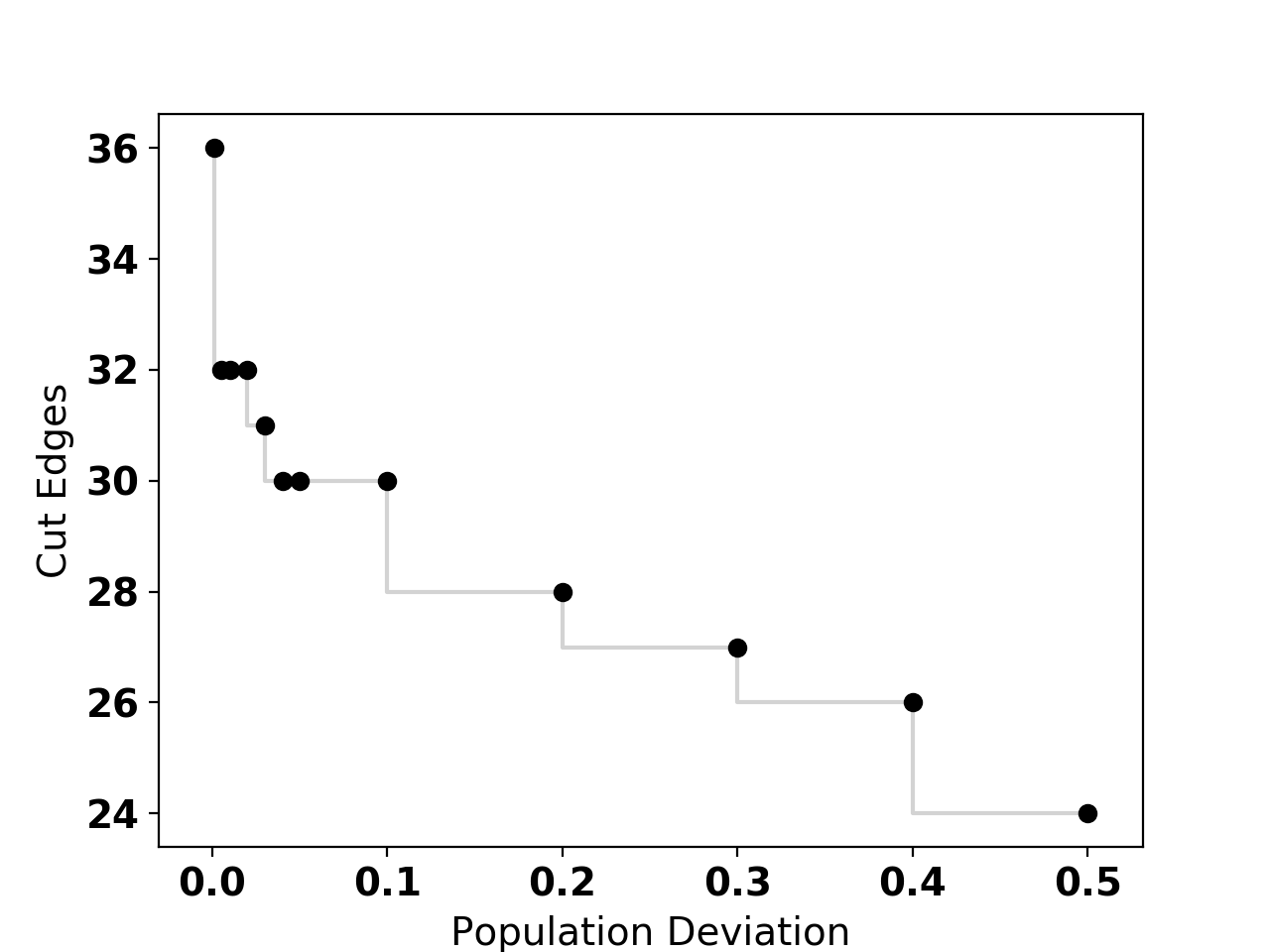}
      \caption{\label{fig:arkansas_opt_edges_chart} Arkansas}
    \end{subfigure}
\caption{ \label{fig:integer_program_cut_edges_chart} The relationship between allowed population deviation and minimum cut edges in districting plans built out of Iowa counties (left) and Arkansas counties (right).  The black dots represent resulting runs of our example integer program for various values of population deviation.  The gray line shows a lower bound on the minimum number of cut edges.  For example, the minimum value is not known for deviations between 0.1 and 0.2 (i.e., 10\% and 20\%), but the value must be between 27 and 29 in Iowa and between 28 and 30 in Arkansas.}    
\end{figure}

\begin{figure} 
\centering
    \begin{subfigure}[ht]{0.3\textwidth}
      \includegraphics[width=\textwidth, trim={7cm 0cm 8cm 1.6cm},clip]{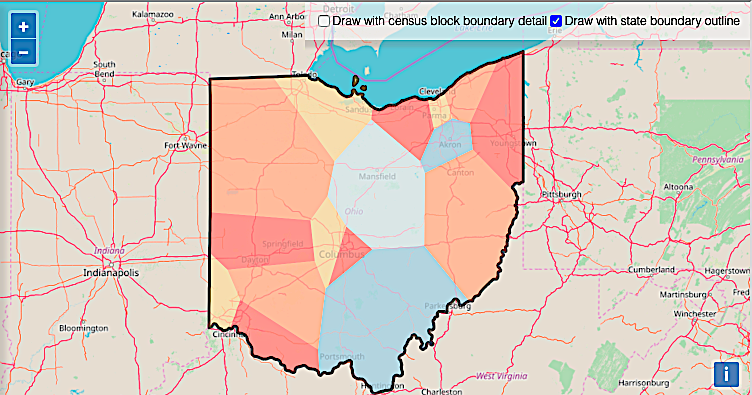}
      \label{fig:ohio1}
    \end{subfigure}\ \ \ \ 
        \begin{subfigure}[ht]{0.3\textwidth}
      \includegraphics[width=\textwidth, trim={7cm 0cm 8.7cm 1.8cm},clip]{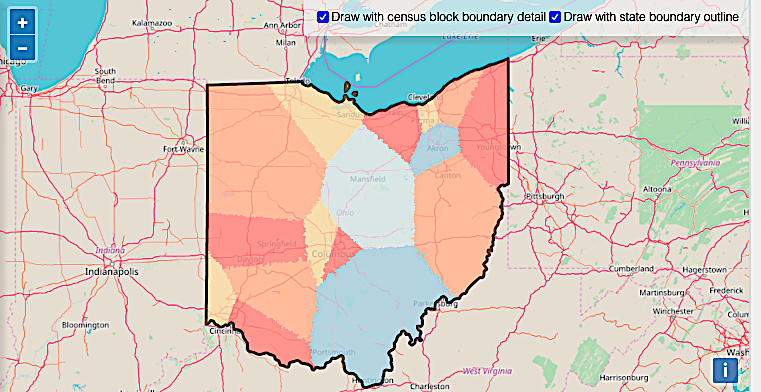}
      \label{fig:ohio2}
    \end{subfigure}\\
    \begin{subfigure}[ht]{0.3\textwidth}
      \includegraphics[width=\textwidth, trim={1.5cm 0cm 1.5cm 1.6cm},clip]{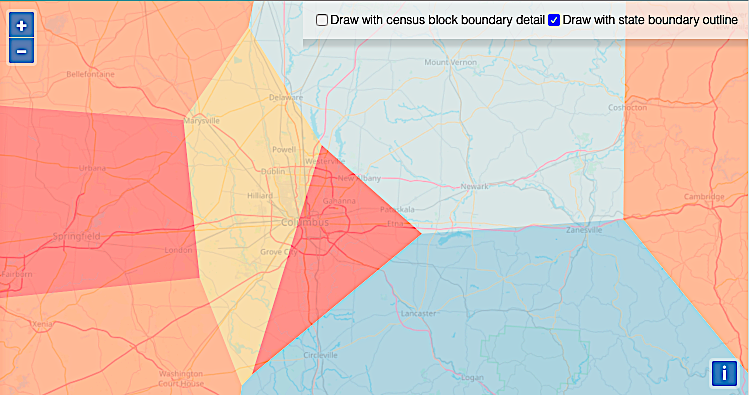}
    \end{subfigure}\ \ \ \ 
        \begin{subfigure}[ht]{0.3\textwidth}
      \includegraphics[width=\textwidth, trim={1.5cm 0cm 1.5cm 1.6cm},clip]{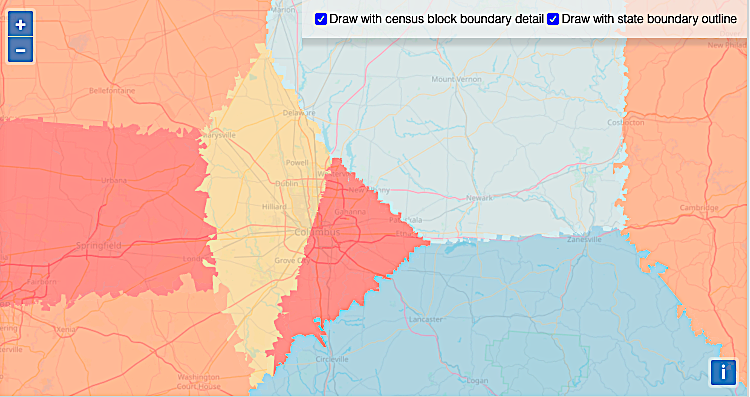}
    \end{subfigure}
\caption{\label{fig:refine_geometry} The figures on the left are power diagrams for Ohio (generated as described in \cite{cohen2018balanced}) and the figures on the right show how these geometric districts can be adjusted to respect the boundaries of census blocks while still maintaining population balance and contiguity.  Figures courtesy of Philip Klein.}
\end{figure}

\clearpage
\section{Generating All Plans: Enumeration}\label{sec:enum}

A natural algorithmic strategy in redistricting is simply to enumerate \emph{all} valid plans.  That is, given a list of rules determining which plans are valid, the computer is tasked with generating a list of every possible compliant plan.    
In this section, we will explain why 
enumeration is impossible in practical terms.

If we could enumerate all plans, we would have a straightforward optimization algorithm: score all possible plans to identify the best one (this is called a \emph{brute force} algorithm).  This approach to optimization is \emph{exact} because it considers every possible alternative, so it is guaranteed to find the best one.  Random plan generation is also straightforward with enumeration: from the set of all valid plans, select one uniformly at random, or one weighted by some desired probability distribution. Because of its conceptual advantages, enumeration has been proposed as a strategy to identify and evaluate plans for decades \cite{garfinkel1970optimal,rossiter1980,oloughlin1982,jenkins1972}.

If enumeration is so powerful, why is it not used broadly in redistricting?  
There are two key issues.  First is the sheer number of ways we can draw district lines, making the list of valid plans unfathomably large in  practice. 
For this reason, in practice it can only be used on very small instances.  Put differently, redistricting famously suffers from \emph{combinatorial explosion}: as the problem gets larger, the number of valid solutions increases exponentially, quickly exceeding the practical limits of computing power and data storage.

\begin{table}
\begin{center}
\begin{tabular}{ |c|r|} 
 \hline
 $n$ &\# plans  \\
 \hline
 1 & 1 \\ 
 2 & 2 \\ 
 3 & 10 \\
 4 & 117 \\ 
 5 & 4,006 \\ 
 6 & 451,206 \\
 7 & 158,753,814  \\ 
 8 & 187,497,290,034 \\ 
 9 & 706,152,947,468,301 \\
 \hline
\end{tabular}
\end{center}
\caption{The number of ways to divide an $n\times n$ grid into $n$ contiguous parts with $n$ units each~\cite{Ruiter2010OnJS,Harris2010CountingNT}\label{table:numberplans}.}
\end{table}

To get a sense of how quickly the number of valid solutions increases, consider the simple problem of partitioning an $n\times n$ grid into $n$ equal-sized districts~\cite{Ruiter2010OnJS,Harris2010CountingNT}.  The number of partitions as a function of $n$ is given in Table \ref{table:numberplans}, for small values of $n$, and shown 
in Figure~\ref{fig:complete-enum} for the $n=3$ case.
Even for this relatively simple redistricting problem, these numbers quickly become too large for enumeration to be practical.  This combinatorial explosion is not unique to grids:  Enumerating plans for actual states is out of the question in nearly any context. For example, the number of ways to build four congressional districts out of the 99 counties of Iowa is estimated to be about $10^{24}$ (or a trillion trillions)~\cite{fifield2019essential}, but the exact number is not known.  This estimate is tiny compared to the number of plans that can be built from the finer units like precincts or census blocks that are typically used.  At the precinct level, the still-small problem of dividing 250 contiguous Florida precincts into two districts has approximately $5 \times 10^{39}$ different valid solutions~\cite{fifield2019essential}, which in turn is minuscule compared to the full problem of dividing Florida's roughly 6,000 precincts into its 27 congressional districts.

Even when the full list of plans for a given geographic area is small enough to store on a computer, we have to consider the amount of time it takes to \emph{generate} such a list.  
That is, not only is the list of plans extremely long, each plan on the list can take a long time to find. 

Another problem arises from combinatorial explosion. For the $6\times 6$ grid with four districts, recall that plans can have anywhere from 12 to 28 cut edges. Complete enumeration shows that over 93\% of all these plans have 21-28 cut edges, putting them in the worst half of that range. Like other districting issues, this imbalance only accelerates as the size of the problem grows. In a full-sized problem, more than 99.999\% of balanced, contiguous plans are so wildly shaped that they would never be considered in practice\footnote{This is another matter of counting: there are more winding lines than straight lines,
so there are far more non-compact than compact plans.
See \cite{deford2019recombination} for more discussion.}. So if you are trying to use the enumeration to get an overview of possibilities, you may not get a very good picture if you simply weight them all equally.

Given that enumeration is neither computationally tractable nor sufficient for understanding real-world redistricting problems, we need other strategies for generating and assessing plans.

\section{Generating Many Plans:  Sampling}\label{sec:random_gen}

Enumeration is an example of an \emph{exhaustive search} technique, in which we visit every corner of the space of districting plans to get a complete understanding. But as we have discussed, computational impediments make enumeration impractical when processing real-world data.  For this reason, most algorithms related to redistricting generate and analyze a relatively small set of districting plans, essentially targeting their search.
With this motivation in mind, in this section we introduce \emph{sampling} algorithms, whose job is to produce a short but useful list of options. 

Enumeration algorithms are \emph{deterministic} in nature, meaning that every time the same piece of code is run we receive the same result.  In contrast, sampling algorithms tend to be \emph{randomized}, meaning that they have the ability to make different decisions every time they are run.  Randomized---also known as \emph{stochastic} or \emph{nondeterministic}---algorithms can be extremely powerful in their simplicity and efficiency.  

We will focus on \emph{random plan generators}, randomized algorithms that generate samples of redistricting plans given a fixed piece of geography.  These methods produce output that can either be analyzed on its own or used as a subroutine for other algorithms in redistricting.  
For example, optimization algorithms designed to extract high-quality plans frequently use random plans as starting points and then employ a number of strategies to improve the quality of the starting plan to shape it into the optimized output. 

There are several crucial questions to ask when evaluating random plan generators:
\begin{itemize}
\item \textbf{What is the \emph{distribution} of the generated plans?} \\ By nature, sampling methods only generate a subset of possible plans.  For this reason, we must understand both the distributional design and unintentional biases of these methods.  Your first thought might be to sample from a \emph{uniform} probability distribution, in which all valid districting plans are equally likely to be included in the sample.\footnote{An efficient algorithm probably does not exist that can draw uniform samples efficiently~\cite{najt}.}  But some sampling methods might instead be weighted towards more compact plans, or tilted towards a particular partisan balance.  Whether intentional or unintentional, this weighting can have substantial consequences if sampling is used to summarize the population of alternative plans.
\item \textbf{Can the sampler generate any possible plan?} \\
Even though sampling might not be uniform, we might want to know whether there is \emph{some} nonzero probability of generating every possible plan.  Some sampling methods restrict their consideration to plans with certain shapes or other properties, which make it easier to traverse the space of plans but may unintentionally exclude plans relevant to a given redistricting task.
\item \textbf{Do the samples accurately capture priorities and constraints?} \\ Redistricting rules can be complex, placing many restrictions on the properties of acceptable plans.  It can be difficult to customize a new sampling algorithm to each set of rules and regulations.  \emph{Winnowing}, in which samples are generated using a first method and then non-compliant plans are discarded, can repair a sampler after the fact, but few if any plans may be left in an ensemble after this cleanup step, and it can limit control over the probability distribution.
\end{itemize}

\subsection{Generating Plans from Scratch}\label{sec:random_scratch}

Many random plan generators start from a blank slate, taking as input the parameters of a redistricting task: the desired number of districts, the population of each census unit, and the adjacency of these units (i.e., which units share a border).  The algorithm then outputs a random plan that assigns these units to districts or describes where to draw the lines between districts.

\subsubsection{Random Assignment and Rejection Sampling}\label{sec:randomassignment}

Perhaps the simplest approach to generating a plan is the random assignment algorithm, which is one of several approaches implemented in the BARD redistricting software package \cite{altman2011bard}.  This algorithm divides a region into $k$ districts by randomly and independently assigning each unit a district label from $1$ through $k$.  Figure~\ref{fig:random} shows a typical random assignment; unsurprisingly, it does not satisfy any of the familiar constraints, such as contiguity or population balance.  One way to rectify this problem is to 
repeatedly generate candidate plans, discarding invalid plans until a valid one is produced.  This is our first encounter with the tactic called {\em rejection sampling},
shown in Algorithm~\ref{alg:randomunitassignment}.

This is not very efficient.  In fact, it is so unlikely that random assignment of census units results in a valid plan that we would expect to discard an astronomically large number of proposed candidate plans before finding a single valid one.

\begin{algorithm} 
  \begin{algorithmic}[1]
	\For{each census unit $i$}
	    \State $\textrm{District assignment}(\textrm{unit }i)\gets$\textsc{Random}$(1,2,\ldots,k)$
	\EndFor
  \end{algorithmic} 
  \caption{Random-Unit Assignment}
  \label{alg:randomunitassignment}
\end{algorithm}

\begin{algorithm} 
  \begin{algorithmic}[1]
	\While{plan is invalid}
	    \State $\textrm{Plan}\gets$\textsc{Random-Unit Assignment}
	\EndWhile

  \end{algorithmic} 
  \caption{Random-Unit Assignment with Rejection}
  \label{alg:rejectionsampling}
\end{algorithm}

Random assignment is the easiest algorithm to analyze theoretically and has the favorable property that \emph{every} possible plan can be generated with some nonzero (but vanishingly small) probability, but most of the analysis simply reveals that it is ineffective. Instead, in practice it is desirable for samplers to produce valid plans with reasonably high probability.  In the remainder of this section, we will stay attentive to the question of rejection rate.

\subsubsection{Flood Fill and Agglomeration}\label{sec:flood_fill}

A widely discussed family of practical plan-generating algorithms employs a \emph{flood fill} strategy.  These algorithms \emph{grow} districts from seed units by gluing together adjacent units until the districts reach the desired population.  Many flood fill variants have been proposed, including \cite{rossiter1980, oloughlin1982, cirincione2000assessing, magleby2018new,vickrey1961prevention,thoreson1967computers}; we highlight a few examples below.

\begin{algorithm} 
  \begin{algorithmic}[1]
  \For{$\textrm{each district $i$ }$}
  \State seed$\gets$\textsc{Random}$(\textrm{unassigned census unit})$
  \State $\textrm{District assignment(seed)}\gets i$
    \While{$\textrm{Population(district $i$} \textrm{)}\leq$ target\_population}
    \State spread$\gets$\textsc{Random}($\textrm{unassigned}$ \textsc{Neighbor}$(\textrm{district $i$}))$
    \If{$\textrm{Pop(district $i$)} + \textrm{Pop(spread)} \leq$ target\_population}
    \State $\textrm{District assignment(spread)}\gets i$
    \EndIf\EndWhile
    \EndFor
  \end{algorithmic} 
  \caption{District-by-District Flood Fill}
  \label{alg:districtbydistrictflood}
\end{algorithm}

Many flood fill algorithms build one district at a time, as 
outlined in Algorithm \ref{alg:districtbydistrictflood}; Figure \ref{fig:flood_fill} shows an example.  In this case, 
a single unit is selected as a ``seed'' to start growing a district (the red county in the example in Figure \ref{fig:flood_fill}).  Then, the algorithm iteratively glues units onto its side until the district reaches a desired size. At each step there are multiple options for which unit to add next (pink counties in the example in Figure \ref{fig:flood_fill}); the simplest way to make this decision might be to choose randomly, which is the version in the pseudocode for Algorithm \ref{alg:districtbydistrictflood}.

Rather than making this decision completely randomly, we can make strategic choices that encourage the partially-built district to have particular properties. For example, one variant 
preferentially chooses units that lie within the \emph{bounding box} of the currently growing district to improve compactness of the plans \cite{cirincione2000assessing}; Figure \ref{fig:flood_fill_bb} illustrates this heuristic on our running example.  Another variant also suggested in \cite{cirincione2000assessing} preferentially chooses units that belong to census tracts or counties that are already included in the growing district (see Figure~\ref{fig:flood_fill_county}).

\begin{algorithm} 
  \begin{algorithmic}[1]
  \For{$\textrm{each district $i$ }$}
    \State $\textrm{seed } i\gets$\textsc{Random}$(\textrm{unassigned building-block unit})$
    \State $\textrm{District assignment(seed $i$)}\gets i$
  \EndFor
    \While{$\textrm{some district is still underpopulated}$}
    \State $\textrm{district } j\gets$\textsc{Random}$(\textrm{underpopulated district})$
    \State spread $\gets$\textsc{Random}($\textrm{unassigned}$ \textsc{Neighbor}$(\textrm{district $j$}))$
    \If{$\textrm{Population(district $j$)} + \textrm{Population(spread)} \leq$ target\_population}
    \State $\textrm{District assignment(spread)}\gets j$
    \EndIf
    \EndWhile
  \end{algorithmic} 
  \caption{Whole Plan Flood Fill}
  \label{alg:multidistrictflood}
\end{algorithm}

A challenge with flood fill algorithms is that they can get stuck. Some districts grow in such a way that does not leave enough space for the remaining districts (see Figure~\ref{fig:flood_fill_fails}).  Typically at this point, the plan is rejected and the algorithm starts over, repeating this process until a valid plan is generated, although a few algorithms can \emph{adjust} initially-invalid plans until they are valid (see Section~\ref{sec:refine}).  Though this refinement strategy has a lower rejection rate, these adjustments are often quite computationally intensive and time-consuming. 

A second approach to flood fill 
builds all the districts simultaneously, as  
depicted in Algorithm \ref{alg:multidistrictflood}. Rather than building one complete district, fixing it, and moving on to the next district, this algorithm grows all the districts in the state in parallel.  In each iteration, the algorithm now has to make two decisions:  which district to grow, and which adjacent unit to add onto that district.  A primary advantage of this approach is that all the districts are treated symmetrically.  In contrast, in the one-at-a-time strategy from Algorithm \ref{alg:districtbydistrictflood}, the shape of the first district drawn might be quite different than the shape of the last district because as the algorithm proceeds there are fewer options for how to grow a district outward.

One variant of the whole-plan flood fill method suggested in \cite{liu2016pear} starts from $k$ seed units along the boundary of the state.  Another variant chooses one seed from each of $k$ predefined \emph{zones} across the state, promoting a more uniform distribution of the growing districts \cite{liu2016pear, rossiter1980}.  Figure \ref{fig:flood_fill_seeds} compares these options.  Such variations may be designed to reduce the rejection rate or to tailor the sampling distribution (e.g., to generate samples with more compact plans).    

Another method with a similar flavor to flood fill is an iterative merging approach~\cite{chen2013unintentional,chen2016loser}: at each step, a geographic unit is chosen at random and merged with an 
adjacent unit to form a new aggregate unit.  This process is repeated until the number of aggregate units is equal to the desired number of districts.  These resulting pieces (composed of many of the original census units merged together) may have unbalanced populations, so the algorithm might then make small refinements until the populations are balanced (see Section~\ref{sec:refine}).  The merging process is efficient, and by choosing the closest unit to merge at each step
(measured by distance between the units' centroids), this method promotes compactness. The population-rebalancing process, however, can be inefficient and may degrade compactness. This flood fill variant has been used many times in recent redistricting litigation, including, for example, the 2018 League of Women Voters v. Commonwealth of Pennsylvania case~\cite{2018league}.

For many flood fill variants, every valid plan has a \emph{chance} of being generated in theory.  It is not clear, however, how \emph{likely} some plans are to be generated in practice, and there has been little focus in the literature on the distribution of plans from which flood fill algorithms sample.  

We can empirically demonstrate the non-uniformity of the flood fill method.  Figure~\ref{fig:6_violin} depicts the distribution of the \emph{number} of cut edges across sampled plans generated by different flood fill variants\footnote{For this demonstration, each flood fill variant was run 300,000 times.  In our implementation, the success rate of flood fill on a $6 \times 6$ grid ranged between 5 and 10 percent for the standard and whole-plan variants and was close to 40 percent for the bounding-box variant.} versus complete enumeration, which is possible because we can check all 442,791 ways to partition a $6 \times 6$ grid into four equal-sized districts.  Figure~\ref{fig:flood_fill_6x6} shows the corresponding distributions of \emph{where} these cut edges occur most frequently across the samples.  

Recall that low numbers of cut edges indicate that the districts have short boundaries.  If a flood fill method sampled uniformly among all valid plans, a large sample of generated plans would be expected to have a nearly identical distribution of cut edges as the full enumeration. 
These experiments demonstrate that---as we would expect---flood fill does not uniformly sample from the set of valid plans and that different flood fill variants give different distributions of generated plans.  

This is our first example of a non-uniform distribution over plans (i.e., some districting plans are more likely to appear than others). 
We have already seen that non-uniformity can be desirable or even necessary for a useful sampling method, but 
it could easily be the case that innocuous modeling decisions significantly affect the behavior of the resulting samples.  This \emph{distributional design} question comes into play if we want to perform statistical calculations, e.g., comparing a human-drawn plan to the ``average'' computer-drawn plan generated using a sampling method.  Without an understanding of the sampling distribution, a districting plan being ``typical'' or being an ``outlier'' holds little means.

\subsection{Generating Plans from Plans}\label{sec:random_step}

A different sampling strategy generates a random plan by editing an existing plan, rather than starting from a blank slate.  This strategy could be \emph{perturbative}, for example generating a new plan by wiggling the boundaries of an old plan, or it could take larger steps, for example merging several districts in an existing plan and then redistricting that region using a method from Section~\ref{sec:random_scratch}.  In either case, we are likely to ``see'' part of the previous plan in the edited plan, but if we repeat this process enough times in a \emph{random walk}, this \emph{autocorrelation} decreases:  after many steps, we should 
see a plan that has little in common with the initial one.

There are many reasons why random walks might be preferable to generating a completely new plan for each sample.  As we have discussed, the space of potential plans is huge and includes many undesirable examples.  If we find a good plan, we might want to see if we can edit it to find others (or to make it even better) rather than obliterating it.  
Furthermore, editing steps are often more efficient to implement than generating a plan from scratch. On the other hand, suppose each time we generate a new plan, we do so by making a tiny perturbation in the boundary of a plan we have already generated (cf.\ the ``flip'' method described below).  Then, we will need many, many steps of this random walk before the plans we generate look significantly different from one another.  This ``explore--exploit'' trade-off---between the potential for large \emph{exploring} steps to find a completely new, effective plan and the potential for small perturbative changes, \emph{exploiting} a good plan to make it better---is a typical one in randomized search. 

To introduce some terminology, the new plan generated by editing an existing plan corresponds to a \emph{step} of the algorithm.  The set of plans that can be generated in one step from some Plan A define the \emph{neighbors} of Plan A in the state space.   A \emph{random walk} is a process for taking a sequence of these steps, where at each step a neighboring plan is generated and chosen to be the next plan from which to continue.  Random walk algorithms are motivated and discussed in more detail in DeFord and Duchin's chapter in \cite{book}.  Here we briefly describe them mostly as points of comparison to other district generation methods.

\subsubsection{Flip and Recombination}\label{sec:flip}\label{sec:recom}

Perhaps the most minimalistic way  to generate a new plan from an existing plan is to change the district assignment of a single unit at a time.  For instance, if a census block in District A lies on the boundary between District A and District B, changing the assignment of this block so that it is in District B results in a slightly different districting plan.  We proceed if this simple edit known as \emph{flipping} preserves the key property that Districts A and B are contiguous.

\begin{algorithm} 
  \begin{algorithmic}[1]
  \State $\textrm{flip unit} \gets$ \textsc{Random}$(\textrm{boundary unit})$
  \State $\textrm{district $i$} \gets$ $\textrm{District assignment}(\textrm{flip unit})$
  \State $\textrm{district $j$} \gets \textrm{District assignment}(\textrm{neighboring unit not in district $i$})$
  \If{reassigning flip unit from $i$ to $j$ results in a contiguous and population-balanced plan}
  \State $\textrm{District assignment}(\textrm{flip unit}) \gets \textrm{district $j$}$
  \EndIf
  \end{algorithmic} 
  \caption{Basic Flip Step}
  \label{alg:flipalgorithm}
\end{algorithm}

Flip-step algorithms change the assignments of randomly chosen units along district boundaries; they are used widely in the redistricting literature \cite{nagel1965simplified,browdy1990simulated,fifield2015new,ricca2008local,oloughlin1982}.  Algorithm \ref{alg:flipalgorithm} details a single flip step, illustrated in Figure \ref{fig:flipstepexample}.  Specifically, a candidate flip unit is chosen randomly from geographic units on the boundaries of two or more districts.  The assignment of this candidate unit is then \emph{flipped} from its current district to that of a neighboring district, unless doing so results in an invalid plan (e.g., districts becoming discontiguous or resulting population deviation larger than allowed).  A single flip step creates a new plan that is nearly---but not completely---identical to the previous plan, so typically this step is iterated many times (up to millions or billions) to generate the next plan in a random walk.  

Modifications to the flip step can be made to anticipate and fix issues that appear with the naive version.  For instance, to promote plans with balanced population, \cite{nagel1965simplified} proposes swapping the assignments of two units on either side of a shared boundary, rather than just flipping a unit from one side to another.  We can call this a \emph{swap} step.  To take larger coherent steps than a single-unit flip, some techniques~\cite{fifield2015new} flip clusters of contiguous units along the same boundary.  Introducing a probabilistic weighting can promote compactness or any other desired priority.

Another method of stepping from one one valid plan to a neighboring valid plan is to merge two or more neighboring districts and re-partition them into new districts, keeping the rest of the plan the same; see Figure~\ref{fig:recom_example} for an example.  
This strategy has been named \emph{recombination} in recent work (see \cite{deford2019recombination} for a survey).

\subsection{Comparing Samplers}\label{sec:compare}

We conclude this section by comparing several different sampling methods for generating four-district plans from the $10 \times 10$ grid and the Iowa counties.  
Again, as we did for the flood fill variants, we compare the distributions of both the number of cut edges (Figures~\ref{fig:10_violin}-\ref{fig:iowa_violin}) and location of where the cut edges occur most frequently (Figures~\ref{fig:flood_fill_10grid_exp}-\ref{fig:flood_fill_iowa_exp}).  
This time, since the problem is too big for complete enumeration, we use our samplers:  standard flood fill,  bounding box flood fill, a random walk using flip steps, a random walk using recombination steps, iterative merging, and a power diagram method (described below in Section~\ref{sec:geometric}).\footnote{As we will see below, power diagrams are \emph{geometric} partitions in which district boundaries may split census units. We assign each of these split units to the district with the most populous share of the unit, rejecting any plan that fails to maintain contiguity and population balance.  This last method helps illustrate the blurry line between sampling and optimization.}

For both the grid and Iowa, the samples generated by flip-step walks have substantially more cut edges than those generated by the other variants.  The samplers that tend to generate more compact plans also tend to have cut edges more highly concentrated closer to the center and very few cut edges near the corners and boundary.  Interestingly, the edges near the higher populated Iowa counties are also substantially more frequently cut than the less populated counties (see Figure~\ref{fig:iowa_population_map}).  Again we see that for both the $10 \times 10$ grid and Iowa, the bounding-box flood fill variant tends to generate plans that have fewer cut edges than the standard flood fill variant.

Though samples derived from recombination walks, iterative merging, and power diagrams tend to have more compact plans (fewer cut edges) than the other variants, the power diagram samples in particular contain only a small number of \emph{distinct} solutions (for the grid, there was only \emph{one} unique solution in a sample of 100,000).  This illustrates that even algorithms with some randomness might not generate a particularly diverse, let alone \emph{representative}, sample.

\section{Seeking ``Best'' Plans: Optimization}\label{sec:optimization}

Perhaps the most obvious application of computation to redistricting involves \emph{optimization} of districting plans.  Optimization algorithms \emph{extremize} (maximize or minimize) an \emph{objective function} while satisfying some set of \emph{constraints}.  In the context of redistricting, an optimization algorithm might be designed to maximize measures such as the Polsby--Popper or Reock compactness scores (see \cite{polsby-popper} and \cite{reock}) or the number of competitive districts, or to minimize measures such as population deviation, the number of county splits, or the number of cut edges.  The objective function might be one of these single measures, or it can be a composite that combines several.  Accompanying the objective function, the constraints likely include legal requirements, such as contiguity and population balance. Other requirements such as VRA compliance can be difficult to express as formal mathematical constraints.

Unfortunately, redistricting optimization problems are not easily solved in practice.  
Nearly any way of phrasing optimization for redistricting suggests that it belongs to a class of problems called \emph{NP-hard}. 
For this reason, we should not expect optimization to extract the \emph{best possible} solution to a redistricting problem. 

Complexity limitations have been known throughout the history of  algorithmic redistricting:  In 1965~\cite{nagel1965simplified}, Nagel acknowledged that their proposed algorithm ``will not guarantee that the criterion is as low as mathematically possible, though it should be low enough to satisfy the political and judicial powers that be.'' 

A reasonable expectation of optimization algorithms is that they can identify \emph{good enough} plans and make improvements to proposed plans.  With this looser goal in mind, in this section we describe several algorithmic optimization techniques that have been proposed for redistricting.

\subsection{Clustering and Voronoi Approaches}\label{sec:geometric}

Most methods in this chapter construct plans by assigning labels to a fixed set of census or other geographical units.  A different class of methods for sampling and designing plans ignores these unit boundaries in favor of drawing lines directly on a map of the underlying geography.  These \emph{geometric} methods typically lead to plans with compact boundaries, although sometimes this compactness comes at the cost of other redistricting criteria that are harder to express geometrically.

Note that while there is a huge-but-finite number of plans when we construct them out of census units, there is an \emph{infinite} number of ways we can draw geometric dividing lines on a map.  For this reason, we cannot expect these algorithms to have a positive probability of generating every possible plan.  Rather, they often make local decisions intended to promote generation of attractive plans (e.g., that the boundaries between districts must be straight lines) but might assemble these local decisions in a stochastic fashion.

A central example of a geometric approach to redistricting draws inspiration from another part of computer science. A common task in statistics and machine learning involves \emph{clustering} data based on proximity. 
Clustering methods are often geometric in nature: their job is to find the dividing lines between different groups of data points.  Along these lines we can cluster units on a map into districts based on geographic and other considerations.

For example, a ``splitline" algorithm~\cite{splitline} repeatedly divides regions into two subregions, starting with the entire state and ending with districts, at each step identifying a line that evenly splits the population (see Figure~\ref{fig:splitline}). The result is a fairly compact districting plan.  
A sampling variant might randomly draw from the set of valid splitlines in each bipartitioning step, while an optimization variant might choose the shortest splitline each time.

One of the simplest and most common approaches to clustering uses \emph{Voronoi diagrams}, such as the ones illustrated in Figure~\ref{fig:Voronoi_diagrams}.  In a Voronoi diagram, we choose a set of $k$ points on the map to be district \emph{hubs}.  Then, the map is divided into regions based on proximity:  The \emph{Voronoi cell} associated with a particular district hub is the set of points on the map closer to that hub than to any other hub.   That means that the cells are built for efficiency in distance terms, which obviously promotes compactness.  This in addition to straight-line boundaries
and convex shapes has made the Voronoi approach to redistricting attractive to several teams of researchers \cite{dobrin2005review,svec2007applying,miller2007problem,ricca2008local,novaes2009solving,levin2019automated}.  

Note that the Voronoi process is only deterministic after the location of the hubs has been fixed (see Figure~\ref{fig:Voronoi_diagrams});  a reasonable question to ask when using Voronoi diagrams for redistricting is where to place the $k$ hubs to optimize the quality of the diagram. 
The most popular formulation is called a \emph{$k$-means problem}, seeking to place district hubs to minimize the average squared distance between a resident and their district's hub~\cite{guest2019gerrymandering, altman2011bard,cohen2018balanced}. 
A simple and often effective algorithm for $k$-means is \emph{Lloyd's algorithm} \cite{forgy1965cluster,lloyd1982least}, detailed in Algorithm \ref{alg:kmeansalgorithm}, which alternates between moving each district hub to the average location of that district's residents (the \emph{centroid}) and drawing new districts with those hubs, then iterating until this process converges, i.e., the changes get arbitrarily small.

\begin{algorithm} 
  \begin{algorithmic}[1]
   \State $\textrm{Identify $k$ initial district hubs: hub $1$, hub $2$, \, ..., \, hub $k$ (also called \emph{means}).}$
   \While{$\textrm{process has not yet converged}$} 
   \For{each geometric point $i$}
	    \State $\textrm{District assignment}(\textrm{point }i)\gets \textrm{District assignment}(\textrm{hub closest to point $i$})$
	\EndFor
	\For{each district $j$}
	    \State $\textrm{hub $j$}\gets \textrm{Centroid(district $j$)}$
	\EndFor
    \EndWhile
  \end{algorithmic} 
  \caption{Lloyd's $k$-Means Algorithm}
  \label{alg:kmeansalgorithm}
\end{algorithm}

Optimal placement for the hubs is famously NP-hard, or likely to be computationally intractable, and Lloyd's algorithm often converges on a {\em local optimum} (the best in its neighborhood) rather than a \emph{global optimum}.

There are several fundamental challenges for Voronoi-type algorithms in this setting: first, we must decide on a notion of distance. Should we measure distance to the nearest hub based on geographic distance, travel time, or something else?  Moreover, these Voronoi diagrams have lines that cut across building blocks like census blocks and precincts.  
Finally, the algorithm so far is completely targeted to distance minimization and does not balance population.  

A few of the issues above can be addressed.  For example,  \emph{power diagrams} are generalizations of Voronoi diagrams in which each hub also has an associated weight~\cite{fryer2011measuring,cohen2018balanced}.\footnote{The power diagram cell for a hub $h$ with weight $w_h$ and distance function $d$ is the set of points $x$ such that $d(x,h)^2-w_h\leq d(x,h')^2-w_{h'}$ for every other hub $h'\neq h$.}  In one power diagram implementation based on a modified Lloyd's algorithm, Cohen-Addad, Klein, and Young \cite{cohen2018balanced} construct districting plans that are population-balanced and compact.  
Moreover, all of the cells are convex polygons, and they have at most six sides on average. 

Although the districts induced by power diagram cells are balanced and compact, they still face the issue of units.
To achieve population balance, the polygons often split units through their centroids, where their entire population is assumed to be located.
This means that assigning census units based on these idealized polygon districts is not straightforward.  Modifying these plans to respect unit boundaries may ultimately require sacrificing compactness, population balance, and even contiguity. We discuss refinement issues more generally in  Section~\ref{sec:refine}.

\subsection{Metaheuristics and  Random Walk Variants}\label{sec:metaheuristics}

When perfect optimization is elusive, computer scientists often turn to \emph{heuristics}, which accept approximate or local solutions instead of exact or global solutions.
In practice, a good heuristic can often identify strong solutions quickly.  While some heuristics are specialized to redistricting, \emph{metaheuristics} are general strategies 
that can be applied ``out of the box" to optimization problems drawn from many different domains. Various computational redistricting methods have adapted well-known metaheuristic algorithms to map-drawing.

Many common metaheuristics employ random walks of the kind discussed in Section~\ref{sec:random_step}, which explore the \emph{state space} (the set of all valid plans) by starting with some plan and making an edit.  Since we begin with a plan and compare it to neighbors, we can also call this strategy a \emph{local search}.  A basic local search algorithm known as \emph{hill climbing} is illustrated in Figure~\ref{fig:optima}.  In each step, the algorithm considers  replacing the current plan with a proposed neighbor, and proceeds with the replacement if the new plan has a better score.  
For example, the neighborhood of a plan A might be composed of all plans that can be created from Plan A by a single flip step, swap step, or recombination step (see Section~\ref{sec:random_step}).
In addition, we now need an objective function and a rule for determining acceptance of each neighbor based on its score.

Careful engineering is required to design an effective local search algorithm.  If the objective functions take a lot of time to evaluate or if the algorithm has to evaluate many neighboring plans at each step, it can take a lot of time to carry out a single step of local search.  On the other hand, if plans do not have many neighbors, it could take many steps before a poorly-performing plan is improved to an acceptable level.

Several local-search variants have been used to design districting plans:
\begin{itemize}
\item 
\textbf{Hill climbing}, as already mentioned, only accepts improvements until it reaches a plan whose neighbors are all worse, which is necessarily a local optimum. (This is used in \cite{altman2011bard, ricca2008local,kim2011optimization}) Hill climbing often gets stuck in local optima, as illustrated in Figure \ref{fig:optima}.  
To improve the likelihood of success, hill climbing algorithms often call a sampler to draw many different random starting plans and restart the process several times, keeping the best-performing local optimum among the different runs.  
\item 
\textbf{Simulated annealing} (used in \cite{altman2011bard,fifield2015new,ricca2008local,browdy1990simulated, jacobs2018partial})  attempts to avoid local optima by sometimes allowing moves to worse scores.  Inspired by certain physical processes, this stochastic algorithm maintains an additional  \emph{temperature} parameter that starts ``hot" (high parameter) and ``cools" (decreasing the parameter) slowly over the course of successive steps.  The probability of transitioning to a worse neighboring plan is controlled by the temperature: while the temperature is hot, worse plans are accepted and the algorithm can explore the state space; and as the temperature cools, hill climbing kicks in and the algorithm can improve to a high-quality plan.
\item
\textbf{Tabu search} (used in \cite{altman2011bard,bozkaya2003tabu,ricca2008local,jin2017spatial}) keeps a memory of the plans that it has recently visited and avoids returning to these already-visited plans.  This strategy encourages broader exploration of the state space by preferring unvisited neighbors.  
\item 
\textbf{Evolutionary algorithms} (used in \cite{altman2011bard, liu2016pear}) draw inspiration from biology to quickly create a diverse collection of plans. In this technique, a \emph{population} of plans evolves over the course of the algorithm by a combination \emph{mutating}, or taking a basic random walk step, and \emph{crossover}, which combines two 
plans in a more drastic move to generate one or more \emph{child} plans with traits of both \emph{parents}.
\end{itemize}

The crossover method that we implemented in our evolutionary algorithm, depicted in Figure~\ref{fig:evol_crossover}, is based on the approach in \cite{liu2016pear}.  Two parent plans are drawn from the population and their \emph{common refinement} (the regions resulting from overlaying the two parent plans) is calculated, as in Figure~\ref{fig:evol_refine}.  The common refinement likely has many more regions than the desired number of districts, so the next step is to merge these smaller regions together (using a similar method as~\cite{chen2013unintentional,chen2016loser}) until there is a correct number of districts (see Figure~\ref{fig:evol_merge}).  At this point the districts may have unbalanced populations, so the final step is to make small adjustments to balance out the the population (see Figure~\ref{fig:evol_adjust}).  

\subsubsection{Comparing Metaheuristics}

To offer a coarse comparison of these optimization methods, we will apply several of them to seek the minimum number of cut edges in an Iowa congressional districting plan.  We will apply hill climbing, simulated annealing, and an evolutionary algorithm together with a flip step random walk.  In each case we will allow a population deviation of up to 5\% from ideal district size.   In Figure~\ref{fig:opt_combined} we show the outcomes of running the same suite of metaheuristic algorithms with the same starting plan two different times.

Some details of these comparisons are found in the figure captions, but here are a few themes.  Hill climbing quickly (within a few hundred steps) finds a local minimum in each run, while simulated annealing fluctuates but eventually outperforms the strictly greedy method.  One risk with simulated annealing is that it may pass up a promising solution early in hopes of finding something better, only to end at a poor-quality solution.

To illustrate the evolutionary algorithm, we show the maximum score (light green) and minimum score (dark green) over the population of ten plans at each step.  (The same starting plan from hill climbing and simulated annealing is included in the starting population for the evolutionary runs.)

The relative performance of these metaheuristic approaches, however, depends heavily on user choices. They could always have been run for longer, or cleverly implemented and tuned.  Nonetheless we hope to have illustrated some of the issues and tradeoffs with basic implementations.
In these short and untuned runs, each method identified plans with a small number of cut edges, but none of them found the global minimum.  Figure~\ref{fig:five_pct_iowa_opt} shows an example of a plan with only 29 cut edges and less than 5\% population deviation, which we will show below to be the true minimum.

Metaheuristics are largely agnostic to the particular objective function in a redistricting problem.  This is both a positive and negative aspect of these algorithms:  On the one hand, they are easily adapted to the particularities of a given state or set of district criteria, but on the other hand they are unable to leverage structure in a specific objective function that might make optimization faster.

\subsection{Integer Programming}\label{sec:int_prog}

Somewhere between applied mathematics and computer science is the discipline of {\bf operations research}, which is built around approaches to difficult optimization problems, from maximizing profit while satisfying demand to scheduling tasks on a computer to maximize throughput.  Starting from these problems as central examples, these fields have developed a taxonomy to classify optimization problems based on their objective functions and constraints.  Once we recognize a problem within that taxonomy, we can leverage appropriate general-purpose strategies that solve similar problems efficiently.

Within this taxonomy, many redistricting problems can be understood as \emph{integer programs}, 
which can be written in the following form:
$$
\begin{array}{rl}
\mathrm{minimize}_x\ & f(x)\\
\textrm{subject to}\ & g(x)\geq0\\
& x \in\mathbb Z^n.
\end{array}
$$
Here, $x$ denotes the set of \emph{decision variables} used to encode districting plans.  The function $f(x)$ gives the value of the objective function evaluated at $x$.  The function $g(x)$ encapsulates constraints on $x$ (e.g., that the population of each district must not deviate substantially from the ideal); we can think of $g(x)$ as \emph{vector-valued}, meaning that we can enforce more than one constraint at a time.  Finally, the constraint $x\in\mathbb Z^n$ is mathematical notation denoting that $x$ is a tuple of $n$ \emph{integers}; that is, the unknown in an integer problem is a list of $n$ numbers without fractional parts.  The notation above is extremely generic: Nearly any computational problem whose output consists of integers can be written in this form.\footnote{In a standard trick, an inequality constraint like $2x\le 5$ can be written as $-2x+5\ge 0$ and an equality constraint like $x=10$ can be written as $x\ge 10, -x\ge -10$.  We can put all three together in vector form as $(-2x+5,x,-x)\ge (0,10,-10)$.
This lets us use $g(x)\geq 0$ as a canonical form for systems of equalities and inequalities.}

Not unlike the metaheuristics in Section~\ref{sec:metaheuristics}, algorithms for solving integer programs attempt to cut down the state space of possible solutions to a manageable size by using the constraints and bounding the objective in various ways.  For example, a common strategy is \emph{relaxation}, whereby constraints are removed from an integer program to make it easier; if one of the relaxed constraints is violated in the resulting solution, it is added back to the integer program and solving is restarted~\cite{hojati1996optimal}.  Similarly, \emph{branch-and-bound} algorithms might drop the integer constraint, resulting in a much easier (and typically convex) problem to solve algorithmically as well as a bound on the best possible objective value; then, various variables are pinned to nearby integers until the solution satisfies all the constraints~\cite{buchanandraft,mehrotra1998optimization}.

Integer programming algorithms and local search metaheuristics are similar in the sense that both navigate the space of feasible solutions while looking to improve an objective function.  The main difference is that integer programming algorithms typically (but not always) aim to extract a \emph{global} optimum via carefully-designed bounding and search strategies.  That is, the heuristics used in integer programming are generally
conservative, ordering potential $x$ values to try based on their likelihood of solving the integer program but never throwing one away until it can safely be ruled out. 
For this reason, we can be confident in the output of integer programming tools, but they can take an extremely long time to terminate.

Integer programming formulations have a long history in  redistricting, going back again to the 1960s and the work of Hess, Weaver, and collaborators~\cite{hess1965nonpartisan,buchanandraft,birge1983redistricting,mehrotra1998optimization,norman2003kentucky,caro2004school,borodin2018big,jacobs2018partial,hojati1996optimal, cohen2018balanced}.  Below, we give an example of how one could phrase redistricting as an integer program.

\sidebar{Aside: Districting as an Integer Program}
{Suppose we wish to design $d$ districts in a state with $n$ census units; our task is to assign each  unit to one of the $d$ districts.  We can introduce a binary variable $x_{ij}$ for each census unit $i$ (ranging from $1$ to $n$) and district $j$ (ranging from $1$ to $d$).  We interpret $x_{ij}$ as follows:
$$
x_{ij} = \left\{
\begin{array}{ll}
1 & \textrm{ if  unit $i$ is in district $j$}\\
0 & \textrm{ if  unit $i$ is \emph{not} in district $j$.}
\end{array}
\right.
$$
Our goal is to assign each $x_{ij}$ a value of $0$ or $1$ in such a way that satisfies certain constraints and corresponds to the best possible plan for a certain objective function.

We have to add several constraints to make sure that $x$ is reasonable.  Each variable $x_{ij}$ always takes on one of two values, $0$ or $1$.  These are integers, but to avoid nonsensical outputs like $x_{ij}=5$ we additionally constrain $0\leq x_{ij}\leq 1$ for all units $i$ and districts $j$.   
Similarly, we want to make sure to assign each unit to exactly one district.  Consider a single unit $i$.  If we sum the $x_{ij}$ values for all districts $j$, we have computed the number of districts to which unit $i$ was assigned. For example, if there are four districts, then $x_{i1}$+$x_{i2}$+$x_{i3}$+$x_{i4}$ equals the number of districts to which unit $i$ is assigned. 
Hence, we need to enforce the constraint
$$\sum_{j=1}^d x_{ij} = 1 \textrm{,  for all units }i.$$

Next, suppose we want to ensure that every district has population between a lower bound $\ell$ and an upper bound $u$.  We write  $p_i$ for the population of unit $i$; for example, $p_{8}=100,000$ means that there are 100,000 people in unit 8.  Given the variables $x_{ij}$ above, this means that 
$\sum_{i=1}^n p_i x_{ij}$ records the population of district $j$.
(Since $x_{ij}$ has a value of zero if unit $i$ is not in district $j$, the sum excludes the populations of units $i$ that are not assigned to district $j$, leaving behind just the relevant populations that are used to construct the district.)  This gives
$$\ell \leq \sum_{i=1}^n p_i x_{ij} \leq u \textrm{,  for all districts }j.$$

Finally we need to design an objective function.  There are many possible objective functions that are relevant to redistricting, such as minimizing the sum of distances between voters and their district's center~\cite{hess1965nonpartisan,mehrotra1998optimization, buchanandraft}, minimizing the number of counties that are split into different districts~\cite{birge1983redistricting}, or optimizing a measure of compactness~\cite{jacobs2018partial}.  In our example, we minimize the number of cut edges by introducing another variable $x'$, with components $x'_{ab}$ for each pair of adjacent units $a$ and $b$.  These $x'$ variables will encode whether or not a given edge is a cut edge in the assignment:
$$
x'_{ab} = \left\{
\begin{array}{ll}
1 & \textrm{ if units $a$ and $b$ are in different districts}\\
0 & \textrm{ if units $a$ and $b$ are in the same district.}
\end{array}
\right.
$$
The number of cut edges in a plan can then be written as the sum $\sum_{ab}x'_{ab}$, and the objective is to minimize this sum.

These new $x'$ variables require additional constraints.  They too must be constrained to take on integer values between 0 and 1.  We also need to ensure that the $x'_{ab}$ can only take a value of 0 if $a$ and $b$ are \emph{actually} assigned to the same district; otherwise the algorithm would assign 0 to all of the $x'$ to achieve a minimum objective value of zero.  That is, there must be some district $j$ such that $x_{aj} = 1$ \emph{and} $x_{bj} = 1$ for $x'_{ab}$ to take a value of 0, and otherwise the constraints must require $x_{aj}$ to take a value of (at least) 1.  We achieve this by constraining
$$
\left.
\begin{array}{ll}
x'_{ab} \geq x_{aj}-x_{bj}\\
x'_{ab} \geq x_{bj}-x_{aj}
\end{array}
\right\} \textrm{ for all adjacent units } a \textrm{ and } b \textrm{ and all districts }j
$$
These constraints ensure that $x'_{ab}$ can only equal 0 (reflecting that the edge from $a$ to $b$ is not cut, which happens when $a$ and $b$ are in the same district) if $x_{aj}=x_{bj}$ for all districts $j$.  
(For instance if $x_{a1}=x_{b1}=x_{a3}=x_{b3}=x_{a4}=x_{b4}=0$ while 
$x_{a2}=x_{b2}=1$, this records that $a$ and $b$ are both in district 2.)
If $x_{aj}\neq x_{bj}$ for some $j$, then the inequalities force $x'_{ab}$ to be at least one (indicating a cut edge).

Letting $m$ denote the number of pairs of adjacent units, we can put all these pieces together, arriving at an integer program:
$$
\begin{array}{rll}
\textrm{minimize}_{x'}\ & \sum_{ab}x'_{ab}\\
\textrm{subject to}\ & 0\leq x_{ij} \leq1 & \textrm{ for all units $i$ and districts $j$},\\
& 0\leq x'_{ab} \leq1 & \textrm{ for all adjacent units $a$ and $b$},\\
& \sum_j x_{ij} = 1 & \textrm{ for all units }i,\\
& \ell \leq \sum_i p_ix_{ij}\leq u & \textrm{ for all districts }j,\\
& x'_{ab} \geq x_{aj}-x_{bj} &  \textrm{ for all adjacent units } a \textrm{ and } b \textrm{ and districts }j,\\
& x'_{ab} \geq x_{bj}-x_{aj} & \textrm{ for all adjacent units } a \textrm{ and } b \textrm{ and districts }j,\\
& x\in\mathbb Z^{n\times d}\\
& x'\in\mathbb Z^{m}.
\end{array}
$$
This problem is nothing more than careful, unambiguous mathematical notation for our map-partitioning problem.  Once our problem is written in this form, it can be handed over to powerful \emph{solvers}, pieces of software designed to tackle problems in a specific form efficiently.
}

There are many properties to notice about the problem above. In simple notation, we are able to capture many of the demands of a redistricting problem in a fashion that is easy to communicate to a computer.  More importantly, the objective and constraints are \emph{linear}, a valuable property that can help integer programming algorithms succeed. 
Once expressed as an integer program, our problem can be run through an integer programming solver, software specifically designed to optimize these instances. Many such solvers are available commercially and open source. These solvers tell you when they definitively identify global optima, but 
they may take an extremely long time to do so. 

Our example neglects important criteria for districting plans, some of which are difficult or cumbersome to express in the formalism above.  High up on that list is district contiguity.  
But something is working in our favor here: although contiguity is not explicitly enforced in the constraints of our integer program, because we minimize cut edges, the optimal plans identified by the program tend to be  contiguous.  That is, districts that are split into several parts usually have more cut edges than contiguous districts and so are unlikely to be identified as optimal.  
In Figure~\ref{fig:integer_program_cut_edges}, we see four example outputs of the above integer program using counties as building block units for two different values of population deviation allowance and two different states.  
Three of the identified plans are contiguous and the fourth has only one discontiguous district.  The idea that we dropped the difficult contiguity constraint and got contiguous districts anyway is a successful \emph{relaxation}, which we discuss in general terms below.

When dealing with multiple scores in an optimization framework, one solution $x$ is said to \emph{dominate} another solution $x'$ if it $x$ is  at least as good as $x'$ in each score. In our setting, we can say that plan $P_1$ dominates $P_2$ if it has no more cut edges \emph{and} no greater population deviation.  The plans identified by our integer program lie on the \emph{Pareto frontier}, the set of solutions which are not dominated by any other solutions.   \emph{All} other plans must lie above the curve formed by this frontier.

Notably, we found that the global minimum in Iowa at $\le 5\%$ deviation is 29 cut edges, which beats the local minima identified by metaheuristic runs in Section~\ref{sec:metaheuristics}. 
But beware of two major caveats:  first, Iowa is much smaller than the typical redistricting problem in combinatorial terms, and this program would not have run to completion on any state's precincts or census blocks.  Second, the choice of cut edges as an objective function gave us contiguity more or less "for free."  Explicit contiguity constraints have been included in some approaches~\cite{oehrlein2017cutting, buchanandraft} so that they can optimize other objective functions, at the cost of increased size and complexity for the integer program.

\subsection{Relaxation and Refinement}\label{sec:refine}

We have seen several examples where it is strategic to "relax" constraints, i.e., temporarily loosen them or drop them altogether.  Relaxations can make a difficult problem more tractable, and solutions to these relaxed problems can then be refined, if needed, to make them valid.  Alternatively, these refinement steps can occur at intermediary stages throughout the algorithm.

Given a solution that has been produced by an optimization procedure,
we can refine it in several ways:  passing to discrete units, passing from coarser to finer units, or exchanging units to better meet some goals 
\cite{liu2016pear,chen2013unintentional,chen2016loser,magleby2018new}.
The most common refinement strategy employs local search methods such as flip and swap steps (Section~\ref{sec:random_step}); see \cite{levin2019automated} for an example.  Indeed, many metaheuristic local search methods can be thought of as iterative refinement.

For a coarse-to-fine example, integer programming might be too slow to run at the census block level, so we can first optimize for a plan at the census tract level (with bigger pieces) and then try to break down a small number of tracts to tune the solution to better population balance at the block level.
For a discretization example, the geometric methods in Section~\ref{sec:geometric}  generate districts as polygonal shapes.  To transform these plans into partitions of the census units, \cite{cohen2018balanced} assigns the divided blocks to one of the neighboring districts with attention to population deviation (see Figure~\ref{fig:refine_geometry}).  

Refinements can occur as mid-course adjustments. Consider that agglomerative methods (Section~\ref{sec:flood_fill}) often fail because they build partial plans that have no connected, population-balanced completions.  Instead of restarting the process each time this problem is encountered, a refined procedure can backtrack from near a dead end to look for a sequence of choices that can be completed.  Another example is found in \cite{jacobs2018partial}, where a population-balancing auction is run 
at every iteration of their energy-minimizing algorithm.

Refinements help ameliorate high rejection rates.  There is no guarantee, however, that small refinements can repair invalid plans or correct course effectively in mid-run, and the repair time may end up being longer than the time it takes to run the initial algorithm repeatedly.

\section{Conclusion: The Future of Redistricting Algorithms}

Algorithmic tools are already extremely useful in redistricting.  They can identify promising plans and refine proposed plans.  They can generate samples of many thousands or millions of plans to help contextualize a candidate plan.  

The active field of redistricting algorithm design continues to produce improved techniques, and mapmakers and analysts can make swift use of these advances.  But as with all powerful tools, it is important to understand the limitations of redistricting algorithms in order to use them effectively and responsibly.  In this section we discuss several of these limitations.

In the early days of optimistic outlooks on computational redistricting, it looked like rapid advances in computing would soon overcome the difficulties in the problem.  In 1963, Weaver and Hess \cite{weaver1963procedure} wrote:
\begin{quote}
	No available programs or computer techniques are known which will give a single, best answer to the districting problem, though such a solution seems possible if enough funds and efforts are put to the problem, especially considering the rapid advances in size and sophistication of available computers.
\end{quote}

Real-world redistricting problems, however, are likely to be {\em forever} too complex for computers to solve optimally. Even if there were consensus as to what makes one plan better than another, not only does computational intractability  limit our ability to identify the best solution in current practice, but the underlying reasons might keep that prospect out of reach despite advances in computing.  Instead, we must settle for plans that might be far from optimal.

There are roles both for humans and for computers in redistricting.  Algorithms can efficiently produce potential plans, evaluate their properties, and suggest new ways to divide up a region, but they are limited by the accuracy of the input data, tractability issues, and fidelity of the computational model to the realities of redistricting.  Humans also are subject to the same tractability issues but have better understanding of the populations affected by potential districting plans and the assorted criteria at play when designing voting districts.  Hence, a key piece of the puzzle is how to mediate the relationship between human and machine.  
Subtle issues are at play when designing redistricting systems that citizens and legislators will trust---but not trust \emph{too} much.

\subsection{Abuse and Gaming}
Computers and algorithms do not remove humans from redistricting and therefore do not remove human bias and error from the process.   To the contrary, algorithms sometimes \emph{amplify} human bias, whether intentionally (e.g., an optimization algorithm can be used to maximize the number of seats for a political party) or unintentionally.  One risk of putting unfounded trust in computer-identified plans is that actors can hide their bias behind the justification of a seemingly neutral process.  If the algorithm generates a random plan, a user can repeatedly re-run it until it yields a favorable result.  Similarly, if the output depends on the starting point, then a user can re-run the algorithm from different starting points, looking for a favorable answer.  The more user choices available in an algorithm, the more opportunities to turn the knobs to try to control the answer.  The user can disingenuously defend the  cherry-picked outcome as being neutrally generated by a computer.

Some argue that the opportunities to game the rules would be avoided by the use of optimization: if the \emph{best} plan is mandated, then a user does not have an opportunity to advance an agenda.  Beyond the usual problem of finding some common notion of \emph{best}, there remains the possibility that many dissimilar plans may earn indistinguishably good scores.
An agenda-driven user then can just choose their favorite plan among those tied for best.  

Before placing value on algorithmically-generated districting plans, it is crucial to understand the design decisions and underlying assumptions and simplifications of the algorithm and the effects that these factors have on the resulting outcomes.   It is important to be cautious with techniques that are not accompanied by an explanation of these decisions and, when possible, to replicate and perform sensitivity analyses on techniques before advocating for or building off of them.

\subsection{Best Practices and a Call to Action}
Far more lawyers,  legislators, and everyday citizens are capable of using redistricting software than writing it, which means they must place a degree of trust in developers.  A few concrete steps can begin to counteract this asymmetry in understanding of what is ``under the hood.''
\begin{itemize}
\item Expert consultants and other users of redistricting software should accompany their reports with code and detailed, unambiguous descriptions of the procedures used to arrive at their conclusions. This will help others assess potential bias in their analysis caused by distributional design, under-sampling, instability in computation, or the choice of heuristic.
\item Maps and other datasets used for analysis should be publicly available to make the analysis  reproducible.  
\item Academic and commercial tools for redistricting should be released under open source license to reinforce trust in redistricting procedures.
\end{itemize}

Redistricting problems are not only \emph{too big}, but also \emph{too human}, to be completely addressed by computers.  Algorithms require constraints and objectives to be precisely defined, but in real-world instances this is not straightforward.  Even seemingly simple constraints like contiguity and population balance are not always easy to define: How is geographic adjacency handled for islands and bays? How much population deviation is too much?  How do we define compactness? 
Abstract goals like preserving communities of interest and legal constraints like VRA compliance are nearly impossible to quantify precisely enough for a computer to operationalize.  Many of the legal, social, and political aspects of what makes a valid plan---let alone a good plan---are dependent on \emph{context} and subtleties that are better understood by humans than machines.  Given that there are real, human consequences of redistricting decisions, these complexities should not be entrusted to an algorithm alone. 

Computer technology, mathematical theory, and the political landscape continue to co-evolve.  As technology improves, it holds potential to make the tradeoffs involved in choosing districting plans more transparent, using a wide range of districting possibilities as a tool in assessment.  

Even though algorithmic techniques for redistricting have been used for decades, many of the mathematical and computational aspects of redistricting are not yet fully understood, and existing techniques have considerable room for improvement.  As these areas continue to grow, there are many opportunities for involvement, whether it's analyzing existing algorithmic techniques, replicating published findings, contributing to open-source projects, drawing strong-performing benchmark plans, or designing new redistricting algorithms and analysis tools.

\section*{Acknowledgements}
There are several people who supported the production of this chapter. Daryl DeFord provided coding support for our local search hill climbing and simulated annealing experiments in Section~\ref{sec:metaheuristics}.  Richard Barnes provided Figure~\ref{fig:power_diagrams} using a modification of the algorithms from \cite{cohen2018balanced}.  Phillip Klein provided Figure~\ref{fig:refine_geometry}.  We also acknowledge the generous support of the Prof.\ Amar G.\ Bose Research Grant and the Jonathan M.\ Tisch College of Civic Life. Amariah Becker acknowledges the support of the NSF Convergence Accelerator Grant No. OIA-1937095.

 
\end{document}